\documentclass[12pt]{article}
\usepackage{amsmath}
\usepackage{graphicx,psfrag,epsf, url, hyperref, geometry}
\usepackage{enumerate, tabularx, pdflscape}
\RequirePackage[super,sort&compress,merge]{natbib}
\usepackage{amsthm, tabularx, parnotes, fixmath}
\usepackage[autostyle=true]{csquotes}
\usepackage[english]{babel}
\usepackage{setspace}
\newtheorem{proposition}{Proposition}

\usepackage{booktabs}
\usepackage{threeparttable}

\usepackage{xcolor}
\usepackage{qting}

\usepackage{amssymb}
\usepackage{bm}
\usepackage{setspace}

\usepackage{xcolor}
\usepackage{fancyhdr}
\newcolumntype{C}[1]{>{\centering\let\newline\\\arraybackslash\hspace{0pt}}m{#1}}
\def\bSig\mathbf{\Sigma}

\newcommand{\bX}{\mathbf{X}}

\newtheorem{assumption}{Assumption}

\newcommand{\blind}{0}

\date{}

\begin{document}

\def\spacingset#1{\renewcommand{\baselinestretch}%
{#1}\small\normalsize} \spacingset{1}

\if0\blind
{
  \title{\LARGE\bf Nothing to See Here? A non-inferiority approach to parallel trends}
  \author{Alyssa Bilinski, PhD\thanks{
    Corresponding author: \href{mailto:alyssa_bilinski@brown.edu}{alyssa\_bilinski@brown.edu}. We thank Samantha Burn, David Cutler, Monica Farid, John Giardina, Joshua Kalla, Michael McWilliams, Arman Oganisian, Jonathan Roth, Joshua A.~Salomon, Pedro Sant'Anna, Kosali Simon, and Jos\'{e} Zubizaretta for helpful comments. This work was funded in part by the National Institute of General Medical Sciences (1R35GM155224, AB) and Agency for Healthcare Research and Quality (R01HS028985, LH). The content is solely the responsibility of the authors and does not necessarily represent the official views of funders.} \\
    Departments of Health Services, Policy, and Practice \& Biostatistics \\
    Brown University
    \vspace{10pt} \\ 
    Laura Hatfield, PhD \\
    Statistics and Data Science Department \\
    National Opinion Research Center}
  \maketitle
} \fi

\if1\blind
{
  \bigskip
  \bigskip
  \bigskip
  \begin{center}{\LARGE\bf Parallel Trends in an Unparalleled Pandemic \\ Difference-in-differences for infectious disease policy evaluation}
  \end{center}
  \medskip
} \fi

\begin{abstract}
Difference-in-differences is a popular method for observational health policy evaluation. It relies on a causal assumption that in the absence of intervention, treatment groups' outcomes would have evolved in parallel to those of comparison groups.
Researchers frequently look for parallel trends in the pre-intervention period to bolster confidence in this assumption. 
The popular ``parallel trends test" evaluates a null hypothesis of parallel trends and, failing to find evidence against the null, concludes that the assumption holds. 
This tightly controls the probability of falsely concluding that trends are not parallel but may have low power to detect non-parallel trends. 
When used as a screening step, it can also introduce bias in treatment effect estimates. 
We propose a non-inferiority/equivalence approach that tightly controls the probability of missing large violations of parallel trends measured on the scale of the treatment effect. 
Our framework nests several common use cases, including linear trend tests and event studies. We show that our approach may induce no or minimal bias when used as a screening step under commonly-assumed error structures, and absent violations, can offer a higher-power alternative to testing treatment effects in more flexible models. 
We illustrate our ideas by re-considering a study of the impact of the Affordable Care Act's dependent coverage provision.
\end{abstract}

\vfill

\newpage
\spacingset{1.5} 
\section{Introduction}
\label{s:intro}
Difference-in-differences (DiD) is a popular design for impact evaluation in observational settings with longitudinal data in fields ranging from health policy and economics to education and political science.\citep{ben-michael_trial_2021,lechner_estimation_2010,renson_identifying_2023,roth_whats_2023,tchetgen_universal_2023} 
To conduct causal inference with DiD, we assume that in the absence of an intervention, the treated groups' average potential outcomes would have evolved in parallel with those of the comparison groups (the ``parallel trends assumption''). 

Most authors employing DiD investigate whether outcomes were parallel prior to the intervention,
reasoning that parallel trends in the pre-intervention period increase confidence in the untestable counterfactual parallel trends assumption. \citep{ryan_now_2019} 
We reviewed all 51 DiD papers published in the \emph{Journal of the American Medical Association} (\emph{JAMA}) and \emph{JAMA Internal Medicine} during 2018-2022 and observed two common approaches to evaluating pre-intervention trends (Table S1). 
First, about half ($n=27, 53\%$) plotted outcomes in treatment and comparison groups over time and concluded that they looked parallel.
For instance, one paper noted, ``Visual inspection...revealed small, non-significant differences in pre-intervention episode spending for episodes at [treated] versus comparison hospitals.'' \citep{navathe_association_2018}

Second, following \emph{JAMA} guidance, many papers conducted statistical tests for parallel trends. \citep{dimick_methods_2014} 
The most common approach ($n=18, 35\%$) was to fit a linear regression model, test whether a pre-intervention linear time slope differed between treated and comparison groups, and if $p > 0.05$ for the test, conclude that trends were parallel.  
For instance, one paper stated, ``We directly examined for this possibility by fitting a model containing a treatment indicator, a continuous time variable, the interaction of these 2 variables, and all patient- and hospital-level covariates, restricted to the pre-regulation period...We considered parallel trends as being present if the interaction term from this model was not significant.'' \citep{kahn_association_2019} 

A related approach ($n=9, 18\%$) was to fit an ``event study'' regression with coefficients for the differential difference between treated and comparison units in each period relative to an omitted reference period. 
If trends are indeed parallel prior to the intervention, coefficients in the pre-period should all be close to zero and lack a trend.
Thus, authors typically evaluate the collection of point estimates and their 95\% point-wise confidence intervals.
For instance, one paper wrote, ``We tested the parallel trends assumption ... through event study analyses, which is recommended when evaluating health policies. ... Small and statistically non-significant estimates before adoption suggest that the parallel trends assumption was satisfied.''\citep{abouk_association_2019}

Although testing attempts to make evaluation of pre-intervention parallel trends more formal and systematic than visual inspection, there are two key problems with these conventional tests of pre-intervention trends.  
First, they test a null hypothesis of no difference, thus tightly controlling the probability of falsely declaring non-parallel trends (Type I error) but failing to control the probability of missing non-parallel trends (Type II error).  
Previous authors have argued that these tests are often under-powered, leading to unwarranted confidence in the parallel trends assumption by conflating low power with no violation of parallel trends.\citep{freyaldenhoven_pre-event_2019, kahn-lang_promise_2020}
Conversely, when the sample size is large, these tests may flag even trivial trend differences as statistically significant.\citep{freyaldenhoven_pre-event_2019, roth_pretest_2022}
Although DiD assumes that counterfactual trends would have been parallel, in practice, we may tolerate trend differences that are ``small enough,'' as determined by context-specific knowledge \citep{rambachan_more_2023}. 

Second, using a test for parallel trends as a screening step (i.e., to decide whether to present DiD results) can distort the treatment effect estimates in studies that pass such a test.
Previous work has noted that when trends are truly divergent, a testing step may disproportionately admit cases where random error makes the pre-period trend difference unusually small, thereby exacerbating bias in event study coefficients.\citep{roth_pretest_2022, daw_matching_2018}

{Our paper makes three main contributions to the literature on evaluating and communicating DiD's sensitivity to the parallel trends assumption. First, we propose a non-inferiority/equivalence approach to testing for parallel trends in the pre-intervention period.
This evaluates evidence against the null hypothesis that there are meaningful differences in trends, thus addressing the Type I/Type II error problem.\citep{blackwelder_proving_1982, wellek_testing_2010}
Other authors have proposed non-inferiority/equivalence approaches for event studies \citep{dette_testing_2024} and for balance and placebo tests in other designs outside of DiD. \citep{hartman_equivalence_2018,hartman_equivalence_2021} 
We present a general framework that shifts focus from measuring violations to measuring their impact on the treatment effect: comparing treatment effect estimates from a reduced model (which assumes parallel trends) and an expanded model (which allows for non-parallel trends). 
This accommodates a range of common expanded models, including both linear time trends\citep{mora_alternative_2019, tazhitdinova_difference--differences_2023, strezhnev_group-specific_2024} and event studies. \citep{dette_testing_2024, roth_pretest_2022} 
It also allows us to specify our non-inferiority/equivalence threshold on the scale of the treatment effect and bound potential bias in our treatment effect estimator. 
We therefore connect to other sensitivity analysis methods that use observed differences in pre-intervention outcome evolution to construct sensitivity bounds on treatment effect estimates.\citep{rambachan_more_2023, ye_negative_2022} 
Our approach also shares conceptual underpinnings with e-values, which ask, ``How big would a violation of the causal assumption need to be to meaningfully change my effect estimate?''\citep{vanderweele_sensitivity_2017} 

Second, we extend prior work by establishing conditions under which our testing procedure, if used as a screening step, might introduce bias. \citep{dette_testing_2024, roth_pretest_2022}
Leveraging past literature on model selection, \citep{clogg_statistical_1995, liu_assessing_2009} we show that with i.i.d. normal errors, unit-heteroskedastic normal errors, or time-invariant error correlation, our procedure does not add bias to reduced model treatment effect estimators.
Under other error structures (e.g., autocorrelation), we describe conditions under which test-induced bias is small in magnitude.

Third, we examine the power of our procedure and show that when trends are indeed parallel or nearly parallel, we have high power to pass a non-inferiority test when our threshold is anchored to a treatment effect for which the overall study is well-powered.}

The rest of the paper proceeds as follows. 
In Section~\ref{s:model}, we introduce our notation, target estimand, estimation procedure, and framework for evaluating pre-intervention parallel trends. 
We then present a non-inferiority/equivalence testing framework, formulated in terms of reduced and expanded models, and demonstrate how this nests several common approaches, including event study formulations.
In Section~\ref{s:power}, we characterize conditions under which our estimator introduces no or small bias if employed as a screening step
and consider the power of non-inferiority and equivalence tests to detect meaningful violations of parallel trends.
Section~\ref{s:simulations} demonstrates the performance of our approach in a simulation study.
Section~\ref{s:application} applies our ideas to re-analyze a study of the Affordable Care Act's effect on dependent insurance coverage rates of young people and presents an empirical simulation based on that application.
We conclude in Section~\ref{s:discuss} with a summary of our findings and recommendations for practice.

\section{Testing Framework}\label{s:model}

\subsection{DiD setup, assumptions, and estimator}
{We begin with a canonical DiD setup, in which we observe $i=1,\ldots,n$ units during time periods, $t=1,\ldots,T$, where units are in two groups: a treated group of $n_1$ units (i.e., the set $\mathcal{N}_1$) drawn from a treated population and a comparison group of $n_0$ units drawn from a comparison population ($\mathcal{N}_0$).}
At time $T_1$, an intervention begins for the treated group only.
Let $G_i = \mathbb{I}(i \in \mathcal{N}_1)$ be an indicator of the treated group.
Let $Y_{it}(d)$ denote the potential outcome for unit $i$ at time $t$ under treatment condition $d$, where $d=0$ indicates no treatment and $d=1$ indicates treatment. 
The distinction between the (actual) treatment group and (hypothetical) treatment condition enables us to consider counterfactual outcomes, 
such as the untreated potential outcome of a unit in the treated group during the post-intervention period, $Y_{it}(0) \mbox{ for } i \in \mathcal{N}_1, t \geq T_1$, where $t \geq T_1$ includes $\{T_1,\ldots,T\}$.
Let $y_{it}$ be the realized outcome for unit $i$ at time $t$. 

Our causal target quantity is {the average effect of treatment on the treated (ATT) over all the post-intervention periods},
\begin{align*}
ATT = \frac{1}{T - T_1 + 1} \sum_{t = T_1}^T \mathbb{E}\left(Y_{it}(1)-Y_{it}(0) | G_i=1\right)\;.
\end{align*}
To identify this, we assume that the expected untreated potential outcomes of the two groups would have evolved in parallel, absent an intervention. 
Formally,  
{\small
\begin{align*}
& \underset{\text{average pre-intervention difference}}{\underbrace{\frac{1}{T_1-1} \sum_{t = 1}^{T_1-1} \left( \mathbb{E}(Y_{it}(0)\mid G_i=1) - \mathbb{E}(Y_{it}(0)\mid G_i=0)\right)}} = \underset{\text{average post-intervention difference, absent treatment}}{\underbrace{\frac{1}{T-T_1+1} \sum_{t = T_1}^T \left(\mathbb{E}(Y_{it}(0) \mid G_i=1) - \mathbb{E}(Y_{it}(0)\mid G_i=0)\right)}}
\;.
\end{align*}
}
We combine this with other standard DiD assumptions: 
1) no anticipation, which requires that units do not respond to the intervention before it begins, i.e., $y_{it}=Y_{it}(0) \; \forall \; i$ when $t < T_1$; and 
2) the stable unit treatment value assumption, which rules out interference or spillovers and hidden levels of treatment.\citep{lechner_estimation_2010} 

Then, we can re-write the identified ATT in terms of observable quantities, 
{\small
{\begin{align*}
{ATT} = \frac{1}{T-T_1+1} \left[\sum_{t = T_1}^T \mathbb{E}(y_{it} \mid G_i=1) - \mathbb{E}(y_{it}\mid G_i=0)\right]- \frac{1}{T_1-1} \left[\sum_{t = 1}^{T_1-1} \mathbb{E}(y_{it}\mid G_i=1) - \mathbb{E}(y_{it}\mid G_i=0)\right]\;.
\end{align*}}
}

To estimate this, we could plug in sample averages for each of the expectation terms above.
{However, in practice, it is more common to estimate the ATT using regression, particularly using a two-way fixed effects (TWFE) estimation approach,
\begin{align}\label{eq:constrained_basic}
y_{it} = \beta G_i \mathbb{I}(t \geq T_1) + \alpha_i + \gamma_t  + \epsilon_{it}\;,
\end{align}
\noindent where $\alpha_i$ is a unit fixed effect, $\gamma_t$ is a time fixed effect, and $\epsilon_{it}$ is idiosyncratic mean-zero error.
The coefficient $\beta$ from this model corresponds to the ATT identified above.\citep{lechner_estimation_2010}

In this paper, we use a TWFE model that includes a treatment effect at \emph{each} post-intervention time,
\begin{align}\label{eq:constrained}
y_{it} = \sum_{k=T_1}^T \beta_{k} G_i \mathbb{I}(t = k) + \alpha_i + \gamma_t  + \epsilon_{it}\;.
\end{align}
The coefficients $\beta_{k}$ capture the difference between treated and comparison groups at each post-intervention time relative to the average difference in the pre-intervention period. 
The \emph{average} of these coefficients, $\beta = \frac{1}{T - T_1 + 1} \sum_{k = T_1}^T \beta_{k}$, corresponds to the ATT.\citep{lechner_estimation_2010}
The ATTs from Eq.~(\ref{eq:constrained_basic}) and (\ref{eq:constrained}) are equivalent in the balanced panel characterized here and used throughout this paper, but 
by using the latter, we ensure that expanded models introduced in the next section will identify differential pre-trends using only pre-intervention data.\citep{wolfers_did_2006}}

Recent research has highlighted that the correspondence between the ATT identified by parallel trends and the $\beta$ coefficient  from a TWFE model does not hold when treatment timing is staggered and treatment effects are heterogeneous\citep{callaway_differenceindifferences_2021,goodman-bacon_differenceindifferences_2021,sun_estimating_2021, athey_design-based_2022, de_chaisemartin_two-way_2020, imai_use_2021} or we condition parallel trends on covariates and include them in the model.\citep{griffin_moving_2021,santanna_differenceindifferences_2023,zeldow_confounding_2021} 
{Nonetheless, we introduce our approach in the simple case where the ATT \emph{can} be estimated with TWFE, then discuss some extensions in later sections (defining additional variables as needed).} 
Next, we turn to assessing whether the parallel trends assumption required for DiD seems plausible in a particular context.

\subsection{Traditional versus non-inferiority/equivalence parallel trends tests}
As noted in our literature review, biomedical researchers often investigate differential trends by specifying a model that includes a linear time trend difference between the treated and comparison groups and fitting this model only to pre-intervention data. 
For example, if our analytic model is Eq.~(\ref{eq:constrained}), we might fit the following model to pre-intervention data:
\begin{equation}\label{eq:pre-period_trend}
y_{it} = \alpha_i + \gamma_t + \theta G_i t  + \epsilon_{it}\;,
\end{equation}

The coefficient $\theta$ captures the differential trends between groups, and parallel trends in the pre-intervention period imply $\theta = 0$.
Thus, a common practice is to test 
\begin{equation*}
H_0: \theta = 0 \quad \mbox{ versus } \quad H_A: \theta \neq 0\;.
\end{equation*}
If $p > 0.05$ for this test, researchers conclude that trends are parallel and report $\beta$ estimated with the model in Eq.~(\ref{eq:constrained}).\citep{dimick_methods_2014} 

Although the objective is to provide evidence of parallel pre-intervention trends, as discussed above, this conflates a lack of evidence against the null with evidence of parallel trends.
However, the challenge of wanting to ``prove the null'' is not unique to tests of pre-intervention trends in DiD.  
Drug and device manufacturers often wish to show that their novel treatment is equivalent or non-inferior to the standard of care,
and statistical tests have been developed for this purpose.\citep{wellek_testing_2010} 
{For these tests, we select a threshold $\delta$ that represents the maximum difference we can tolerate and then formulate a test for evidence against differences larger than this threshold. In the DiD context, we might test either of the following sets of hypotheses:
\begin{align*}
\mbox{Non-inferiority: } & H_0: \theta \geq \delta \quad \mbox{ versus } \quad  H_A: \theta < \delta \\
\mbox{Equivalence: } & H_0: |\theta| \geq \delta \quad \mbox{ versus } \quad H_A: |\theta| < \delta\;.
\end{align*}
This allows us to say whether ``large'' violations of parallel trends can be ruled out with some level of statistical certainty. \citep{hahn_understanding_2012, hartman_equivalence_2018}}

{Non-inferiority tests use the same test statistics as traditional tests but different cutoffs. 
For example, with a standard assumption that $\hat{\theta}\sim N\left(\theta, \sigma^2_{\hat{\theta}}\right)$, for a two-sided Wald test, we calculate the test statistic $w = \frac{\hat{\theta}}{\hat{\sigma}_{\hat{\theta}}}$. 
A traditional two-sided test rejects the null if $w>z_{1-\alpha/2}$ or $w < z_{\alpha/2}$. 
The one-sided non-inferiority test above rejects if $w < z_\alpha + \frac{\delta}{\hat{\sigma}_{\hat{\theta}}}$,
and the equivalence test if both $w < z_\alpha + \frac{\delta}{\hat{\sigma}_{\hat{\theta}}}$ and $w > z_{1-\alpha} - \frac{\delta}{\hat{\sigma}_{\hat{\theta}}}$.\citep{wellek_testing_2010} 
We can therefore use the standard two-sided 95\% confidence interval to understand the values of $\delta$ that would lead us to reject the null in non-inferiority tests at the 2.5\% level.}

However, this approach introduces a new challenge: how do we decide what difference in trends we can tolerate?
On the scale of the slope of the differential linear time trend, there is no clear analog of the ``clinically meaningful difference'' thresholds used in testing drugs and devices.
Therefore, we next develop a framework for testing how differential trends impact our treatment effect estimates,
which allows us to specify our threshold on the scale of the treatment effect itself.

\subsection{Non-inferiority/equivalence tests on the treatment effect scale}

We introduce our testing framework on the scale of the treatment effect using a pair of models: one reduced and one expanded.
Suppose the reduced model is Eq.~(\ref{eq:constrained}). One possible expanded model is
\begin{equation}\label{eq:general}
y_{it} = \sum_{k=T_1}^T \beta_{k}^{(e)} G_i \mathbb{I}(t = k)  + \alpha^{(e)}_i + \gamma_t^{(e)} + \theta G_i t  + \epsilon^{(e)}_{it}\;,
\end{equation}
where superscript $(e)$'s distinguish this (e)xpanded model's parameters from those of the reduced model. 
This model contains a linear trend difference, $\theta G_i t$, as in the model used to test for parallel pre-intervention trends in Eq.~(\ref{eq:pre-period_trend}).
However, we now have all the post-period treatment effect coefficients in the model and intend to fit it on all the data. (This will recover an unbiased treatment effect under an assumption of ``parallel growth.'' \citep{mora_alternative_2019})

Denote the average of the post-period coefficients from this model as $\beta^{(e)} = \frac{1}{T - T_1 + 1} \sum_{k = T_1}^T \beta^{(e)}_{k}$. 
To the extent that this differs from $\beta$ derived from Eq.~(\ref{eq:constrained}), it is because of non-zero $\theta$;
thus, $\beta - \beta^{(e)}$ measures the impact of a differential linear trend on the treatment effect. 
We can therefore select a threshold that quantifies the maximum difference in the treatment effect that would imply substantive equivalence. 
That is, we specify our hypotheses as
\begin{align} 
 \mbox{Non-inferiority: } & H_0: \beta - \beta^{(e)} \geq \delta \quad \mbox{ versus } \quad H_A: \beta - \beta^{(e)} < \delta\; \label{eq:att_non-inf_hypoth}\\
 \mbox{Equivalence: } & H_0: \left|\beta - \beta^{(e)}\right| \geq \delta \quad \mbox{ versus } \quad H_A: \left|\beta - \beta^{(e)}\right| < \delta\;. \label{eq:att_equiv_hypoth}
\end{align}

If we reject the null using this procedure, we have established that non-parallel trends are not likely to meaningfully change our treatment effect. In this case, the difference in treatment effect estimates is simply a scaling of the differential slope $\hat{\theta}$.
\begin{proposition}[Reduced vs. expanded model estimators (linear trend difference)]\label{prop:estimator}
\textit{The difference between ordinary least squares (OLS) ATT estimators corresponding to model specifications in Eqs.~(\ref{eq:constrained}) and (\ref{eq:general}) is a linear transformation of the differential trends parameter estimate $\hat{\theta}$ from Eq.~(\ref{eq:general}):}
\begin{align}
\label{eq:vio}
\hat{\beta} - \hat{\beta}^{(e)} = \left(\frac{1}{T-T_1+1} \sum_{t=T_1}^T t - \frac{1}{T_1-1} \sum_{t=1}^{T_1-1} t \right)\hat{\theta} = \frac{T}{2}\hat{\theta}.
\end{align}
\end{proposition}

Therefore, the impact of differential linear trends depends on the magnitude of the slope difference $\hat{\theta}$ and the length of the study period. (See derivation in Appendix A.) 
However, a differential linear trend is not the only way to expand the model to accommodate deviations from parallel trends. 

\subsection{Reduced/expanded model testing framework}\label{ss:reduced_expanded}

We next generalize our approach to accommodate a broader array of reduced and expanded model specifications. 
Collect the parameters of the reduced model into the $p$-vector $\bm{{\beta}} = \begin{bmatrix}\beta_1, \ldots ,\beta_p \end{bmatrix}'$
and denote the corresponding $nT \times p$ design matrix $\mathbf{X}$.
If the reduced model is correctly specified, the ATT of interest is an average of a subset $\mathcal{K}$ of these parameters where $|\mathcal{K}| = K$: $\beta = \frac{1}{K}\sum_{k \in \mathcal{K}} \beta_k$. The TWFE specification in Eq.~(\ref{eq:constrained}) discussed above will be our main example of a reduced model in this text.

The expanded model includes the same covariates and parameters as the reduced model as well as additional terms. We denote additional parameters in the expanded model as $\bm{\theta} = \begin{bmatrix} \theta_{1},\ldots,\theta_{q} \end{bmatrix}'$, with $\mathbf{Z}$ the corresponding $nT \times q$ design matrix. 
We can concatenate the design matrix for the expanded model as
$\mathbf X^{(e)} = \begin{bmatrix} \mathbf X & \mathbf Z \end{bmatrix}$. Let $\bm{\beta}^{(e)}= \begin{bmatrix} \beta_1^{(e)},\ldots,\beta_p^{(e)} \end{bmatrix}'$ indicate parameters corresponding to shared predictors across models.
The corresponding ATT of interest for this model is an average of the subset $\mathcal{K}$ of these: $\beta^{(e)} = \frac{1}{K}\sum_{k \in \mathcal{K}} \beta^{(e)}_k$.
Finally, collect the outcomes and error terms into $nT$-vectors $\mathbf{y}$, $\bm \epsilon$, and $\bm {\epsilon}^{(e)}$.
We can then write reduced and expanded models as

\begin{equation}\label{eq:gen_reduced}
\textbf{Reduced: } \mathbf{y} = \mathbf{X} \bm{\beta}  + \bm{\epsilon}\end{equation}
\begin{equation}\label{eq:gen_expanded}
\textbf{Expanded: }\mathbf{y} = \mathbf{X} \bm{\beta}^{(e)} + \mathbf{Z} \bm{\theta}  + \bm{\epsilon}^{(e)}
\end{equation}

For instance, to represent the TWFE model in Eq.~(\ref{eq:constrained}), we set
$\bm{{\beta}} = \begin{bmatrix}\beta_{T_1}, \ldots,\beta_T, \alpha_1, \ldots, \alpha_n, \gamma_1, \ldots, \gamma_T \end{bmatrix}'$. To represent the expanded differential linear trend model in Eq.~(\ref{eq:general}), let $\bm{\theta} = \theta$, with $\mathbf{Z}$ as an $nT$-vector of $G_it$ values (i.e., $q=1$ additional parameter in the expanded model). We assume that $q\geq 1$ (i.e., the expanded model adds at least one parameter) and both models are identified (i.e., $\begin{bmatrix} \bX & \mathbf Z \end{bmatrix}$ has full column rank).

Using this general framework, we avoid having to derive the equivalent of Eq.~(\ref{eq:vio}) for each set of models. In the following section, this framework will allow us to characterize test properties. To estimate variance and conduct tests, we assume that the expanded model is correctly specified and that the more restrictive reduced model may be correctly specified only if $\bm{\theta} = \mathbf{0}$. We briefly discuss implications of expanded model misspecification in Section \ref{s:power}.    

\subsubsection{Test statistics in reduced/expanded model framework}

Assuming a Gaussian error structure, a test comparing coefficients from the reduced and expanded models has the following form. 
\begin{proposition}[Reduced vs. expanded model estimators (Gaussian errors)]
\label{prop:test_inf}
Assume reduced and expanded models as in Eq.~(\ref{eq:gen_reduced}) and Eq.~(\ref{eq:gen_expanded}) and that the expanded model is correctly specified, with $\bm \epsilon^{(e)} \sim N(0, \bm \Omega)$, where $\bm \Omega$ is an $nT \times nT$ matrix. Let $\mathbf V = {\left(\bX'\bX\right)}^{-1}$ and denote $\mathbf{V}^{(e)}$ analogously for the expanded model. The difference between OLS estimators $\hat{\beta}_k$ and $\hat{\beta}_k^{(e)}$ is:
\begin{align*}
\hat{\beta}_k - \hat{\beta}_k^{(e)} \sim N \left(\beta_k - \beta_k^{(e)}, {\bm \Sigma}_{k,k} + {\bm \Sigma}^{(e)}_{k,k} - 2{\bm \Sigma}^*_{k,k}\right),
\end{align*}
where ${\bm \Sigma} = \mathbf V \bX' {\bm \Omega} \bX \mathbf V$,  ${\bm \Sigma}^{(e)} = \mathbf V^{(e)} \bX^{(e)'} {\bm \Omega} \bX^{(e)} \mathbf V^{(e)}$, ${\bm \Sigma}^* = \mathbf V \bX' \bm \Omega \bX^{(e)} \mathbf V^{(e)}$, and ${\mathbf A}_{k,k}$ indicates the entry in the $k$th row and $k$th column of the matrix $\mathbf{A}$.
\end{proposition}
The proof, given in Appendix B, uses model comparison methods developed in other contexts to derive the covariance of the difference in these coefficients across model specifications.\citep{clogg_statistical_1995, liu_assessing_2009} The result allows us to define the following procedure, using OLS and standard normal-based variance estimators:
\begin{enumerate}
    \item Estimate $\bm{\hat{\beta}}$ and $\bm{\hat{\beta}}^{(e)}$ using OLS and ${\hat{\mathbf \Sigma}}$, ${\hat{\mathbf \Sigma}}^{(e)}$, ${\hat{\mathbf \Sigma}}^*$ under chosen error assumptions (using residuals from the expanded model to construct the latter three quantities).
    \item Create a linear combination of parameter estimates, $\hat{\beta}-\hat{\beta}^{(e)}$, and the corresponding standard error of the difference.
    \item Conduct a non-inferiority Wald test on $\hat{\beta}-\hat{\beta}^{(e)}$, the difference in ATT estimates between reduced and expanded models, using the hypotheses defined in Eq.~(\ref{eq:att_non-inf_hypoth}) (for non-inferiority) or Eq.~(\ref{eq:att_equiv_hypoth}) (for equivalence).
\end{enumerate} 
We walk through detailed implementation of this procedure, accounting for heteroskedastic errors, clustering, and survey weights, in Appendix B and provide R code for readers (\url{https://github.com/laura-hatfield/NonInfParTren/}).

{We can gain some intuition about this procedure by considering the special case of independent and identically distributed errors.}

\begin{proposition}[Reduced vs. expanded model estimators ($i.i.d.$ errors)]\label{prop:test}
\textit{Assume reduced and expanded models as in Eq.~(\ref{eq:gen_reduced}) and Eq.~(\ref{eq:gen_expanded}) and that the expanded model is correctly specified, with $\epsilon_{it}^{(e)} \overset{i.i.d.}{\sim} N\left(0,\sigma^2_{(e)}\right)$.
Recall that $\beta = \frac{1}{K} \sum_{k \in \mathcal{K}} \beta_k$ and $\beta^{(e)} = \frac{1}{K} \sum_{k \in \mathcal{K}} \beta_k^{(e)}$ are the parameters of interest.  The difference between the corresponding OLS ATT estimators is:}
\begin{align}
\hat{\beta} - \hat{\beta}^{(e)} \sim N \left(\beta - \beta^{(e)}, 
{\sigma}^2_{\hat{\beta}^{(e)}} - {\sigma}^2_{\hat{\beta}}\right),
\end{align}
\textit{where $\sigma^2_{\hat{\beta}^{(e)}}$ is the variance of $\hat{\beta}^{(e)}$ (corresponding to the expanded model), and $\sigma^2_{\hat{\beta}}$ is the variance of $\hat{\beta}$ (corresponding to the potentially misspecified reduced model but defined using common error variance, $\sigma^2_{(e)}$).}
\end{proposition}

The proof, in Appendix B, illustrates how the variance of the difference between $\hat{\beta}$ and $\hat{\beta}^{(e)}$ is less than the variance of $\hat{\beta}^{(e)}$.
This occurs because $Var(\hat{\beta}) = Cov\left(\hat{\beta}, \hat{\beta}^{(e)}\right)$, and thus 
testing the difference between treatment effect estimates subtracts off the shared component of the two models, thereby reducing the variance compared to estimating the treatment effect in the expanded model. 

In the special i.i.d. case, the lack of covariance between the difference $\hat{\beta}-\hat{\beta}^{(e)}$ and the estimated coefficient from the reduced model $\hat{\beta}$ implies that conditioning on $\hat{\beta}-\hat{\beta}^{(e)}$ will not bias the reduced model estimator. 
We build on this in the following section, discussing the potential of a non-inferiority/equivalence test to induce bias in ATT estimators.
We also discuss the relationship between the power of the overall study and the power of these non-inferiority/equivalence tests.

Overall, this test procedure resembles a Hausman specification test, except that the Hausman test uses a null hypothesis of a correct reduced model specification (rather than a null that only the expanded model is correctly specified).\citep{hausman_specification_1978, clogg_statistical_1995} 

\subsection{Functional form selection}
Within this reduced/expanded framework, researchers have broad latitude to select functional form. Table \ref{tab:model_menu} gives several examples, using the TWFE model in Eq.~(\ref{eq:constrained}) as the reduced model and several possible expanded models.

\subsubsection{Linear time trends}
Three specifications in Table \ref{tab:model_menu} add differential linear time trends.
The first is simply Eq.~(\ref{eq:general}), which allows treated groups to have a differential linear time trend with slope $\theta$. 
A second model allows each unit its own linear trend, $\theta_i$. (In a balanced panel, this produces the same overall ATT estimate as the first.) 
The third allows differential linear time trends in groups of units defined by something other than treatment; units with the same value of a covariate (e.g., rural counties) have slope $\theta_{\ell(i)}$ for groups $\ell=1,\ldots,L$.

\subsubsection{Event studies}

{Another popular approach to testing parallel trends involves event study models,
also shown in Table \ref{tab:model_menu}.
Adding to our reduced model treatment group-specific time fixed effects $\theta_k$ for $k=1,...,T_1-2$ at each pre-intervention time relative to the final pre-intervention time yields the conventional event study specification:
\begin{equation}\label{eq:event-study}
\textbf{Expanded: }y_{it} = \sum_{k=T_1}^T \beta_{k}^{(e)}  G_i \mathbb{I}(t = k)  + \alpha_i + \gamma_t^{(e)}  + \sum_{k=1}^{T_1 - 2}\theta_k G_i \mathbb{I}(t = k) + \epsilon_{it}^{(e)}\;.
\end{equation}
If the parallel trends assumption holds exactly, all the $\theta_k$ will be zero.
As noted in our literature review, it is common to conclude that trends are parallel if the confidence intervals all cover zero. 
Joint F-tests may offer further formalization of this test, evaluating a collection of $\theta_k$.

However, to adopt a non-inferiority testing framework on the scale of the treatment effect, we must clarify how we believe the pre-intervention coefficients inform us about parallel outcome evolution into the post-intervention period, absent intervention.\citep{dette_testing_2024, rambachan_more_2023} 
Models that add differential linear trends extrapolate pre-intervention trend differences into the post-intervention period, which is why their impact on the treatment effect depends on both the magnitude of differential slopes and the length of the study period (Proposition \ref{prop:estimator}).
The situation is more complicated for event study models.
If we believe the pre-period $\theta_k$ represent transient shocks, we might choose $\delta$ to be the largest treatment effect impact we can tolerate and test $H_0: \theta_k \geq \delta$. For instance, Rambachan and Roth\citep{rambachan_more_2023} use the maximum pre-period differential change to construct sensitivity bounds, and Dette and Schumann\citep{dette_testing_2024} suggest formulating a non-inferiority test on the maximum or mean of the pre-intervention coefficients.
The reasoning is that we are looking for evidence against shocks large enough to substantially impact our treatment effect,
if they were to strike in the post-intervention period.
If we instead believe the pre-period $\theta_k$ represent single-period differential trends, we might formulate our test based on extrapolating these trend differences into the post-intervention period (as in Proposition \ref{prop:estimator} and Eq.~(\ref{eq:vio})).
In this case, we are looking for evidence against differential trends large enough to impact our treatment effect, 
if they were to persist into the post-intervention period.

Tests on specific pre-intervention coefficients and tests on extrapolations into the post-intervention period can both be cast in terms of reduced vs expanded model effect estimates as in Proposition \ref{prop:test_inf}.  
For example, suppose we want to test the average pre-intervention coefficient $\frac{1}{T_1-2} \sum_{k=1}^{T_1-2}\theta_k$. \citep{dette_testing_2024}  
If we test $H_0: \beta-\beta^{(e)} \geq \delta$, where $\beta$ is from the reduced model in Eq.~(\ref{eq:constrained}) and $\beta^{(e)}$ is from the expanded event study model in Eq.~(\ref{eq:event-study}), this is equivalent to evaluating $H_0: \frac{1}{T_1-1} \sum_{k=1}^{T_1-2}\hat{\theta}_k \leq -\delta$, which can be scaled to conduct our test of interest (see Appendix C for derivation).  
In this case, $\beta^{(e)}$ is estimated relative only to the reference period, rather than to the average of all pre-intervention periods. 
The choice of $\delta$ determines whether and how a trend difference is assumed to be extrapolated into the post-intervention period. 
In Appendix C, we also show other tests of event study coefficients can be estimated in the form required by Proposition \ref{prop:test_inf}. 
This crosswalk will be useful for understanding the implications of conditioning estimates on event study non-inferiority tests in the following section.}

\subsubsection{Other specifications}

{As alternative expanded specifications, we could add differential time fixed effects for units with the same covariate value/cluster membership or some combination of differential linear time trends and differential time fixed effects.\citep{bailey_war_2015}
Another strategy assumes that trends are parallel only for units with the same covariate values (i.e., conditional parallel trends).\citep{callaway_differenceindifferences_2021}
To the extent that these can be encoded in an expanded model using regression, they are further examples of our reduced/expanded model testing framework.
However, in addition to regression-based estimators for conditional parallel trends, \citep{zeldow_confounding_2021} 
other popular techniques use propensity score weighting or doubly robust estimators; the latter are beyond the scope of this paper.\citep{roth_whats_2023, santannazhao2020}}

\section{Post-test bias and power}\label{s:power}

\subsection{Post-selection estimation and inference}\label{ss:bias}
Do we abandon our DiD study if we fail a test of parallel trends?
Usually not. 
Rather, we may selectively report results if trends are ``parallel enough'' or present expanded model results if not. 
However, previous work has highlighted that distortions may arise from using tests of parallel pre-intervention trends as such a screening step, specifically when presenting effect estimates from event study models.\citep{roth_pretest_2022}
{We build on these results to show that non-inferiority/equivalence tests, when used as a screening step, may introduce no or minimal bias in the reduced model estimator but substantial bias in the expanded model estimator. In the following sections, we show these results in detail.}

\subsubsection{Reduced model}

Distortions in the reduced model estimator depend on the covariance between the reduced model treatment effect estimator and the difference between reduced and expanded model treatment effect estimators. 
We first consider the case of no covariance (and thus, no distortion), then expand to more general cases.

{Given that the expanded model may also be misspecified, we introduce an explicit correct model specification.
Following the pattern above, let $\mathbf{\gamma} = \begin{bmatrix}\gamma_1,\ldots,\gamma_r\end{bmatrix}'$ denote the additional parameters and $\mathbf{W}$ their corresponding design matrix.
Then the correct model is, 
\begin{align}\label{eq:correct}
\textbf{Correct: \: } \mathbf{y} = \mathbf{X} \bm{\beta}^{(w)} + \mathbf{Z} \bm{\theta}^{(w)} + \mathbf{W} \bm{\gamma}  + \bm{\epsilon}^{(w)} \;,
\end{align}
which we use to define the ``no covariance condition.''}

\begin{assumption}[No covariance condition]\label{assump:no_cov}
\emph{Assume that the correct model specification follows Eq.~(\ref{eq:correct}), with $\bm \epsilon^{(w)}\sim N(0, \bm \Omega)$, where $\bm \Omega$ is an $nT \times nT$ matrix, and that the reduced and expanded models are specified as in Eq.~(\ref{eq:gen_reduced}) and Eq.~(\ref{eq:gen_expanded}), with $\hat{\beta}$ and $\hat{\beta}^{(e)}$ denoting corresponding OLS ATT estimators. Further assume that the combination of error structure and model specifications yields $Cov\left(\hat{\beta}, \hat{\beta} - \hat{\beta}^{(e)}\right) = 0$.}
\end{assumption}

\bigskip

Assumption \ref{assump:no_cov} holds straightforwardly in the i.i.d. case when the expanded model is correctly specified, following Proposition \ref{prop:test}. {In Appendix D, we also show that the expanded model need not be correctly specified for Assumption \ref{assump:no_cov} to hold in the i.i.d case.} 
{We also characterize conditions under which Assumption \ref{assump:no_cov} holds in the generalized error setup from Proposition \ref{prop:test_inf}. For example, Assumption \ref{assump:no_cov} holds for linear and event study tests if errors are independent across units, but heteroskedastic (i.e., $\epsilon_{it}\sim N(0, \sigma_i^2)$ with all $\epsilon_{it}$ independent) or error correlation is constant within units. In these cases as well, the expanded model need not be correctly specified for Assumption \ref{assump:no_cov} to hold. When Assumption \ref{assump:no_cov} holds, conditioning on testing does not induce bias in the reduced model estimator.} 

\begin{proposition}[Reduced model test-induced distortions under Assumption \ref{assump:no_cov}]\label{prop:no_bias}
\textit{Under Assumption \ref{assump:no_cov}, if we conduct a non-inferiority test with a threshold $\delta$, rejecting the null if $\hat{\beta} - \hat{\beta}^{(e)} < \delta^*$, where $\delta^*= z_{\alpha}{\sigma}_{\hat{\beta}-\hat{\beta}^{(e)}} + \delta$,
then there is no distortion in the reduced model ATT induced by conditioning on the test result}:  
\[\mathbb{E}\left(\hat{\beta} \middle | \hat{\beta} - \hat{\beta}^{(e)} < \delta^* \right) - \mathbb{E}\left(\hat{\beta} \right)= 0\] and likewise,
\[Var\left(\hat{\beta} \middle | \hat{\beta} - \hat{\beta}^{(e)} < \delta^* \right) - Var\left(\hat{\beta} \right)= 0.\]
\end{proposition}
The proof, given in Appendix E, derives {from the assumption that the covariance between $\hat{\beta}$ and $\hat{\beta} - \hat{\beta}^{(e)}$ is 0, and for multivariate normally distributed variables, zero covariance implies independence.}
In Appendix E, we extend this result to show that it holds when we formulate the test as an equivalence test (Corollary 2). 

Note that this does not imply that $\hat{\beta}$ is unbiased, only that conditioning whether to report $\hat{\beta}$ on a non-inferiority test does not add bias, and that results of the non-inferiority test still bound its bias.  
Therefore, {if we pass a non-inferiority test, it is appropriate to present $\hat{\beta}$ from the reduced model and the results of the test of $\hat{\beta}-\hat{\beta}^{(e)}$ as a bound on the bias.}  These results do not require the threshold to be pre-specified or explicit, and testing does not introduce distortions in variance.

{However, there may be test-induced bias in other circumstances, particularly when errors are autocorrelated over time. We formalize this in the following proposition.}

\begin{proposition}[Reduced model test-induced bias]
\label{prop:bias_non_iid}
Assume setup and reduced and expanded model estimators as in Proposition \ref{prop:test_inf}.  If we conduct a non-inferiority test with a threshold $\delta$, rejecting the null if $\hat{\beta} - \hat{\beta}^{(e)} < \delta^*$, where $\delta^*= z_{\alpha}{\sigma}_{\hat{\beta}-\hat{\beta}^{(e)}} + \delta$, then the expectation of the reduced model ATT estimator conditional on passing the test may differ from its unconditional expectation: 
\begin{align*}
\mathbb{E}\left(\hat{\beta} \middle | \hat{\beta} - \hat{\beta}^{(e)} < \delta^* \right) - \mathbb{E}\left(\hat{\beta}\right) &= -
\frac{Cov\left(\hat{\beta}, \hat{\beta} - \hat{\beta}^{(e)}\right)}{{\sigma}_{\hat{\beta}-\hat{\beta}^{(e)}}}
\frac{\phi\left(z_{\alpha} + \frac{\delta - \beta + \beta^{(e)}}{
        {\sigma}_{\hat{\beta}-\hat{\beta}^{(e)}}}\right)}
     {\Phi\left(z_{\alpha} + \frac{\delta - \beta + \beta^{(e)}}{
        {\sigma}_{\hat{\beta}-\hat{\beta}^{(e)}}}\right)},
\end{align*}
where $\phi$ and $\Phi$ are the probability density function and cumulative distribution function of a standard normal, respectively.
\end{proposition}

Proposition \ref{prop:bias_non_iid} indicates that test-induced distortions depend on the covariance between the reduced model estimator and the difference estimator as well as the choice of threshold (proof in Appendix E). {Still, even with error autocorrelation producing non-zero $Cov \left(\hat{\beta}, \hat{\beta}-\hat{\beta}^{(e)}\right)$, bias may be small if the covariance term is small and/or thresholds are not too strict. It may also be conservative under positive error autocorrelation with positive treatment effects. We will explore this further via simulation in Section \ref{s:simulations}.} 

{Last, note that conditioning on the test outcome leads to a smaller variance for $\hat{\beta}$. The variance of a truncated distribution is necessarily smaller than that of the original distribution. Although the truncation applies to the distribution of $\hat{\beta} - \hat{\beta}^{(e)}$, the joint normal relationship between this and $\hat{\beta}$ implies that the truncation also reduces the conditional variance of $\hat{\beta}$, following from standard properties of the truncated multivariate normal distribution \citep{roth_pretest_2022}.
}

\subsubsection{Expanded model}
By contrast, testing may introduce bias in the expanded model estimator, even when it does not induce bias in the reduced model estimator. This occurs in part because, although the formula of the bias appears similar to that above, the covariance between $\hat{\beta}^{(e)}$ and $\hat{\beta} - \hat{\beta}^{(e)}$ may be sufficiently large to produce meaningful distortions,  even when the covariance between $\hat{\beta}$ and $\hat{\beta} - \hat{\beta}^{(e)}$ is 0. Formally, we quantify test-induced bias:

\begin{proposition}[Expanded model test-induced bias]
\label{prop:bias_expanded}
Assume setup and reduced and expanded model estimators as in Proposition \ref{prop:test_inf}. If we conduct a non-inferiority test with a threshold $\delta$, rejecting the null if 
$\hat{\beta} - \hat{\beta}^{(e)} < \delta^*$, where $\delta^*=  z_{\alpha}{\sigma}_{\hat{\beta}-\hat{\beta}^{(e)}} + \delta$, then the expectation of the expanded model ATT estimator conditional on passing the test may differ from its unconditional expectation:
\begin{align*}
\mathbb{E}\left(\hat{\beta}^{(e)} \middle | \hat{\beta} - \hat{\beta}^{(e)} < \delta^* \right) - \mathbb{E}\left(\hat{\beta}^{(e)}\right) &=
-\frac{Cov\left(\hat{\beta}^{(e)}, \hat{\beta} - \hat{\beta}^{(e)}\right)}{{\sigma}_{\hat{\beta}-\hat{\beta}^{(e)}}}
\frac{\phi\left(z_{\alpha} + \frac{\delta - \beta + \beta^{(e)}}{
        {\sigma}_{\hat{\beta}-\hat{\beta}^{(e)}}}\right)}
     {\Phi\left(z_{\alpha} + \frac{\delta - \beta + \beta^{(e)}}{
        {\sigma}_{\hat{\beta}-\hat{\beta}^{(e)}}}\right)}. 
        \end{align*}
\end{proposition}

The proof is provided in Appendix E, and we also illustrate this phenomenon in simulations in Section \ref{s:simulations}. Similar to the prior proposition, the expression above characterizes the case when expanded model estimates (e.g., an event study) are presented after passing a test. In this case, bias declines with a larger magnitude of $\delta$.  An analogous distortion occurs if expanded model results are reported only after failing a test (e.g., adding differential linear trends), but with an opposite impact of threshold stringency. (Presenting expanded model results after failing a test would also necessitate assessing plausibility of a different identifying assumption; we leave further consideration of this to future work.) 

This result suggests that although an expanded model may reduce bias by adjusting for non-parallel trends, these reductions in bias may be offset by test-induced bias. 
Indeed, we will see such scenarios in our simulation study (Section \ref{s:simulations}).
Overall, {we may be wary of reporting the results from the expanded model after conducting a test}.
{As above, when distortions in variance occur from a testing procedure, using unconditional variance would remain conservative.}

\subsection{Power of non-inferiority tests}
Recall that we pass a traditional test when we have high uncertainty (i.e., low power).
Thus, we may worry that switching to a non-inferiority framework will make passing too difficult.  
In this section, we characterize the power of our tests. 
We consider the power of a non-inferiority test when trends truly are parallel (i.e., $\bm \theta= \mathbf{0}$) and compare this test's power to that of a treatment effect test under the expanded model in the simple case of i.i.d. errors.  We will explore power in more complex scenarios in simulations in Section \ref{s:simulations}.

\begin{proposition}[Non-inferiority difference-in-differences power]\label{prop:DiD.NI.heuristic.alt}
\textit{Assume setup and reduced and expanded model estimators as in Proposition \ref{prop:test}. If a non-inferiority test has power $p$ evaluated under an asymptotic normal approximation to rule out violations of parallel trends equal to or larger than $\beta^*_k$ (i.e., to reject $H_0: \beta_k- \beta_k^{(e)} \geq \beta_k^*$, with $\beta_k^*>0$) in a Wald test at level $\alpha$, and assuming no violation exists $(\bm \theta = \mathbf{0})$, then $p > p_e$, where $p_e$ is the power to detect $\beta_k^{(e)} = \beta_k^*$ (likewise evaluated) in a one-sided Wald test at level $\alpha$ in an expanded model.}
\end{proposition}

In other words, the power of a non-inferiority test with a given threshold will exceed power to detect an effect as large as the threshold with an expanded model (proof in Appendix F).
We also show in Appendix F that this may exceed power to detect an effect in the reduced model. 
This does not imply that non-inferiority tests and tests of treatment effects are equivalent tasks.

Rather, this comparison helps contextualize power and illustrates that non-inferiority tests may still provide useful information even when the expanded model is under-powered for a treatment effect the size of the chosen threshold. 

Power declines as the threshold becomes stricter, as we are trying to rule out smaller (harder to detect) violations.  
It is also lower for equivalence formulations (see Appendix F).\citep{dette_equivalence_2018, wellek_testing_2010} 
Last, power to rule out trend differences exceeding a threshold decreases as the true trends diverge, and violations approach the threshold value.

\section{Simulations}\label{s:simulations}

\subsection{Methods}
We conducted a simulation study to demonstrate the empirical performance of non-inferiority and equivalence tests.
Using Eq.~(\ref{eq:general}), we generated observations for 60 units at $T=25$ time points, assigning half to treatment beginning at $T_1 = {21}$ (i.e., 5 treated periods).
We drew unit fixed effects from a standard normal distribution and errors from normal distributions with two different variance structures.
The first (denoted ``Independent, heteroskedastic'') assumed errors were independent across the 25 time points but each unit had its own variance parameter.
The second (denoted ``AR(1), $\rho = 0.2$'') assumed errors were independent across units but followed a first-order autoregressive process with correlation parameter $\rho = 0.2$ across time within each unit. 

We assumed a constant treatment effect and included scenarios either with no violation of parallel trends (i.e., $\theta = 0$) or a linear violation of parallel trends that increased the expected value of the reduced model treatment effect by 50\%. 
For ease of interpretation, we scaled the treatment effect and violation in terms of ``TX80,'' the effect size that a TWFE estimator had 80\% power to detect under no violation of parallel trends.

For each scenario, we fit the reduced model in Eq.~(\ref{eq:constrained}) 
and compared it to two expanded models.
One included a linear trend difference, as in Eq.~(\ref{eq:general}),
and the other used an event study specification, as in Eq.~(\ref{eq:event-study}). 
We estimated the treatment effect using all three models and tested for a violation by comparing the reduced vs. expanded model results with both non-inferiority and equivalence tests.
For these tests, we considered one threshold that was more lax (100\% of TX80) and one that was stricter (20\% of TX80).  
Across 100,000 replications, we present results in terms of empirical power/Type I error (percentage of treatment effect tests and non-inferiority/equivalence tests meeting statistical significance across different violation scenarios) and bias (both in estimated treatment effects and incremental bias from conditioning on tests). 

\subsection{Results}
The results of the simulation study are summarized in Table \ref{tab:sims_dgp}.
Non-inferiority tests strictly controlled the probability of missing a violation that exceeded the threshold (i.e., making a Type I error).
In scenarios with a violation of 50\% TX80 and a threshold of 20\% TX80, the probability of passing a non-inferiority or equivalence test remained below 5\%.
(See the ``Power'' results in the columns labeled ``NI'' and ``EQ'' in scenarios marked with double daggers.)

When the true violation was smaller than the threshold (as in the scenarios with a violation of 50\% TX80 and a threshold of 100\% TX80 or no violation), power depended on the strictness of the threshold.
For example, in the scenario with no violation and independent, heteroskedastic errors, the probability of passing a non-inferiority test based on a linear expanded model declined from 88\% at the more lax threshold (100\% TX80) to 12\% at the stricter threshold (20\% TX80).
However, at the more lax threshold, the power of the non-inferiority test was greater than that of the expanded model (52\%) to detect the treatment effect.
Such differences were more pronounced for expanded models with linear differential trends than event study models.  

For models with independent, heteroskedastic errors, there was no test-induced bias in the reduced model treatment effect estimator.
However, conditional on passing a test, there could be substantial test-induced bias in results from the expanded model (range: 8-157\%), which worsened with a stricter threshold. There was also test-induced bias in the expanded model estimator conditional on not passing, though opposite in sign and larger at more lenient thresholds. 
In several scenarios, test-induced bias exceeded misspecification bias in the reduced model estimator, meaning that after conditioning on the test result, the expanded model estimator was more biased than the reduced model estimator.

With autocorrelated errors, we observed some test-induced bias in the reduced model treatment effect estimator conditional on passing a test, ranging from 0 to -2\% of TX80 for linear models and -3 to -9\% for event study models and increasing with stricter thresholds. Nevertheless, test-induced bias in the reduced model estimator was still smaller in magnitude than that in the expanded model estimator conditional on passing (range 8\%-128\%). Conditional on not passing, there was less test-induced bias in the expanded model estimator only when the violation substantially exceeded the threshold (and thus the test rarely passed, meaning the conditioning event almost always occurred).

\section{Application to the ACA Dependent Coverage Provision}
\label{s:application}
We applied our approach to re-analyze the impact of the United States (US) 2010 Patient Protection and Affordable Care Act (ACA) on young adults' health insurance coverage. 
The ACA was enacted March 23, 2010 by the US Congress. 
Beginning September 23, 2010, it required commercial insurers to offer health insurance coverage to dependents up to age 26 on a parent or guardian's plan. 
Previously, the federal rules only required coverage on a parent's plan up to age 18.
As a result, nearly a third of people aged 19-25 went without health insurance.\citep{cohen_2012_health}
Several studies have assessed the impact of this provision using DiD, comparing coverage among people newly eligible to join parents' plans to coverage among other young people not affected by the policy change.\citep{akosa_antwi_effects_2013,barbaresco_impacts_2015,sommers_affordable_2013, cantor_early_2012}

\subsection{Methods}
\subsubsection{Data}
We used replication code and data provided by authors of one of these studies.\citep{akosa_antwi_effects_2013}
From their extract of the Survey of Income and Program Participation (SIPP), we used monthly self-reported insurance coverage in the following categories:
any insurance, dependent coverage on a parent's plan, employer-sponsored insurance for oneself, privately purchased individual insurance, and government insurance (e.g., Medicare, Medicaid).
As in the original analysis, we compared coverage among people aged 19-25 years (the treated group) to that of people aged 16-18 and 27-29 years (the comparison groups), omitting people aged 26 years because of their ambiguous treatment status. 
The study period was Aug 2008 through Nov 2011.

\subsubsection{Models}
As in the original authors' specification, we first fit a linear probability model for each insurance type separately,
\begin{equation}\label{eq:aca_model}
\textbf{Reduced: }y_{it} = \sum^{40}_{k=20}\beta_{k} G_i \mathbb{I}(t = k) + \mathbf{X}_{it}\bm{\beta}_X + \alpha_{s(i)} + \gamma_t + \epsilon_{it}\;, 
\end{equation}
where $y_{it}$ was a binary indicator for person $i$ having coverage at month $t$,
$G_i$ was a binary indicator for the treated group (i.e., age 19-25), $\mathbf{X}_{it}$ was a row vector of covariates, and
$s(i)$ indexed the $i$th person's state. 
The vector of covariates $\mathbf{X}_{it}$ contained 
age (categorical, encoding treatment status), 
gender (binary), 
race/ethnicity (categorical), 
marital status (binary),
student status (binary), 
household income as a proportion of the federal poverty line (continuous), 
and squared household income as a proportion of the federal poverty line.

We next fit an expanded model with a differential linear time trend in the treated group $\theta G_i t$:
\begin{equation}\label{eq:aca_exp_model}
\textbf{Expanded: }y_{it} = \sum^{40}_{k=20}\beta_{k}^{(e)} G_i \mathbb{I}(t = k) + \mathbf{X}_{it}\bm{\beta}^{(e)}_X + \alpha_{s(i)}^{(e)} + \gamma_t^{(e)} + \theta G_i t + \epsilon_{it}^{(e)}\;
\end{equation}

From both the reduced and expanded models, we focused on implementation effects,
$\beta = \frac{1}{14} \sum^{40}_{k=27} \beta_{k}$ and $\beta^{(e)}= \frac{1}{14} \sum^{40}_{k=27} \beta_{k}^{(e)}$, which 
average the coefficients from implementation (Oct 2010) to the end of the study (Nov 2011).
To determine a non-inferiority threshold for the difference in the estimated effects from the two models, we use estimated treatment effects from an earlier analysis of the ACA's dependent coverage provision.\citep{cantor_early_2012}
That study's outcomes were slightly different, but the significant results ranged from a 2.1 percentage point decrease in ``private, self or spouse'' coverage to a 5.3 percentage point increase in ``private, non-spouse dependent''. 
Thus, we sought to rule out changes greater than or equal to $|2.1|$ or $|5.3|$.  For the stricter threshold, we also considered a non-inferiority benchmark, denoted $2.1^*$, which adjusted non-inferiority tests based on the sign of the expected treatment effect (i.e., ruling out violations $>2.1$ for any health insurance and dependent coverage and $<-2.1$ for others). 

Following the original authors, we used normal-based robust standard errors, clustered at the state level,
and weighted regression with person weights from SIPP for all the models.
In tests of the difference between the treatment effects from the reduced and expanded models, we implemented the inference described in Proposition~\ref{prop:test_inf}.

Finally, we also replicated the original authors' tests for parallel pre-intervention trends by fitting a model with differential linear trends to data from the pre-period only, that is, from Aug 2008 to just before enactment in Feb 2010:
\begin{equation}
y_{it} = \mathbf{X}_{it}\bm{\beta}_X + \alpha_{s(i)} + \gamma_t + \theta G_i t + \epsilon_{it}\;
\end{equation}

Two differences from the authors' original model merit mention.
First, the original model used interactions between treatment group and three time periods (pre-ACA, enactment, and implementation), whereas we saturated the model with coefficients for the treated group in each post-enactment month $\beta_k$ and added corresponding month-year fixed effects (therefore omitting linear time trends, which had minimal impact per Tables S3 and S4).
We used this saturated specification to avoid fitting a pre-intervention trend to time heterogeneity in treatment effects; as the authors only fit differential trends by treatment status to pre-intervention data, they avoided this concern.
The authors' original specification also included an interaction term between state-month unemployment (continuous) and treatment group.
However, we omitted this variable because of its collinearity with differential linear time trends in our reduced vs expanded testing strategy (see Appendix G).

\subsubsection{Empirical simulations}

To understand the magnitude of violations we could have ruled out, as well as potential distortions introduced by testing, we also generated simulations based on the SIPP data.
We first fit the expanded model in Eq.~(\ref{eq:aca_exp_model}) on the any health insurance outcome.
Setting $\theta=0$, we used the fitted model to generate predicted values, $\hat{p}_{it}$, and residuals, $\hat{u}_{it}$. 
We considered two data-generating processes:
\begin{enumerate}
\item \emph{Normal with heteroskedastic errors across state clusters:} We estimated sample residual variance, $\hat{\sigma}_s$, by state and then simulated synthetic datasets as follows:
\begin{align*}
{y}_{it}^{sim} &= \hat{p}_{it} + \theta G_i t + \epsilon_{it}^{sim},\\
\epsilon_{it}^{sim} &\stackrel{ind}{\sim} N\left(0, \hat{\sigma}_{s(i)}^2\right)
\end{align*}
where $s(i)$ indicated the $i$th person's state. 
We considered $\theta$ corresponding to either no violation or a linear violation approximately a third of the treatment effect.
\item \emph{Cluster-resampling wild bootstrap:} For simulations that preserved intra-cluster correlation, both across individuals and over time, we bootstrapped the data set by sampling with replacement at the state cluster level.  For each state draw, we then set the values of the component units: 
\begin{align*}
{y}_{it}^{sim}&= \hat{p}_{it} + \theta G_i t + q_{s(i)} \hat{u}_{s(i)t},
\end{align*}
where $q_s$ was a cluster-level random variable taking 1 with probability 0.5 and -1 with probability 0.5 (i.e., randomly flipping the signs of the errors by cluster). We considered the same $\theta$ as above.
\end{enumerate}

We ran 15,000 replications of each scenario, each time generating data from the specified data-generating process and fitting the reduced (Eq.~(\ref{eq:aca_model})) and expanded (linear trend difference, Eq.~(\ref{eq:aca_exp_model})) models. 
Following the estimation and testing workflow of our earlier simulations, we estimated the treatment effect in each and tested for the violation (reduced vs. expanded) using both non-inferiority and equivalence tests, with thresholds of $\{1^*, 2.1^*, |5|\}$, with $1^*$ added for an even more stringent threshold. 
We calculated power/Type I error of non-inferiority/equivalence tests and bias (in estimated treatment effects and incremental bias from conditioning on tests).

\subsection{Results}
\subsubsection{Parallel trends tests}
Akosa Antwi and co-authors plotted the proportion with any insurance in treated and comparison groups over time (Figure 1 of their paper) and reported ``generally a similar pattern prior to the ACA passage.''\citep{akosa_antwi_effects_2013} 
To extend the visual investigation of pre-trends, we plotted differences in coverage between treated and comparison groups for each of the insurance coverage outcomes (Figures S2 and S3).
Table S2 replicates the pre-intervention trend tests from the original paper's Appendix Table A1, showing no statistically significant non-parallel trends.
However, absent the interaction between unemployment and treatment group, there were statistically significant differential trends in the dependent coverage outcome in the pre-period (Figure S4).

\subsubsection{Treatment effects}
The original authors found statistically significant increases in any ($+3.2\%$) and dependent ($+7.0\%$) coverage, significant decreases in employer ($-3.1\%$) and individual ($-0.8\%$) coverage, and a small, non-significant decrease in government coverage ($-0.3\%$) (see their Table 2).\citep{akosa_antwi_effects_2013}
We replicated those results in Tables S3 and S4. 

Our model formulation yielded very similar results, shown in Table \ref{tab:reg_results} in rows with Model type ``Original''.  
Adding differential linear trends to this model (rows with Model type ``+ trend'') substantially reduced the estimated impacts on dependent, employer, and individual outcomes, and made the nearly zero effect on government coverage more positive (though not statistically significant). 
Applying our non-inferiority approach, all outcomes except dependent coverage passed at the most generous $|5.3|$ threshold.  
Both any health insurance and individual coverage passed at the $2.1^*$ threshold, and at the strictest $|2.1|$ threshold, only individual coverage passed. With the inclusion of the unemployment/treatment interaction, the treatment effect estimates from the expanded model had substantially higher variance, and only 3 of the prior 7 tests passed (Table \ref{tab:reg_results}). 

\subsection{Empirical simulation results}

In simulations, we found high power to rule out violations greater than $|5|$ (88-100\%) and low-to-moderate power to rule out those greater than 2.1 (21-60\%) (Table \ref{tab:emp_sim_results}).  
Non-inferiority tests controlled type I error at approximately $5\%$ for simulations where violations exceeded the threshold.  
{We observed $\leq$|1\%| incremental test-induced bias in reduced model treatment effect estimators for scenarios with independent errors and $\leq$|6\%| in bootstrap-based simulations that preserved the cluster correlation structure.}  
For expanded model estimators, test-induced bias could be substantial, up to 80\%.

\section{Discussion}\label{s:discuss}
Conventional guidance in the medical literature suggests testing a null hypothesis of no differential pre-intervention trend when conducting DiD.\citep{dimick_methods_2014}
This practice obscures meaningful violations by tightly controlling Type I error, the probability of incorrectly detecting violations, rather than Type II error, the probability of missing them. 
By contrast, our non-inferiority/equivalence framework informs researchers about the magnitude of violations (and their impacts on treatment effects) that can be ruled out. 
Our general framework for testing reduced versus expanded models enables flexible relaxations of the parallel trends assumption, allowing us to specify the non-inferiority/equivalence threshold on the scale of the treatment effect itself.
We can even avoid committing to a single threshold by using the relationship between hypothesis tests and confidence intervals to rule out values outside a 95\% confidence interval.
We characterized conditions under which our procedure, if used as a pre-screening step, may introduce no or minimal bias in the reduced model ATT estimator.
Finally, we showed that our strategy may have higher power than tests of treatment effects in the expanded model under no violation.

\subsection{Implementing our framework}

{For applied DiD studies, we therefore recommend the following implementation of our framework: 
\begin{enumerate}[1.]
\item Specify models.
    \begin{enumerate}[a.] 
        \item The reduced model should encode plausible causal assumptions, such as parallel trends, and encode a treatment effect $\beta$.
        \item The expanded model should encode plausible relaxations of the assumptions and encode an analogous treatment effect $\beta^{(e)}$. 
    \end{enumerate}
  Table \ref{tab:model_menu} suggests a variety of expanded models using familiar models from the literature.
\item Perform a non-inferiority/equivalence test.
\begin{enumerate}[a.]
  \item Using the study's power and notes below as a guide, choose a threshold $\delta$ for the largest change in the treatment effect that would still imply substantive equivalence. Alternatively, construct a 95\% confidence interval around $\hat{\beta}-\hat{\beta}^{(e)}$ and present values outside the range as ``ruled out'' impacts on the treatment effect.
  \item Using results from Proposition \ref{prop:test_inf} or \ref{prop:test}, conduct a test that accounts for across-model dependence in the parameters, or examine the bounds of the corresponding confidence interval.
  \begin{enumerate}[i.]
  \item If the test rejects the null (or the ruled-out range is sufficient), present $\hat{\beta}$ and the 95\% CI on $\hat{\beta}-\hat{\beta}^{(e)}$ as a bound on the bias, evaluating and noting risk of model selection-related distortions.
  \item If the test fails to reject the null, reconsider the design, comparison group, and/or model specification, evaluating and noting risk of model selection-related distortions.
  \end{enumerate}
\end{enumerate}
\end{enumerate}}

\subsection{Threshold choice}

{As noted above, we can avoid committing to a single value of $\delta$ by using the relationship between hypothesis tests and confidence intervals to determine the range of impacts that we can rule out.
The 95\% confidence interval bounds on $\hat{\beta} - \hat{\beta}^{(e)}$ represent the magnitudes that can be ruled out at a $2.5\%$ error rate.  
For example, we could interpret a 95\% confidence interval on $\hat{\beta}-\hat{\beta}^{(e)}$ of $(-1, 4)$ as follows:
``We can rule out (at the 2.5\% level) violations that would reduce the treatment effect by more than $1$ or increase it by more than $4$.''  
This is the procedure recommended by Hartman and Hidalgo in the related setting of wanting to provide evidence to support the appropriateness of a regression discontinuity design.\citep{hartman_equivalence_2021}}

{However, even if using a confidence interval, researchers will benefit from context to help guide their evaluation of this range. Other researchers may prefer to test a specific threshold $\delta$, which introduces the challenge of selecting this value.
We can use the power of the overall study to help determine a threshold.
For instance, we might say that we wish to rule out differential trends that would change our treatment effect by some fraction of the effect size our study is powered to detect.
In our simulation study, we found that when using a threshold equal to the treatment effect for which the reduced model had 80\% power (i.e., lines with Threshold=100 in Table \ref{tab:sims_dgp}), non-inferiority tests had moderate-to-high power when trends were truly parallel.
Even in the presence of a violation equal to half the treatment effect for which we were powered, both non-inferiority and equivalence tests with a linear trend expanded model retained reasonable power.
However, the power of non-inferiority and especially of equivalence tests was reduced when the expanded model was an event study specification.

Another strategy to inform our threshold selection is to examine related effect estimates from the literature, as we did in our applied analysis in Section~\ref{s:application}. We can take advantage of the greater power of non-inferiority tests compared to equivalence formulations if we know the plausible direction and magnitude of the treatment effect.}
{Suppose we have an expected effect of $\delta^{*}$ (e.g., based on previous studies) and we are most concerned with differential trends that would lead us to erroneously conclude that the intervention had a true effect. 
Then a non-inferiority test of $H_0: \beta-\beta^{(e)} \geq \delta^*$ if $\delta^*>0$ (or $H_0: \beta-\beta^{(e)} \leq \delta^*$ if $\delta^*<0$) can rule out differential trends that would lead us to estimate a treatment effect of $\delta^*$ in the reduced model when there is truly none (in the expanded model).
}

\subsection{Limitations}

Our proposed approach has several limitations.
Although we show good statistical properties for reduced model estimates following a test, this only enables researchers to bound bias under specific assumptions, not guarantee unbiased effect estimates.
Our approach also requires researchers to impose parametric functional form restrictions on trend differences.  
Furthermore, when conditioning on test results, there is a risk of distortions beyond what we explored. {For example, highly autocorrelated data, observed in some contexts\citep{bertrand_how_2004}, could drive instances of significant test-related bias in reduced model specifications. This may merit further study and exploration in specific applications.} Conducting multiple tests (e.g., on event study coefficients) could also inflate false discovery rates, for which researchers could apply standard corrections.
{In addition, although we give bias results for the impact of pre-test conditioning, we have not provided results for variance outside of special (``no covariance condition'') cases, noting only that any distortions from using unconditional variance estimates would be conservative. 
We leave this problem to future work.
}
{Last, for equivalence tests, our Wald-based approach is conservative;\citep{wellek_testing_2010} future work could extend recent innovations that improve equivalence test power to the DiD context.\citep{dette_equivalence_2018}} 

Despite these, we believe this non-inferiority formulation is practical for a wide range of clinical and policy applications. 
It uses the familiar tools of regression and testing while avoiding the pitfalls of conventional parallel trends testing by providing more transparent bounds on the likely impacts of non-parallel trends on DiD estimates.

\section*{Conflicts of interest}

The authors declare no potential conflict of interests.

\section*{Supporting information}
Appendices, Tables, and Figures referenced in Sections 2-5 are available with this paper.

\pagebreak

\bibliography{wileyNJD-AMA}

\pagebreak

\begin{table*}
\centering
\caption{Possible expanded model specifications.\label{tab:model_menu}}
\begin{tabular*}{0.7\textwidth}{@{\extracolsep\fill}lrl@{\extracolsep\fill}}
\toprule
\multicolumn{2}{l}{\textbf{Model}} & \textbf{Parameterization} \\
\midrule
\multicolumn{2}{l}{Reduced} & $y_{it} = \sum_{k=T_1}^T \beta_{k} G_i \mathbb{I}(t = k) + \alpha_i + \gamma_t + \epsilon_{it}$ \\
\multicolumn{2}{l}{Expanded} & $y_{it} = \sum_{k=T_1}^T \beta^{(e)}_{k} G_i \mathbb{I}(t = k) + \alpha^{(e)}_i + \gamma^{(e)}_t + \epsilon^{(e)}_{it} + \square$ \\
\midrule
\multicolumn{3}{l}{Linear time trends} \\
   & treatment group & $ \square = \theta G_i t $ \\
   & unit  & $ \square = \theta_i t $ \\ 
   & covariate group$^1$ & $ \square = \theta_{\ell(i)} t $ \\ 
\midrule
\multicolumn{3}{l}{Differential time fixed effects} \\
   & event study$^2$ &      $\square = \sum_{k=1}^{T_1-2} \theta_k G_i \mathbb{I}(t = k)$ \\
   & covariate group$^1$ &      $\square =\sum_{k=1}^{T_1-2} \theta_{\ell(i)k} \mathbb{I}(t = k)$  \\
\bottomrule
\end{tabular*}
\begin{tablenotes}
\centering
\item $^1$ $\ell(i)$ indicates the covariate group to which unit $i$ belongs.
\item $^2$ An event study has differential time fixed effects by treatment status.
\end{tablenotes}
\end{table*}

\clearpage 

\begin{table*}[!ht]
\centering
\begin{threeparttable}
\footnotesize
\caption{Simulation results.\label{tab:sims_dgp}}
\tabcolsep=0pt
\begin{tabular*}{\textwidth}
{@{\extracolsep{\fill}}p{1in}p{.5in}p{.6in}p{.7in} p{.3in}p{.3in}p{.3in}p{.3in} | p{.3in}p{.35in}p{.3in}p{.35in}p{.35in}@{\extracolsep{\fill}}}
\toprule
& & & & \multicolumn{4}{c}{\emph{Power}} & \multicolumn{5}{c}{\emph{Bias (\% of TX80)}}  \\ 
\cline{5-8} \cline{9-13}\\
Error & Model & Violation\tnote{$^\dagger$} & Threshold\tnote{$^\dagger$} & NI & EQ & R & E & R & Test (R) | Pass & E & Test (E) | Pass & Test (E) | Fail \\
\midrule
Independent, & Linear & 0 & 100 & 88 & 75 & 80 & 52 & 0 & 0 & 0 & 8 & -55\\
heteroskedastic &  &  & 20 & 12 & 0 & 80 & 52 & 0 & 0 & 0 & 56 & -8\\
 &  & 50 & 100 & 39 & 38 & 99 & 53 & 50 & 0 & 0 & 33 & -21\\
 & &  & 20& 0\tnote{$^\ddagger$}   & 0\tnote{$^\ddagger$}  & 99 & 53 & 50 & - & 0 & - & 0\\

Independent, & Event & 0 & 100 & 40 & 0 & 80 & 25 & 0 & 0 & 0 & 64 & -43\\
heteroskedastic & study  &  & 20 & 8 & 0 & 80 & 25 & 0 & 0 & 0 & 124 & -11\\
 &  & 50 & 100 & 21 & 0 & 99 & 30 & 50 & 0 & 12 & 91 & -25\\
 &  &  & 20  &  2\tnote{$^\ddagger$} & 0\tnote{$^\ddagger$} & 99 & 30 & 50 & 0 & 12 & 157 & -4\\

AR(1) & Linear & 0 & 100 & 87 & 74 & 80 & 53 & 0 & 0 & 0 & 8 & -53\\
$\rho = 0.2$ &  & & 20 & 12 & 0 & 80 & 53 & 0 & -2 & 0 & 55 & -8\\
 &  & 50 & 100 & 38 & 37 & 99 & 53 & 50 & -1 & 0 & 33 & -20\\
 &  &   & 20 & 0\tnote{$^\ddagger$} & 0\tnote{$^\ddagger$} & 99 & 53 & 50 & - & 0 & - & 0\\

AR(1) & Event &  0 & 100 & 50 & 5 & 80 & 32 & 0 & -3 & 0 & 43 & -43\\
$\rho = 0.2$  & study &  & 20 & 9 & 0 & 80 & 32 & 0 & -6 & 0 & 99 & -9\\
 & &  50 & 100 & 26 & 4 & 99 & 39 & 50 & -5 & 12 & 67 & -23\\
 & & &  20 & 2\tnote{$^\ddagger$} & 0\tnote{$^\ddagger$} & 99 & 39 & 50 & -9 & 12 & 128 & -3\\
\bottomrule
\end{tabular*}
\begin{tablenotes}[flushleft, leftmargin=0pt]
\item \emph{Simulations vary (n = 100,000 per scenario): (1) residual error structure (heteroskedastic across clusters and independent or AR(1) with $\rho = 0.2$); (2) the expanded model (linear trend difference or event study); (3) the true violation magnitude; (4) the threshold for a non-inferiority test.}  
\item \emph{Power indicates power for: (1) a non-inferiority test (NI); (2) an equivalence test (EQ); (3) the reduced model (R); or (4) the expanded model (E).}
\item \emph{Bias indicates percentage bias in: (1) the reduced model treatment effect estimator (R); (2) incremental bias conditional on passing a non-inferiority test (Test (R) | Pass); (3) the expanded model treatment effect estimator (E); (4) incremental bias conditional on passing (Test (E) | Pass) or failing (Test (E) | Fail) a non-inferiority test.}
\item {$^\dagger$} Violation magnitude and threshold are given as a percentage of the treatment effect for which the reduced model has 80\% power under no violation (TX80).  
\item {$^\ddagger$} Violation exceeds threshold; power indicates Type I error (controlled at $< 5\%$).
\end{tablenotes}
\end{threeparttable}
\end{table*}

\begin{table}
\centering
\begin{threeparttable}
\footnotesize
\caption{\label{tab:reg_results}Effects of the ACA dependent coverage provision on insurance coverage.} 
\begin{tabular*}{\textwidth}{@{\extracolsep\fill}lrrrccc@{\extracolsep\fill}}
\toprule%
Outcome & Model & Effect (95\% CI) & Diff (95\% CI) & Rule out $|2.1|$? & Rule out $2.1^*?$ & Rule out $|5.3|?$ \\
\hline
Any & Original & 2.9 (1.3, 4.4) &  &  & \\
 & +  trend & 3.4 (0.9, 5.9) & -0.6 (-2.8, 1.6) & No & Yes{$^\dagger$} & Yes{$^\dagger$}\\
Dependent & Original & 7.0 (5.6, 8.3) &  &  & \\
 & +  trend & 3.6 (1.2, 6.1) & 3.3 (0.9, 5.8) & No & No & No\\
Employer & Original & -3.2 (-4.3, -2.1) &  &  & \\
 & +  trend & -1.5 (-4.0, 1.1) & -1.7 (-4.1, 0.7) & No & No & Yes\\
Individual & Original & -0.8 (-1.2, -0.4) &  &  & \\
 & +  trend & -0.2 (-1.4, 1.0) & -0.6 (-1.6, 0.4) & Yes{$^\dagger$} & Yes{$^\dagger$} & Yes\\
Government& Original & -0.4 (-1.6, 0.7) &  & & \\
 & +  trend & 1.4 (-0.7, 3.6) & -1.9 (-3.8, 0.0) & No & No & Yes\\
\bottomrule
\end{tabular*}
\begin{tablenotes}[flushleft, leftmargin=0pt]
\item \emph{Treatment effects were estimated by fitting the models in Eqs.~(\ref{eq:aca_model}) and (\ref{eq:aca_exp_model}), each representing the differential change, on the percentage point scale, averaged over the post-implementation period. Diff = difference in treatment effect in reduced versus expanded model; CI = confidence interval}
\item {$^\dagger$} Indicates rule out in the main specification without the time*unemployment interaction, but not in the model with the interaction.
\end{tablenotes}
\end{threeparttable}
\end{table}

\begin{table*}
\centering %
\small
\caption{Empirical simulation results.\label{tab:emp_sim_results}}
\begin{tabular*}{\textwidth}{@{\extracolsep\fill}llcccc@{\extracolsep\fill}}
\toprule%
& & \multicolumn{2}{c}{\emph{No violation}} & \multicolumn{2}{c}{\emph{Minor violation (1.1\%)}} \\
 & & Normal & Bootstrap & Normal & Bootstrap\\
\hline
Non-inferiority & Rule out $|5|$  & 95 & 100 & 88 & 97\\
test power &  Rule out $2.1$ & 47 & 60 & 21 & 26\\
&Rule out $1$ & 20 & 23 & 5\tnote{$^\dagger$} &5\tnote{$^\dagger$} \\ 
\hline
Reduced model \\
 & ATT & 3.4 & 3.4 & 4.5 & 4.5\\
& Bias (\%) & 0 & 0 & 31 & 31\\
\emph{Incremental}& | Rule out $|5|$& 0 & 0 & 0 & 0\\
\emph{test-induced} & | Rule out $2.1$ & 0 & -2 & 1 & -4\\
\emph{bias (\%)}& | Rule out $1$& 1 & -3 & 1 & -6\\ \hline

Expanded model \\
& ATT &   3.4 & 3.4 & 3.4 & 3.4\\
& {Bias (\%)} & 0 & 0 & 0 & 0\\
\emph{Incremental}& | Rule out $|5|$   & 0 & 0 & 8 & 2\\
\emph{test-induced}& | Rule out $2.1$   & 33 & 18 & 55 & 36\\
\emph{bias (\%)}& | Rule out $1$ & 55 & 38 & 80 & 58\\
\\
& | Cannot rule out $|5|$ & 1 & 1 & -61 & -64\\
& | Cannot rule out $2.1$ & -30 & -28 & -15 & -12\\
& | Cannot rule out $1$  & -15 & -12 & -5 & -3\\
\bottomrule%
\end{tabular*}
\begin{tablenotes}[flushleft, leftmargin=0pt]
\item \emph{This table reports results from empirical simulations ($n=15,000$ per scenario) from normal and bootstrap-based data-generating processes. The top section describes non-inferiority test power over different thresholds. The bottom two sections display the ATT in the reduced and expanded models, percentage bias, and incremental percentage bias conditional on passing (| Rule out) or failing (| Cannot rule out) non-inferiority tests at different thresholds.}
\item {$^\dagger$} Violation exceeds threshold; thus, power column represents Type I error (controlled at $\alpha = 0.05$).
\end{tablenotes}
\end{table*}

\end{document}


\begin{titlepage}
\title{Supplement for ``Nothing to See Here? A non-inferiority approach to parallel trends"}
\author{Alyssa Bilinski, PhD and Laura A. Hatfield, PhD}
\date{}
\maketitle
\thispagestyle{empty}
\end{titlepage}
\clearpage

\renewcommand{\theequation}{S\arabic{equation}}

\begin{table}
\caption{{Literature review of \emph{JAMA} and \emph{JAMA IM} difference-in-differences (DiD) studies.}  We summarize all DiD studies published in the \emph{Journal of the American Medical Association} (\emph{JAMA}) and \emph{JAMA Internal Medicine} (\emph{IM}) from 2018 through 2022. Studies were identified by searching the journals' archives for the term difference(s)(-)in(-)difference(s). They are described in terms of whether they were research letters ({RL}), whether they mentioned the parallel trends (PT) assumption, what tests were used to assess parallel trends (a linear slope test, a joint-F test for pre-intervention deviations, or other), and whether outcome plots or event study (E-S) plots were shown.}  
\label{tab:jama_studies1}
\begin{tabularx}{1\textwidth}{lllllll}
\toprule
\textbf{RL} & \textbf{Mention PT} & \textbf{Slope} & \textbf{Joint F} & \textbf{Other} & \textbf{Plot} & \textbf{E-S plot} \\
\toprule
\multicolumn{7}{p{0.9\linewidth}}{(1) \emph{Association Between COVID-19 Lockdown Measures and Emergency Department Visits for Violence-Related Injuries in Cardiff, Wales}} \\
\multicolumn{7}{p{0.9\linewidth}}{10.1001/jama.2020.25511} \\
Yes & No & No & No & No & Yes & No \\
\midrule
\multicolumn{7}{p{0.9\linewidth}}{(2) \emph{Association Between Hospital Voluntary Participation, Mandatory Participation, or Nonparticipation in Bundled Payments and Medicare Episodic Spending for Hip and Knee Replacements}} \\
\multicolumn{7}{p{0.9\linewidth}}{10.1001/jama.2021.10046} \\
\multicolumn{7}{p{0.9\linewidth}}{``There were nondivergent trends in episodic spending across hospital groups during the period before starting the bundled payment program."} \\
Yes & Yes & No & Yes & No & No & No \\
\midrule
\multicolumn{7}{p{0.9\linewidth}}{(3) \emph{Association Between State-Mandated Protocolized Sepsis Care and In-hospital Mortality Among Adults With Sepsis}} \\
\multicolumn{7}{p{0.9\linewidth}}{10.1001/jama.2019.9021} \\
\multicolumn{7}{p{0.9\linewidth}}{``We directly examined for this possibility by fitting a model containing a treatment indicator, a continuous time variable, the interaction of these 2 variables, and all patient- and hospital-level covariates, restricted to the preregulation period...We considered parallel trends as being present if the interaction term from this model was not significant. In cases in which there were parallel trends, we simplified the comparative interrupted time series model to a difference-in-differences model by excluding the term for the interaction of the treatment indicator with the continuous time variable."} \\
No & Yes & Yes & No & No & Yes & No \\
\midrule
\multicolumn{7}{p{0.9\linewidth}}{(4) \emph{Association Between the Experimental Kickoff Rule and Concussion Rates in Ivy League Football}} \\
\multicolumn{7}{p{0.9\linewidth}}{10.1001/jama.2018.14165} \\
Yes & No & No & No & No & No & No \\
\midrule
\multicolumn{7}{p{0.9\linewidth}}{(5) \emph{Association Between the Implementation of a Population-Based Primary Care Payment System and Achievement on Quality Measures in Hawaii}} \\
\multicolumn{7}{p{0.9\linewidth}}{10.1001/jama.2019.8113} \\
\multicolumn{7}{p{0.9\linewidth}}{``We estimated the risk-standardized probability of achieving the primary outcome, which indicated no significant difference between the groups in trends before intervention (eFigures 2-4 in the Supplement)."} \\
No & Yes & No & No & No & Yes & No \\
\midrule
\end{tabularx}
\end{table} 

\clearpage

\begin{tabularx}{1\textwidth}{lllllll}
\midrule
\textbf{RL} & \textbf{Mention PT} & \textbf{Slope} & \textbf{Joint F} & \textbf{Other} & \textbf{Plot} & \textbf{E-S plot} \\
\midrule
\multicolumn{7}{p{0.9\linewidth}}{(6) \emph{Association of a Beverage Tax on Sugar-Sweetened and Artificially Sweetened Beverages With Changes in Beverage Prices and Sales at Chain Retailers in a Large Urban Setting}} \\
\multicolumn{7}{p{0.9\linewidth}}{10.1001/jama.2019.4249} \\
\multicolumn{7}{p{0.9\linewidth}}{``Analyses focused on the years 2016 and 2017 because the parallel trends assumption (ie, that the preintervention trend in the outcome is similar for the treatment and control locations) held for beverage volume sales in Philadelphia compared with Baltimore during 2016 but did not hold from January 1, 2014, to December 31, 2015, based on generalized estimating equations using a continuous time variable, the locations, and the interaction between the 2 (eAppendix 1 A.4.a in the Supplement)."} \\
No & Yes & Yes & No & No & Yes & No \\
\midrule
\multicolumn{7}{p{0.9\linewidth}}{(7) \emph{Association of Coronary Artery Bypass Grafting vs Percutaneous Coronary Intervention With Memory Decline in Older Adults Undergoing Coronary Revascularization}} \\
\multicolumn{7}{p{0.9\linewidth}}{10.1001/jama.2021.5150} \\
No & No & No & No & No & No & No \\
\midrule
\multicolumn{7}{p{0.9\linewidth}}{(8) \emph{Association of Hospital Participation in a Medicare Bundled Payment Program With Volume and Case Mix of Lower Extremity Joint Replacement Episodes}} \\
\multicolumn{7}{p{0.9\linewidth}}{10.1001/jama.2018.12345} \\
\multicolumn{7}{p{0.9\linewidth}}{``The assumption of parallel trends under our difference-in-differences method was tested using a generalized linear regression of volume on a BPCI market indicator, time variable, and the interaction, using a Wald test that did not indicate divergent secular trends during the pre-BPCI period ($P = .92$) (eTable 1 in the Supplement)."} \\
No & Yes & Yes & No & No & Yes & No \\
\midrule
\multicolumn{7}{p{0.9\linewidth}}{(9) \emph{Association of Hospital Participation in Bundled Payments for Care Improvement Advanced With Medicare Spending and Hospital Incentive Payments}} \\
\multicolumn{7}{p{0.9\linewidth}}{10.1001/jama.2022.18529} \\
\multicolumn{7}{p{0.9\linewidth}}{``Visual inspection (eFigure 2 in the Supplement) and statistical tests (eTable 3 in the Supplement) revealed small, nonsignificant differences in preintervention episode spending for episodes at BPCI-A vs comparison hospitals. Nevertheless, we adjusted for these differences in preintervention trends to produce a conservative estimate of the association of BPCI-A participation with changes in episode spending (eTable 4 in the Supplement)."} \\
No & Yes & Yes & No & No & Yes & Yes \\
\midrule
\end{tabularx}

\clearpage

\begin{tabularx}{1\textwidth}{lllllll}
\midrule
\textbf{RL} & \textbf{Mention PT} & \textbf{Slope} & \textbf{Joint F} & \textbf{Other} & \textbf{Plot} & \textbf{E-S plot} \\
\midrule
\multicolumn{7}{p{0.9\linewidth}}{(10) \emph{Association of Initiation of Basal Insulin Analogs vs Neutral Protamine Hagedorn Insulin With Hypoglycemia-Related Emergency Department Visits or Hospital Admissions and With Glycemic Control in Patients With Type 2 Diabetes}} \\
\multicolumn{7}{p{0.9\linewidth}}{10.1001/jama.2018.7993} \\
\multicolumn{7}{p{0.9\linewidth}}{``This model was based on the counterfactual assumption that if patients who initiated insulin analogs had instead initiated NPH insulin, their changes in hemoglobin A1c level would be similar to the changes observed in the NPH insulin reference group, who were frequency matched based on the propensity score quintile."} \\
No & Yes & No & No & No & No & No \\
\midrule
\multicolumn{7}{p{0.9\linewidth}}{(11) \emph{Association of Medicaid Expansion With 1-Year Mortality Among Patients With End-Stage Renal Disease}} \\
\multicolumn{7}{p{0.9\linewidth}}{10.1001/jama.2018.16504} \\
\multicolumn{7}{p{0.9\linewidth}}{``Mortality rates were similar in expansion and nonexpansion states prior to 2014 and then diverged, with declines in mortality rates beginning in the first 6 months of 2014." ``Pre-2014 trends in outcomes for expansion and nonexpansion states did not differ significantly (eTable 4 in the Supplement)."} \\
No & Yes & Yes & No & No & Yes & No \\
\midrule
\multicolumn{7}{p{0.9\linewidth}}{(12) \emph{Association of Participation in the Oncology Care Model With Medicare Payments, Utilization, Care Delivery, and Quality Outcomes}} \\
\multicolumn{7}{p{0.9\linewidth}}{10.1001/jama.2021.17642} \\
\multicolumn{7}{p{0.9\linewidth}}{``For each claims-based measure, we tested the null hypothesis that OCM and comparison episodes had parallel trends during the 18-month baseline period (eTable 5 in Supplement 1). Difference-in-differences results are not reported for outcome measures for which we rejected the parallel trends assumption ($\alpha = .05$)."} \\
No & Yes & Yes & No & No & No & No \\
\midrule
\multicolumn{7}{p{0.9\linewidth}}{(13) \emph{Association of Real-time Continuous Glucose Monitoring With Glycemic Control and Acute Metabolic Events Among Patients With Insulin-Treated Diabetes}} \\
\multicolumn{7}{p{0.9\linewidth}}{10.1001/jama.2021.6530} \\
\multicolumn{7}{p{0.9\linewidth}}{``The difference-in-differences method assumes outcomes in the exposed group—had they not been exposed to the intervention (ie, counterfactual case)—would be qualitatively similar to the observed outcomes in the unexposed (reference) group" ``No violations of model assumptions (ie, experimental treatment assignment, parallel trends, common shock, and no spillover) were detected."} \\
No & Yes & No & No & No & No & No \\
\midrule
\multicolumn{7}{p{0.9\linewidth}}{(14) \emph{Association of Remote vs In-Person Benefit Delivery With WIC Participation During the COVID-19 Pandemic}} \\
\multicolumn{7}{p{0.9\linewidth}}{10.1001/jama.2021.14356} \\
\multicolumn{7}{p{0.9\linewidth}}{``There was no statistical evidence of differing trends in WIC participation across these states prior to the pandemic ($\beta = .0002; P = .91$)."} \\
Yes & Yes & Yes & No & No & No & No \\
\midrule
\end{tabularx}

\clearpage

\begin{tabularx}{1\textwidth}{lllllll}
\midrule
\textbf{RL} & \textbf{Mention PT} & \textbf{Slope} & \textbf{Joint F} & \textbf{Other} & \textbf{Plot} & \textbf{E-S plot} \\
\midrule
\multicolumn{7}{p{0.9\linewidth}}{(15) \emph{Association of Skilled Nursing Facility Participation in a Bundled Payment Model With Institutional Spending for Joint Replacement Surgery}} \\
\multicolumn{7}{p{0.9\linewidth}}{10.1001/jama.2020.19181} \\
\multicolumn{7}{p{0.9\linewidth}}{``The validity of the difference-in-differences approach was assessed by testing the parallel trends assumption, comparing the slope over time in the pre-BPCI period for BPCI participants and nonparticipants. None of the outcomes had violations of this assumption (eTable 1 in the Supplement)."} \\
No & Yes & Yes & No & No & No & No \\
\midrule
\multicolumn{7}{p{0.9\linewidth}}{(16) \emph{Association of State Medicaid Expansion Status With Low Birth Weight and Preterm Birth}} \\
\multicolumn{7}{p{0.9\linewidth}}{10.1001/jama.2019.3678} \\
\multicolumn{7}{p{0.9\linewidth}}{``This Figure was used to visualize trends across the study period as well as whether trends were parallel prior to the expansion date, which was formally tested using linear indicator variables in the DID and DDD models. The DDD comparison of relative disparities in rates of low birth weight between black and white infants as well as 2 DID comparisons among outcomes in black infants (preterm birth and low birth weight) failed the parallel trends test; however, all other significant DID and DDD comparisons in the primary analyses passed this test (Table 3 and Table 4)."} \\
No & Yes & Yes & No & No & Yes & No \\
\midrule
\multicolumn{7}{p{0.9\linewidth}}{(17) \emph{Association of the Affordable Care Act Dependent Coverage Provision With Prenatal Care Use and Birth Outcomes}} \\
\multicolumn{7}{p{0.9\linewidth}}{10.1001/jama.2018.0030} \\
\multicolumn{7}{p{0.9\linewidth}}{``We detected small but significant differential linear trends prior to 2010 for early prenatal care and neonatal intensive care unit (NICU) admission overall and among unmarried women only (eTable 2 in the Supplement). Estimating our primary difference-in-differences regression as if the policy took effect in July 2009 (post hoc placebo testing), we identified a similar pattern, with significant results for early prenatal care and NICU admission overall and among unmarried women only (eTable 3 in the Supplement)."} \\
No & Yes & Yes & No & No & Yes & No \\
\midrule
\multicolumn{7}{p{0.9\linewidth}}{(18) \emph{Association of the Healthy, Hunger-Free Kids Act With Dietary Quality Among Children in the US National School Lunch Program}} \\
\multicolumn{7}{p{0.9\linewidth}}{10.1001/jama.2020.9517} \\
No & No & No & No & No & No & No \\
\midrule
\end{tabularx}

\clearpage

\begin{tabularx}{1\textwidth}{lllllll}
\midrule
\textbf{RL} & \textbf{Mention PT} & \textbf{Slope} & \textbf{Joint F} & \textbf{Other} & \textbf{Plot} & \textbf{E-S plot} \\
\midrule
\multicolumn{7}{p{0.9\linewidth}}{(19) \emph{Hospital Quality Improvement Interventions, Statewide Policy Initiatives, and Rates of Cesarean Delivery for Nulliparous, Term, Singleton, Vertex Births in California}} \\
\multicolumn{7}{p{0.9\linewidth}}{10.1001/jama.2021.3816} \\
No & No & No & No & No & Yes & No \\
\midrule
\multicolumn{7}{p{0.9\linewidth}}{(20) \emph{Implementation of a Health Plan Program for Switching From Analogue to Human Insulin and Glycemic Control Among Medicare Beneficiaries With Type 2 Diabetes}} \\
\multicolumn{7}{p{0.9\linewidth}}{10.1001/jama.2018.21364} \\
No & No & No & No & No & Yes & No \\ \midrule
\multicolumn{7}{p{0.9\linewidth}}{(21) \emph{Medical Debt in the US, 2009-2020}} \\
\multicolumn{7}{p{0.9\linewidth}}{10.1001/jama.2021.8694} \\
\multicolumn{7}{p{0.9\linewidth}}{{[}Placebo outcome test:{]} ``To assess whether the association between Medicaid expansion and medical debt reflected confounding factors (such as differential economic trends), we conducted the analyses separately using nonmedical debt as the outcome."} \\
No & No & No & No & No & Yes & No \\
\midrule
\multicolumn{7}{p{0.9\linewidth}}{(22) \emph{Pass-Through of a Tax on Sugar-Sweetened Beverages at the Philadelphia International Airport}} \\
\multicolumn{7}{p{0.9\linewidth}}{10.1001/jama.2017.16903} \\
Yes & No & No & No & No & No & No \\
\midrule
\multicolumn{7}{p{0.9\linewidth}}{(23) \emph{Prescription Drug Monitoring Program Mandates and Opioids Dispensed Following Emergency Department Encounters for Patients With Sickle Cell Disease or Cancer With Bone Metastasis}} \\
\multicolumn{7}{p{0.9\linewidth}}{10.1001/jama.2021.10161} \\
\multicolumn{7}{p{0.9\linewidth}}{{[}In main text:{]} ``Parallel trends assumptions before mandate implementation were met (Supplement)." {[}In supplement:{]} ``For a staggered design, the state-of-the-art approach to testing the parallel trend assumption is to conduct an event study analysis." ``All estimated differences were not statistically different from 0, supporting parallel trends in the two outcomes leading up to implementation of any mandate." ``Here, results of the event study analysis suggest parallel trends in all outcome-sample combinations except one....Moreover, our finding that comprehensive mandates were associated with a reduction in MMEs dispensed to patients with SCD represented a reversal of the temporal trend seen over 7-18 months before implementation, rather than a continuation of a pre-existing trend. We thus do not consider this temporal deviation from the parallel trend assumption a threat to the validity of our findings."} \\
Yes & Yes & No & No & Yes & No & Yes \\
\midrule
\end{tabularx}

\begin{tabularx}{1\textwidth}{lllllll}
\midrule
\textbf{RL} & \textbf{Mention PT} & \textbf{Slope} & \textbf{Joint F} & \textbf{Other} & \textbf{Plot} & \textbf{E-S plot} \\
\midrule
\multicolumn{7}{p{0.9\linewidth}}{(24) \emph{Seasonal Influenza Activity During the SARS-CoV-2 Outbreak in Japan}} \\
\multicolumn{7}{p{0.9\linewidth}}{10.1001/jama.2020.6173} \\
Yes & No & No & No & No & No & No \\
\midrule
\multicolumn{7}{p{0.9\linewidth}}{(25) \emph{Trends in US Ambulatory Care Patterns During the COVID-19 Pandemic, 2019-2021}} \\
\multicolumn{7}{p{0.9\linewidth}}{10.1001/jama.2021.24294} \\
\multicolumn{7}{p{0.9\linewidth}}{``Second, 2019 utilization rates were assumed to be reasonable counterfactual control rates for 2020 had the pandemic not occurred. We concluded that this assumption was plausible after visually comparing trends between 2018 and 2019 (Figure 1 and eFigure 7 in the Supplement) and tested this assumption using a placebo test for parallel trends (eTable 3 in the Supplement)."} \\
No & Yes & No & No & No & Yes & No\\
\midrule
\multicolumn{7}{p{0.9\linewidth}}{(26) \emph{Association Between Automotive Assembly Plant Closures and Opioid Overdose Mortality in the United States}} \\
\multicolumn{7}{p{0.9\linewidth}}{10.1001/jamainternmed.2019.5686} \\
\multicolumn{7}{p{0.9\linewidth}}{{[}In main text:{]} ``Prior to plant closures, baseline opioid overdose mortality rates in exposed counties were lower than those in unexposed counties, with no evidence of differential trends in the primary outcomes." {[}In supplement: {]} ``The event study specifications, which also provide a more transparent test of violations of the parallel trends assumption required for causal inference, avoid this problem by indexing the reference point to time since the event (as opposed to calendar time) and by allowing associations to vary over time."} \\
No & Yes & No & No & No & No & Yes \\
\midrule
\multicolumn{7}{p{0.9\linewidth}}{(27) \emph{Association Between Medicaid Expansion and Rates of Opioid-Related Hospital Use}} \\
\multicolumn{7}{p{0.9\linewidth}}{10.1001/jamainternmed.2020.0473} \\
\multicolumn{7}{p{0.9\linewidth}}{``However, we estimated event study regressions comparing changes between expansion and nonexpansion states to assess parallel trends in the prepolicy period. eFigure2 in the Supplement provides statistical data suggesting that the expansion and nonexpansion states mostly changed similarly in preexpansion years, which increases our confidence that the states would have continued to trend similarly were it not for the expansion."} \\
No & Yes & No & No & No & No & Yes \\
\midrule
\multicolumn{7}{p{0.9\linewidth}}{(28) \emph{Association Between State Laws Facilitating Pharmacy Distribution of Naloxone and Risk of Fatal Overdose}} \\
\multicolumn{7}{p{0.9\linewidth}}{10.1001/jamainternmed.2019.0272} \\
\multicolumn{7}{p{0.9\linewidth}}{``We tested the parallel trends assumption, which is necessary for obtaining unbiased estimates using the difference-in-differences framework, through event study analyses, which is recommended when evaluating health policies." ``Small and statistically non-significant estimates before adoption suggest that the parallel trends assumption was satisfied."} \\
No & Yes & No & No & No & No & Yes \\
\midrule
\end{tabularx}

\clearpage

\begin{tabularx}{1\textwidth}{lllllll}
\midrule
\textbf{RL} & \textbf{Mention PT} & \textbf{Slope} & \textbf{Joint F} & \textbf{Other} & \textbf{Plot} & \textbf{E-S plot} \\
\midrule
\multicolumn{7}{p{0.9\linewidth}}{(29) \emph{Association between US state medical cannabis laws and opioid prescribing in the Medicare Part D population}} \\
\multicolumn{7}{p{0.9\linewidth}}{10.1001/jamainternmed.2018.0266} \\
\multicolumn{7}{p{0.9\linewidth}}{``We tested our data for parallel trends in prescribing between ``never-MCL" states and pre-MCL years for states that implement the policy during our study period; we cannot reject the null hypothesis of parallel trends, which supports the use of our models (see online eAppendix eTable 8 in the Supplement)."} \\
No & Yes & Yes & No & No & No & No \\
\midrule
\multicolumn{7}{p{0.9\linewidth}}{(30) \emph{Association of Cigarette Sales With Comprehensive Menthol Flavor Ban in Massachusetts}} \\
\multicolumn{7}{p{0.9\linewidth}}{10.1001/jamainternmed.2021.7333} \\
\multicolumn{7}{p{0.9\linewidth}}{``There were nondivergent trends in state-level sales of menthol and nonflavored cigarette packs per 1000 people in Massachusetts and comparison states during the period before Massachusetts's comprehensive flavor ban."} \\
Yes & Yes & No & No & No & No & No \\
\midrule
\multicolumn{7}{p{0.9\linewidth}}{(31) \emph{Association of Coded Severity With Readmission Reduction After the Hospital Readmissions Reduction Program}} \\
\multicolumn{7}{p{0.9\linewidth}}{10.1001/jamainternmed.2017.6148} \\
\multicolumn{7}{p{0.9\linewidth}}{``Trends in rates of readmission were parallel between control and exposed hospitals before implementation of the HRRP (Figure, B and C)."} \\
Yes & Yes & No & No & No & Yes & No \\
\midrule
\multicolumn{7}{p{0.9\linewidth}}{(32) \emph{Association of County-Level Prescriptions for Hydroxychloroquine and Ivermectin With County-Level Political Voting Patterns in the 2020 US Presidential Election}} \\
\multicolumn{7}{p{0.9\linewidth}}{10.1001/jamainternmed.2022.0200} \\
\multicolumn{7}{p{0.9\linewidth}}{{[}Placebo outcome test:{]} ``We assessed countylevel rates of new prescriptions for hydroxychloroquine and ivermectin (ie, patients with no fills for the medication in the previous 6 months) per 100 000 enrollees and 2 control medications, methotrexate sodium and albendazole (which have similar clinical applications as hydroxychloroquine or ivermectin, respectively, but are not proposed as COVID-19 treatments)....There were no substantive changes in overall prescribing volume for methotrexate or albendazole."} \\
Yes & Yes & No & No & Yes & Yes & No \\
\midrule
\end{tabularx}

\clearpage

\begin{tabularx}{1\textwidth}{lllllll}
\midrule
\textbf{RL} & \textbf{Mention PT} & \textbf{Slope} & \textbf{Joint F} & \textbf{Other} & \textbf{Plot} & \textbf{E-S plot} \\
\midrule
\multicolumn{7}{p{0.9\linewidth}}{(33) \emph{Association of Disability Compensation With Mortality and Hospitalizations Among Vietnam-Era Veterans With Diabetes}} \\
\multicolumn{7}{p{0.9\linewidth}}{10.1001/jamainternmed.2022.2159} \\
\multicolumn{7}{p{0.9\linewidth}}{``we tested the significance of an interaction between BOG status and quarter (indicating that trends in outcomes were not different for BOG and NOG in the six quarters prior to the policy change). We did not examine parallel pre-policy trends for disability compensation payments because these data were available on an annual basis, leaving only two pre-policy measurements."} \\
No & Yes & Yes & No & No & Yes & No \\
\midrule
\multicolumn{7}{p{0.9\linewidth}}{(34) \emph{Association of Medicaid Expansion With Quality in Safety-Net Hospitals}} \\
\multicolumn{7}{p{0.9\linewidth}}{10.1001/jamainternmed.2020.9142} \\
\multicolumn{7}{p{0.9\linewidth}}{``Formal tests of preexpansion trends did not reach statistical significance (eTable 5 in the Supplement)."} \\
No & Yes & Yes & No & No & Yes & No \\
\midrule
\multicolumn{7}{p{0.9\linewidth}}{(35) \emph{Association of Medical and Adult-Use Marijuana Laws With Opioid Prescribing for Medicaid Enrollees}} \\
\multicolumn{7}{p{0.9\linewidth}}{10.1001/jamainternmed.2018.1007} \\
\multicolumn{7}{p{0.9\linewidth}}{{[}In main text:{]} "we performed ``parallel-trend assumption" tests by statistically and graphically comparing the prepolicy trends between medical marijuana states, adult-use marijuana states, and the comparison states." ``Moreover, the ``parallel-trend assumption" tests and falsification tests lent weight to the validity of the methods (Supplement)."} \\
No & Yes & Yes & No & No & Yes & No \\
\midrule
\multicolumn{7}{p{0.9\linewidth}}{(36) \emph{Association of Physician Management Companies and Private Equity Investment With Commercial Health Care Prices Paid to Anesthesia Practitioners}} \\
\multicolumn{7}{p{0.9\linewidth}}{10.1001/jamainternmed.2022.0004} \\
\multicolumn{7}{p{0.9\linewidth}}{``The coefficients in the precontract period did not differ significantly between PMC and non-PMC facilities."} \\
No & Yes & No & No & No & No & Yes \\
\midrule
\multicolumn{7}{p{0.9\linewidth}}{(37) \emph{Association of Scheduled vs Emergency-Only Dialysis With Health Outcomes and Costs in Undocumented Immigrants With End-stage Renal Disease}} \\
\multicolumn{7}{p{0.9\linewidth}}{10.1001/jamainternmed.2018.5866} \\
\multicolumn{7}{p{0.9\linewidth}}{``Our DiD approach accounts for between-group differences, assuming that patients in the scheduled dialysis group would have had utilization and cost trends parallel to those of patients in the emergency-only group had they not received coverage."} \\
No & Yes & No & No & No & Yes & No \\
\midrule
\end{tabularx}

\clearpage

\begin{tabularx}{1\textwidth}{lllllll}
\midrule
\textbf{RL} & \textbf{Mention PT} & \textbf{Slope} & \textbf{Joint F} & \textbf{Other} & \textbf{Plot} & \textbf{E-S plot} \\
\midrule
\multicolumn{7}{p{0.9\linewidth}}{(38) \emph{Association of state opioid prescription duration limits with changes in opioid prescribing for Medicare beneficiaries}} \\
\multicolumn{7}{p{0.9\linewidth}}{10.1001/jamainternmed.2021.4281} \\
\multicolumn{7}{p{0.9\linewidth}}{``Before the start of duration limits in 2016, days of opioid prescribed were parallel in exposed states and control states."} \\
Yes & Yes & No & No & No & Yes & No \\
\midrule
\multicolumn{7}{p{0.9\linewidth}}{(39) \emph{Association of Surprise-Billing Legislation with Prices Paid to In-Network and Out-of-Network Anesthesiologists in California, Florida, and New York: An Economic Analysis}} \\
\multicolumn{7}{p{0.9\linewidth}}{10.1001/jamainternmed.2021.4564} \\
\multicolumn{7}{p{0.9\linewidth}}{{[}Found pre-trends and then:{]} ``Because anticipatory effects would invalidate the difference-in-differences assumption of no preexisting trends, eTable 5 and eFigures 2 and 3 in the Supplement present results of the analyses for California and Florida using the quarter during which the law was introduced to the state legislature."} \\
No & Yes & No & No & No & No & Yes \\
\midrule
\multicolumn{7}{p{0.9\linewidth}}{(40) \emph{Association of Team-Based Primary Care With Health Care Utilization and Costs Among Chronically Ill Patients}} \\
\multicolumn{7}{p{0.9\linewidth}}{10.1001/jamainternmed.2018.5118} \\
\multicolumn{7}{p{0.9\linewidth}}{``We visually and formally checked the assumption of parallel trends in the pre-period for valid difference-in-difference inference." ``In eFigure 2 in the Supplement, we present figures and empirical tests of the parallel trends assumption and find that parallel trends exist in all variables except for total cost of care and outpatient visits in the more than 2 comorbidity sample. Parallel trends were not met in any outcome in the less than 2 comorbidity subsample."} \\
No & Yes & Yes & No & No & Yes & No \\
\midrule
\multicolumn{7}{p{0.9\linewidth}}{(41) \emph{Association of the Comprehensive End-Stage Renal Disease Care Model With Medicare Payments and Quality of Care for Beneficiaries With End-Stage Renal Disease}} \\
\multicolumn{7}{p{0.9\linewidth}}{10.1001/jamainternmed.2020.0562} \\
\multicolumn{7}{p{0.9\linewidth}}{``A core assumption of the difference-in-differences design is that the intervention and comparison populations have parallel trends for a given outcome during the baseline period. Parallel trend tests were conducted for all outcomes at the 5\% level. All outcomes, except catheter use for wave 1 facilities, passed parallel trend tests (eTable 2 in the Supplement). Although statistical trend tests of wave 1 catheter use did not pass, visual inspection of the relative trends to the comparison group appeared parallel. In addition, the coefficient of the difference in trends at baseline, although significant, equalled 0.00046."} \\
No & Yes & Yes & No & No & No & No \\
\midrule
\end{tabularx}

\clearpage

\begin{tabularx}{1\textwidth}{lllllll}
\midrule
\textbf{RL} & \textbf{Mention PT} & \textbf{Slope} & \textbf{Joint F} & \textbf{Other} & \textbf{Plot} & \textbf{E-S plot} \\
\midrule
\multicolumn{7}{p{0.9\linewidth}}{(42) \emph{Changes in Health Care Use and Outcomes After Turnover in Primary Care}} \\
\multicolumn{7}{p{0.9\linewidth}}{10.1001/jamainternmed.2020.6288} \\
\multicolumn{7}{p{0.9\linewidth}}{{[}In main text:{]} ``Modeled pretrends of exposed and unexposed beneficiaries' health care use and health outcomes did not systematically differ prior to PCP exit (eMethods 2 in the Supplement)." {[}In the supplement:{]} ``Treated and control patients see their assigned PCPs at the same rate as illustrated by curves moving in parallel. Both curves slope downward due to mean reversion and patients dying over time."} \\
No & Yes & No & No & No & Yes & Yes \\
\midrule
\multicolumn{7}{p{0.9\linewidth}}{(43) \emph{Changes in Health Care Use Associated With the Introduction of Hospital Global Budgets in Maryland}} \\
\multicolumn{7}{p{0.9\linewidth}}{10.1001/jamainternmed.2017.7455} \\
\multicolumn{7}{p{0.9\linewidth}}{``This approach, which is standard in difference-in-difference studies, relies on the assumption of parallel preintervention trends in Maryland and the control group. However, it could produce biased estimates if preintervention trends differ. Therefore, in a second analysis, we assumed that differences between Maryland and the control group would have continued to change at the preintervention rate in the absence of Maryland's program."} \\
No & Yes & No & No & Yes & Yes & No \\
\midrule
\multicolumn{7}{p{0.9\linewidth}}{(44) \emph{Changes in Hospital Income, Use, and Quality Associated With Private Equity Acquisition}} \\
\multicolumn{7}{p{0.9\linewidth}}{10.1001/jamainternmed.2020.3552} \\
\multicolumn{7}{p{0.9\linewidth}}{``A joint F test to assess differences in preacquisition trends for hospital income and use measures as well as process quality measures did not show a significant difference for any of the 11 measures (eTable 14 in the Supplement)."} \\
No & Yes & No & Yes & No & Yes & No \\
\midrule
\multicolumn{7}{p{0.9\linewidth}}{(45) \emph{Clinical Outcomes After Intensifying Antihypertensive Medication Regimens Among Older Adults at Hospital Discharge}} \\
\multicolumn{7}{p{0.9\linewidth}}{10.1001/jamainternmed.2019.3007} \\
No & No & No & No & No & No & No \\
\midrule
\multicolumn{7}{p{0.9\linewidth}}{(46) \emph{Evaluation of Economic and Clinical Outcomes Under Centers for Medicare \& Medicaid Services Mandatory Bundled Payments for Joint Replacements}} \\
\multicolumn{7}{p{0.9\linewidth}}{10.1001/jamainternmed.2019.0480} \\
\multicolumn{7}{p{0.9\linewidth}}{{[}In the main text:{]} ``Tests of preintervention spending trends between treatment and control showed that differences were not statistically significant (eTable 3 in the Supplement)." {[}In the supplement:{]} ``This table reports model estimates of differential changes in pre-intervention spending at the episode level associated with random assignment of the index hospital into the CJR model. The model was adjusted for age, sex, CMS Hierarchical Condition Category (CMS-HCC) risk score, indicator for hip fracture status, hospital fixed effect, a vector of quarter indicators, and interactions between each quarter and CJR status."} \\
No & Yes & Yes & No & No & Yes & No \\
\midrule
\end{tabularx}

\clearpage

\begin{tabularx}{1\textwidth}{lllllll}
\hline
\textbf{RL} & \textbf{Mention PT} & \textbf{Slope} & \textbf{Joint F} & \textbf{Other} & \textbf{Plot} & \textbf{E-S plot} \\
\hline
\multicolumn{7}{p{0.9\linewidth}}{(47) \emph{Health Care Utilization and Cost Outcomes of a Comprehensive Dementia Care Program for Medicare Beneficiaries}} \\
\multicolumn{7}{p{0.9\linewidth}}{10.1001/jamainternmed.2018.5579} \\
No & No & No & No & No & No & No \\
\hline
\multicolumn{7}{p{0.9\linewidth}}{(48) \emph{Hospital Responses to Incentives in Episode-Based Payment for Joint Surgery: A Controlled Population-Based Study}} \\
\multicolumn{7}{p{0.9\linewidth}}{10.1001/jamainternmed.2021.1897} \\
\multicolumn{7}{p{0.9\linewidth}}{{[}In main text:{]} ``In a test of the assumptions in a difference-in-differences model, we did not observe differential trends in our outcomes in the period before CJR implementation (eTable 2 in the Supplement); nor did we see differential shifts in LEJR volumes (eMethods 4 and eTable 3 in the Supplement)." {[}In the supplement:{]} ``As shown in eTable 2, we found that LEJR spending and risk were increasing differentially in MSAs selected for the CJR program in the pre-period, but statistically we were unable to reject the null hypothesis of no differential pre-period trend for any of our outcomes."} \\
No & Yes & Yes & No & No & Yes & No \\
\hline
\multicolumn{7}{p{0.9\linewidth}}{(49) \emph{Medicare Accountable Care Organization Enrollment and Appropriateness of Cancer Screening}} \\
\multicolumn{7}{p{0.9\linewidth}}{10.1001/jamainternmed.2017.8087} \\
No & No & No & No & No & No & No \\
\hline
\multicolumn{7}{p{0.9\linewidth}}{(50) \emph{Rates of Advanced Imaging by Practice Peers After Malpractice Injury Reports in Florida, 2009-2013}} \\
\multicolumn{7}{p{0.9\linewidth}}{10.1001/jamainternmed.2019.0163} \\
Yes & No & No & No & No & No & Yes \\
\hline
\multicolumn{7}{p{0.9\linewidth}}{(51) \emph{Utilization of Long-Acting Reversible Contraceptives in the United States After vs Before the 2016 US Presidential Election}} \\
\multicolumn{7}{p{0.9\linewidth}}{10.1001/jamainternmed.2018.7111} \\
Yes & No & No & No & No & Yes & No \\
\hline
\end{tabularx}

\clearpage
\newpage

\section{Appendix A. Difference between reduced model and expanded model with a linear slope}\label{app:NI.estimator}

Recall from the main text that we assume a balanced panel with the constrained model specification: 
\begin{align}\label{eq:constrained}
y_{it} = \sum_{k=T_1}^T \beta_{k} G_i \mathbb{I}(t = k) + \alpha_i + \gamma_t  + \epsilon_{it},
\end{align}
\noindent where $y_{it}$ is the outcome for unit $i$ at time $t$, $\gamma_t$ is a time fixed effect, and $\alpha_i$ is a unit fixed effect. 
The $\beta_{k}$ represent differential changes in the treated group relative to the comparison group in each post-intervention period. 
If the model is correctly specified, the average of these, ${\beta} = \frac{1}{T-T_1+1} \sum_{k = T_1}^T {\beta_{k}}$, is the ATT.

When we add a slope difference $\theta$, we obtain the expanded model specification,
\begin{equation}\label{eq:general}
y_{it} = \sum_{k=T_1}^T \beta_{k}^{(e)}  G_i \mathbb{I}(t = k)  + \alpha_i^{(e)} + \gamma_t^{(e)}  +   \theta G_i t + \epsilon^{(e)}_{it},
\end{equation}
in which $(e)$ distinguishes model components from those of the previous model. 

\begin{proposition}[Reduced vs. expanded model estimators (linear trend difference)]\label{prop:estimator}
\textit{The difference between ordinary least squares (OLS) ATT estimators corresponding to model specifications in Eqs.~(\ref{eq:constrained}) and (\ref{eq:general}) is a linear transformation of the differential trends parameter estimate $\hat{\theta}$ from Eq.~(\ref{eq:general}):}
\begin{align}
\label{eq:vio}
\hat{\beta} - \hat{\beta}^{(e)} = \left(\frac{1}{T-T_1+1} \sum_{t=T_1}^T t - \frac{1}{T_1-1} \sum_{t=1}^{T_1-1} t \right)\hat{\theta} = \frac{T}{2}\hat{\theta}.
\end{align}
\end{proposition}

\begin{proof}

The OLS specification from Eq.~(\ref{eq:constrained}) implies sample moment conditions:
\begin{align*}
\frac{1}{n_1}\sum_{i: G_i = 1} y_{ik} &= \hat{\bar{\alpha}}_1 + \hat{\gamma}_k + \hat{\beta}_k, &\: k \in T_1,\ldots,T \\
\frac{1}{n_0}\sum_{i: G_i = 0} y_{ik} &= \hat{\bar{\alpha}}_0 + \hat{\gamma}_k, &\: k \in T_1,\ldots,T \\
\frac{1}{n_1(T_1-1)}\sum_{i: G_i = 1, \: t: t<T_1} y_{it} &= \hat{\bar{\alpha}}_1 + \hat{\bar{\gamma}}_{t<T_1} \\
\frac{1}{n_0(T_1-1)}\sum_{i: G_i = 0, \: t: t<T_1} y_{it} &= \hat{\bar{\alpha}}_0 + \hat{\bar{\gamma}}_{t<T_1},
\end{align*}
where $\hat{\bar{\alpha}}_g$ indicates the average unit fixed effect over units with $G_i = g$ and $\hat{\bar{\gamma}}_{t<T_1}$ indicates the average time fixed effect over $\{1,\ldots,T_1-1\}$. Rearranging these, we obtain:
\begin{align}\label{v1}
\hat{\beta}_k = \frac{1}{n_1}\sum_{i: G_i = 1} y_{ik} - \frac{1}{n_0}\sum_{i: G_i = 0} y_{ik} - \frac{1}{T_1-1}\left(\frac{1}{n_1}\sum_{i: G_i = 1, \: t: t<T_1} y_{it} - \frac{1}{n_0}\sum_{i: G_i = 0, \: t: t<T_1} y_{it}\right)
\end{align}

By contrast, the OLS specification from Eq.~(\ref{eq:general}) implies sample moment conditions:
\begin{align*}
\frac{1}{n_1}\sum_{i: G_i = 1} y_{ik} &= \hat{\bar{\alpha}}_1^{(e)} + \hat{\gamma}_k^{(e)} + \hat{\beta}_k^{(e)} + \hat{\theta}k, &\: k \in T_1,\ldots,T \\
\frac{1}{n_0}\sum_{i: G_i = 0} y_{ik} &= \hat{\bar{\alpha}}_0^{(e)} + \hat{\gamma}_k^{(e)}, &\: k \in T_1,\ldots,T \\
\frac{1}{n_1(T_1-1)}\sum_{i: G_i = 1, \: t: t<T_1} y_{it} &= \hat{\bar{\alpha}}_1^{(e)} + \hat{\bar{\gamma}}_{t<T_1}^{(e)} + \frac{1}{T_1-1} \sum_{t=1}^{T_1-1}\hat{\theta} t\\
\frac{1}{n_0(T_1-1)}\sum_{i: G_i = 0, \: t: t<T_1} y_{it} &= \hat{\bar{\alpha}}_0^{(e)} + \hat{\bar{\gamma}}_{t<T_1}^{(e)}
\end{align*}

Rearranging these, we obtain:
\begin{align}\label{v2}
\hat{\beta}_k^{(e)}  + \hat{\theta}k - \frac{1}{T_1-1} \sum_{t=1}^{T_1-1}\hat{\theta} t &= \frac{1}{n_1}\sum_{i: G_i = 1} y_{ik} - \frac{1}{n_0}\sum_{i: G_i = 0} y_{ik} - \frac{1}{T_1-1}\left(\frac{1}{n_1}\sum_{i: G_i = 1, \: t: t<T_1} y_{it} - \frac{1}{n_0}\sum_{i: G_i = 0, \: t: t<T_1} y_{it}\right)
\end{align}

Noting that the right sides of Eqs.~(\ref{v1}) and (\ref{v2}) are equal, we have:

\begin{align*}
\hat{\beta}_k^{(e)} &= \hat{\beta}_k - \left(k - \frac{1}{T_1-1} \sum_{t=1}^{T_1-1} t\right) \hat{\theta}
\end{align*}

Averaging over post-intervention periods from $T_1$ to $T$ leaves:
\begin{align*}
\hat{\beta} - \hat{\beta}^{(e)} &= \left(\frac{1}{T-T_1+1} \sum_{t=T_1}^T t - \frac{1}{T_1-1} \sum_{t=1}^{T_1-1} t \right)\hat{\theta} \\
&= \left(\frac{T_1+T}{2} - \frac{1+T_1-1}{2}\right)\hat{\theta} =\frac{T}{2}\hat{\theta}
\end{align*}

\end{proof}

If the expanded model is correctly specified but the reduced model is not, reduced model treatment effect bias is larger in magnitude when there is a longer study period or when the slope difference between groups is larger.

\section{Appendix B. Test statistics}

Recall that in our generalized framework, we have a reduced model with $p$ parameters, $\beta_1,\ldots,\beta_{p}$, and an expanded model with an additional $q$ parameters, $\theta_{1},\ldots,\theta_{q}$:
\begin{equation}\label{eq:gen_reduced}
\textbf{Reduced: } \mathbf{y} = \mathbf{X} \bm{\beta}  + \bm{\epsilon}
\end{equation}
\begin{equation}\label{eq:gen_expanded}
\textbf{Expanded: }\mathbf{y} = \mathbf{X} \bm{\beta}^{(e)} + \mathbf{Z} \bm{\theta}  + \bm{\epsilon}^{(e)}
\end{equation}
In this setup, if the reduced model is correctly specified, the ATT of interest is an average of a subset $\mathcal{K}$ of the parameters in $\bm{\beta}$ where $|\mathcal{K}| = K$: $\beta = \frac{1}{K}\sum_{k \in \mathcal{K}} \beta_k$. In the expanded model, the corresponding quantity is: $\beta^{(e)} = \frac{1}{K}\sum_{k \in \mathcal{K}} \beta_k^{(e)}$. We assume that $q \geq 1$ (i.e., the expanded model adds at least one parameter) and that $\begin{bmatrix} \bX & \mathbf Z \end{bmatrix}$ has full column rank.

\begin{lemma}[Form of $\left({\left(\bX^{(e)'}\bX^{(e)}\right)}^{-1}\right)_{1:p,1:p}$]
\label{lemma:variance}
Denote the model matrix of Eq.~(\ref{eq:gen_expanded}) as $\bX^{(e)}$. We can write the upper left $p \times p$ terms of ${{\left(\bX^{(e)'}\bX^{(e)}\right)}^{-1}}$:
\begin{align*}
\left({\left({\bX^{(e)}}'\bX^{(e)}\right)}^{-1}\right)_{{1:p, 1:p}} &= \left({\bX}' \bX - {\bX}'\mathbf{Z}({\mathbf{Z}}' \mathbf{Z})^{-1}{\mathbf{Z}}'\bX\right)^{-1} \\
&=\left({\bX}' \mathbf M_Z \bX \right)^{-1},
\end{align*}
where $\mathbf M_Z = \mathbf I-\mathbf{Z}(\mathbf{Z}' \mathbf{Z})^{-1}\mathbf{Z}'$ denotes the ``annihilator matrix" associated with a projection onto the orthogonal complement of the span of $\mathbf Z$, the added covariates in the expanded model.
\end{lemma}

\begin{proof}
Following prior work,\citep{clogg_statistical_1995} we can write the expanded model matrix as a block matrix including the reduced model matrix and added variables in the expanded model: $\mathbf X^{(e)} = \begin{bmatrix} \bX & \mathbf Z \end{bmatrix}$.  Then, 

\begin{align*}
{{\left(\bX^{(e)'}\bX^{(e)}\right)}}^{-1} &= \begin{bmatrix}
    \bX'\bX & \bX'\mathbf Z \\
    \mathbf Z' \bX & \mathbf Z' \mathbf Z
\end{bmatrix}^{-1} 
\end{align*}

Per standard approaches to inverting block matrices \citep{lu_inverses_2002} and noting that because $\begin{bmatrix} \bX & \mathbf Z \end{bmatrix}$ has full column rank, $\mathbf D = \mathbf Z'\mathbf Z $ and $\mathbf A-\mathbf B \mathbf D^{-1}\mathbf C = \bX' \mathbf M_{Z} \bX$ are invertible:

\begin{align*}
\begin{bmatrix}
    \mathbf A & \mathbf B \\
    \mathbf C & \mathbf D
\end{bmatrix}^{-1} = \begin{bmatrix}
(\mathbf A-\mathbf B \mathbf D^{-1}\mathbf C)^{-1} & -(\mathbf A-\mathbf B \mathbf D^{-1}\mathbf C)^{-1}\mathbf B \mathbf D^{-1} \\
-\mathbf D^{-1} \mathbf C(\mathbf A-\mathbf B \mathbf D^{-1} \mathbf C)^{-1} & \mathbf D^{-1} + \mathbf D^{-1}\mathbf C (\mathbf A-\mathbf B \mathbf D^{-1}\mathbf C)^{-1}\mathbf B \mathbf D^{-1}
\end{bmatrix}
\end{align*}

Applying this to ${\left({\bX^{(e)}}'\bX^{(e)}\right)}^{-1}$ as written above, the upper left entry is $p \times p$:
\begin{align*}
\left({\left(\bX^{(e)'}\bX^{(e)}\right)}^{-1}\right)_{1:p,1:p} &= \left(\bX' \bX - \bX' \mathbf{Z}(\mathbf{Z}' \mathbf{Z})^{-1}\mathbf{Z}' \bX\right)^{-1} \\
&= \left(\bX' \mathbf M_Z \bX \right)^{-1}
\end{align*}

\end{proof}

\begin{lemma}[OLS estimator for $\hat{\bm \beta}^{(e)}$]
\label{lemma:estimator}
The OLS estimator $\hat{\bm \beta}^{(e)}$, the coefficients in the expanded model associated with variables shared in both reduced and expanded models, can be written:
\begin{align*}
\hat{\bm \beta}^{(e)} = \left(\bX' \mathbf M_Z \bX \right)^{-1} \mathbf{X}' \mathbf M_Z \mathbf{y}
\end{align*}
\end{lemma}
\begin{proof}
By construction, the OLS estimator corresponding to Eq.~(\ref{eq:gen_expanded}) is:
\begin{align*}
\begin{bmatrix}
    \hat{\bm \beta}^{(e)} \\
    \hat{\bm \theta}
\end{bmatrix} &= {\left({\bX^{(e)}}'\bX^{(e)}\right)}^{-1} {\bX^{(e)}}' \mathbf y \\
&= \begin{bmatrix}
    \mathbf A & \mathbf B \\
    \mathbf C & \mathbf D
\end{bmatrix}^{-1} \begin{bmatrix} {\bX}' \\ {\mathbf Z}' \end{bmatrix} \mathbf{y} & \text{as defined in Lemma \ref{lemma:variance}}\\
\hat{\bm \beta}^{(e)}  &=  \left({\bX}' \mathbf M_Z \bX \right)^{-1} {\bX}'\mathbf{y} - \left({\bX}' \mathbf M_Z \bX \right)^{-1} {\bX}' \mathbf{Z} \left({\mathbf{Z}}'\mathbf{Z}\right)^{-1} {\mathbf{Z}}' \mathbf{y} \\
&=  \left({\bX}' \mathbf M_Z \bX \right)^{-1} {\mathbf{X}}' \mathbf M_Z \mathbf{y}
\end{align*}
\end{proof}

\begin{lemma}[Covariance between coefficients in reduced and expanded models]
\label{lemma:bias}
\textit{Assume reduced and expanded models per Eq.~(\ref{eq:gen_reduced}) and Eq.~(\ref{eq:gen_expanded}), and the expanded model is correctly specified, with $\epsilon_{it}^{(e)}\overset{i.i.d.}{\sim} N\left(0, \sigma^2_{(e)}\right)$.  Then, $Cov\left(\hat{\beta}_k, \hat{\beta}_j^{(e)}\right) = Cov\left(\hat{\beta}_k, \hat{\beta}_j\right)$ for all $k,j \in \{1,\ldots,p\}$.}
\end{lemma}
\begin{proof}
Our reduced model estimator is:
\begin{align*}
\hat{\bm\beta} &= \left(\bX' \bX\right)^{-1} \bX' \mathbf{y} 
\end{align*}

Per Lemma \ref{lemma:estimator}, the expanded model for coefficients $\{1,\ldots,p\}$ can be written:
\begin{align*}
\hat{\bm \beta}^{(e)} &=  \left(\bX' \mathbf M_Z \bX \right)^{-1} \mathbf{X}' \mathbf M_Z \mathbf{y} \end{align*}

Then, we can find the covariance matrix $Cov \left(\hat{\bm\beta}, \hat{\bm \beta}^{(e)}\right)$, such that $Cov \left(\hat{\bm\beta}, \hat{\bm \beta}^{(e)}\right)_{k,j} =  Cov\left(\hat{\beta}_k, \hat{\beta}_j^{(e)}\right)$: 
\begin{align*}
Cov \left(\hat{\bm\beta}, \hat{\bm \beta}^{(e)}\right) &= Cov \left(\left(\bX' \bX\right)^{-1} \bX' \mathbf{y},  \left(\bX' \mathbf M_Z \bX \right)^{-1} \mathbf{X}' \mathbf M_Z \mathbf{y}\right) \\ 
&= \left(\bX' \bX\right)^{-1} \bX ' Var(\mathbf{y})\left(\left(\bX' \mathbf M_Z \bX \right)^{-1} \mathbf{X}' \mathbf M_Z \right)'  \\
&= \sigma_{(e)}^2\left(\bX' \bX\right)^{-1} \bX ' \mathbf M_Z \mathbf{X} \left(\bX' \mathbf M_Z \bX \right)^{-1} \\
&=  \sigma_{(e)}^2 \left(\bX' \bX\right)^{-1}  \\
&= Var(\hat{\bm \beta}) 
\end{align*}

Because $Cov\left(\hat{\bm\beta}, \hat{\bm \beta}^{(e)}\right)=Var(\hat{\bm \beta})$ (where the latter is the variance-covariance matrix such that $Var(\hat{\bm \beta})_{k,j} = Cov\left(\hat{\beta}_k, \hat{\beta}_j\right)$), we know $ Cov\left(\hat{\beta}_k, \hat{\beta}_j^{(e)}\right)=Cov \left(\hat{\beta}_k, \hat{\beta}_j\right)$ for all $k,j \in \{1,\ldots,p\}$.

\end{proof}

\begin{proposition}[Reduced vs. expanded model estimators (Gaussian errors)]
\label{prop:test_inf}
Assume reduced and expanded models as in Eq.~(\ref{eq:gen_reduced}) and Eq.~(\ref{eq:gen_expanded}) and that the expanded model is correctly specified, with $\bm \epsilon^{(e)} \sim N(0, \bm \Omega)$, where $\bm \Omega$ is an $nT \times nT$ matrix. Let $\mathbf V = {\left(\bX'\bX\right)}^{-1}$ and denote $\mathbf{V}^{(e)}$ analogously for the expanded model. The difference between OLS estimators $\hat{\beta}_k$ and $\hat{\beta}_k^{(e)}$ is:
\begin{align*}
\hat{\beta}_k - \hat{\beta}_k^{(e)} \sim N \left(\beta_k - \beta_k^{(e)}, {\bm \Sigma}_{k,k} + {\bm \Sigma}^{(e)}_{k,k} - 2{\bm \Sigma}^*_{k,k}\right),
\end{align*}
where ${\bm \Sigma} = \mathbf V \bX' {\bm \Omega} \bX \mathbf V$,  ${\bm \Sigma}^{(e)} = \mathbf V^{(e)} \bX^{(e)'} {\bm \Omega} \bX^{(e)} \mathbf V^{(e)}$, ${\bm \Sigma}^* = \mathbf V \bX' \bm \Omega \bX^{(e)} \mathbf V^{(e)}$, and ${\mathbf A}_{k,k}$ indicates the entry in the $k$th row and $k$th column of the matrix $\mathbf{A}$.
\end{proposition}

\begin{proof}
Coefficients in our misspecified reduced model are estimated:
\begin{align*}
\hat{\bm\beta} &= \left(\bX' \bX\right)^{-1} \bX' \mathbf{y} 
\end{align*}

The OLS sampling variance of the misspecified $\hat{\bm\beta}$ coefficient estimator is:
\begin{align*}
\bm \Sigma &= \left(\bX' \bX\right)^{-1} \bX' \bm \Omega \bX \left(\bX' \bX\right)^{-1}
\end{align*}

Similarly, for the expanded model:
\begin{align*}
\begin{bmatrix} \hat{\bm \beta}^{(e)} \\ \hat{\bm \theta} \end{bmatrix} &=  \left({\bX^{(e)}}' \bX^{(e)}\right)^{-1} {\bX^{(e)}}' \mathbf{y}   \\
\bm \Sigma^{(e)} &= \left({\bX^{(e)}}' \bX^{(e)}\right)^{-1} {\bX^{(e)}}' \bm \Omega \bX^{(e)} \left({\bX^{(e)}}' \bX^{(e)}\right)^{-1} 
\end{align*} 

We can take the covariance of $\hat{\bm\beta}$ and ${\begin{bmatrix} {\hat{\bm \beta}^{(e)}}{}' & {\hat{\bm \theta}}' \end{bmatrix}}'$:
\begin{align*}
{\bm \Sigma}^{*} &= Cov\left(\left({\bX}' {\bX}\right)^{-1} {\bX}' \mathbf{y},\left({{\bX}^{(e)}}' {\bX}^{(e)}\right)^{-1} {{\bX}^{(e)}}' \mathbf{y} \right) \\ 
&= \left({\bX}' \bX\right)^{-1} {\bX}' \bm{\Omega} \bX^{(e)} \left({\bX^{(e)}}' \bX^{(e)}\right)^{-1} 
\end{align*}

Extracting entries corresponding to a shared covariate $k$ and applying standard covariance properties ($Var(a-b)=Var(a)+Var(b)-2Cov(a,b)$) completes the proof.
\end{proof}

\begin{proposition}[Reduced vs. expanded model estimators ($i.i.d.$ errors)]\label{prop:test}
\textit{Assume reduced and expanded models as in Eq.~(\ref{eq:gen_reduced}) and Eq.~(\ref{eq:gen_expanded}) and that the expanded model is correctly specified, with $\epsilon_{it}^{(e)} \overset{i.i.d.}{\sim} N\left(0,\sigma^2_{(e)}\right)$.
Recall that $\beta = \frac{1}{K} \sum_{k \in \mathcal{K}} \beta_k$ and $\beta^{(e)} = \frac{1}{K} \sum_{k \in \mathcal{K}} \beta_k^{(e)}$ are the parameters of interest.  The difference between the corresponding OLS ATT estimators is:}
\begin{align}
\hat{\beta} - \hat{\beta}^{(e)} \sim N \left(\beta - \beta^{(e)}, 
{\sigma}^2_{\hat{\beta}^{(e)}} - {\sigma}^2_{\hat{\beta}}\right),
\end{align}
\textit{where $\sigma^2_{\hat{\beta}^{(e)}}$ is the variance of $\hat{\beta}^{(e)}$ (corresponding to the expanded model), and $\sigma^2_{\hat{\beta}}$ is the variance of $\hat{\beta}$ (corresponding to the potentially misspecified reduced model but defined using common error variance, $\sigma^2_{(e)}$).}
\end{proposition}

\begin{proof}

The true variance of $\hat{\beta}$ in the reduced model assuming $i.i.d.$ errors is:
\begin{align*}
Var(\hat{\bm \beta})&=  \left(\bX' \bX\right)^{-1} \left(\bX'  Var(\mathbf{y})  \bX\right) \left(\bX' \bX\right)^{-1}\\
&= \sigma^2_{(e)}\left(\bX' \bX\right)^{-1},
\end{align*}
where $\sigma^2_{(e)}$ corresponds to the expanded model because we assume it to be correctly specified. We denote its $k$th diagonal element as $\sigma^2_{\hat{\beta}_k}$.\footnote{In contrast, the expectation of the OLS estimator for the variance of $\hat{\bm \beta}$, corresponding to the reduced model in Eq.~(\ref{eq:constrained}), is $\sigma^2\left(\bX' \bX\right)^{-1}$, where $\sigma^2$ corresponds to the expectation of the reduced model residual variance estimator. Assuming the expanded model is correctly specified (but reduced model may not be), this is biased by a factor of $\frac{\sigma^2}{\sigma^2_{(e)}}$.} Then: 
\begin{align*}
Var\left(\hat{\beta}_k - \hat{\beta}^{(e)}_k  \right) &= Var\left(\hat{\beta}_k\right) + Var\left(\hat{\beta}^{(e)}_k\right) - 2Cov \left(\hat{\beta}_k, \hat{\beta}_k^{(e)}\right) \\
&= Var\left(\hat{\beta}^{(e)}_k\right) - Var\left(\hat{\beta}_k\right)  & \text{by Lemma \ref{lemma:bias}} \\
&= {\sigma}^2_{\hat{\beta}_k^{(e)}} - {\sigma}^2_{\hat{\beta}_k} & \text{adopting notation above} \\
\end{align*}

The extension to $\hat{\beta}$ follows by averaging over $\hat{\beta}_k$.  
See Clogg et al.~(1995) for further discussion.\citep{clogg_statistical_1995}  
\end{proof}

\subsection{Walk-through}

We apply the testing procedure outlined in Section 2.4.1, evaluating the difference between a linear combination of some $\hat{\beta}_k$ in the reduced model and corresponding $\hat{\beta}_k^{(e)}$ in the expanded model while accounting for clustering, heteroskedastic errors, and survey weights per standard practice.\citep{colin_cameron_practitioners_2015} (Note that in this subsection $\bm{\hat{\beta}}^{(e)}$ is assumed to be of length $p+q$, distinct from its usage in prior sections, to mirror standard regression output.)
\begin{enumerate}
\item Fit the reduced model using weighted linear regression with vector of sampling weights $\bw$. Extract the parameter vector $\bm{\hat{\beta}}=(\hat{\beta}_1,\ldots,\hat{\beta}_p)'$ (of length $p$), design matrix $\bX$, and $\mathbf V = \left(\mathbf X' diag(\mathbf{w}) \mathbf X\right)^{-1}$. 
\item Fit the expanded model. Extract the parameter vector $\bm{\hat{\beta}}^{(e)}$ (which is of length $p+q$), residual vector $\mathbf{\hat{u}}^{(e)}$, design matrix $\bX^{(e)}$, and $\mathbf V^{(e)}$. 
\item Compute within-model cluster robust variance-covariance matrices under the expanded model,\citep{clogg_statistical_1995} assuming a total of $n$ clusters:
\begin{align*}
\hat{\bm \Sigma} &= \frac{n}{n-1}\frac{nT-1}{nT-p-q}\mathbf V \left[ \sum_g {\bX_g}'  \left(\bw_g \circ \mathbf {\hat{u}}^{(e)}_g\right) {\left(\bw_g \circ \mathbf {\hat{u}}^{(e)}_g\right)}' \bX_g \right ] \mathbf V \\
\hat{\bm \Sigma}^{(e)} &= \frac{n}{n-1}\frac{nT-1}{nT-p-q} \mathbf V^{(e)} \left[ \sum_g {\bX^{(e)}_g}'  \left(\bw_g \circ \mathbf {\hat{u}}^{(e)}_g\right) {\left(\bw_g \circ \mathbf {\hat{u}}^{(e)}_g\right)}' \bX^{(e)}_g \right ] \mathbf V^{(e)}
\end{align*}
where $\circ$ indicates element-wise multiplication and $\bX_g$, $\bX^{(e)}_g$, $\mathbf{\hat{u}}_g^{(e)}$, and $\bw_g$ are group-specific subsets of their respective elements.  

Note that $\mathbf{\hat{{u}}}^{(e)}$ must be used in $\hat{\bm \Sigma}$ because we assume the expanded model is correctly specified, but the reduced model may not be. 
\item Compute the covariance between estimators:
$$
\hat{\bm \Sigma}^* = \frac{n}{n-1} \mathbf V \left[ \sum_g \bX_g '  \left(\bw_g \circ \mathbf {\hat{u}}^{(e)}_g\right) \left(\bw_g \circ \mathbf {\hat{u}}^{(e)}_g\right)' \bX^{(e)}_g \right ] \mathbf V^{(e)}
$$
This matrix is $p \times (p+q)$; the top left $p \times p$-submatrix contains covariances between corresponding parameters of the two models.
\item Let $\hat{\kappa}=\mathbf{a}'\bm{\hat{\beta}}$
and $\hat{\kappa}^{(e)}={\mathbf{a}^{(e)}}'\bm{\hat{\beta}}^{(e)}$ denote the linear combinations of parameters from each model that we want to test.
\item Compute the variance of the difference between these two linear combinations, 
$$
{Var}\left(\hat{\kappa}-\hat{\kappa}^{(e)}\right) = 
\mathbf{a}'\hat{\bm \Sigma}\mathbf{a} - 2
\mathbf{a}'\hat{\bm \Sigma}^* \mathbf{a}^{(e)} +
{\mathbf{a}^{(e)}}'\hat{\bm \Sigma}^{(e)}\mathbf{a}^{(e)}
$$ 
\end{enumerate}

We can then use this variance in a Wald test to evaluate hypotheses of interest (e.g., $H_0: {\beta}-{\beta}^{(e)} \geq \delta$).

\section{Appendix C. Event studies}
Recall the event study specification of an expanded model from Table 1:
\begin{equation}\label{eq:event-study}
\textbf{Expanded: }y_{it} = \sum_{k=T_1}^T \beta^{(e)}_{k} G_i \mathbb{I}(t = k) + \alpha^{(e)}_i + \gamma^{(e)}_t + \sum_{\ell=1}^{T_1-2} \theta_\ell G_i \mathbb{I}(t = \ell) + \epsilon^{(e)}_{it}
\end{equation}

Note that $t = T_1-1$ is the omitted reference period, i.e., the corresponding pre-period coefficient is normalized to 0.

\begin{sproposition}[Event study coefficients as the difference between ATT estimators corresponding to reduced and expanded models]\label{sprop:event-study}
\textit{Assume a reduced model as in Eq.~(\ref{eq:constrained}) and expanded model as in Eq.~(\ref{eq:event-study}). Following standard practice, let $\mathcal{K} = \{T_1,\ldots,T\}$ (i.e., the ATT of interest is the average effect over all post-intervention periods, with $\hat{\beta} = \frac{1}{T-T_1+1} \sum_{k = T_1}^T \hat{\beta}_k$ and likewise for $\hat{\beta}^{(e)}$).
Then, when estimated with OLS, $\hat{\beta} - \hat{\beta}^{(e)} = -\frac{1}{T_1-1} \sum_{\ell=1}^{T_1-2}\hat{\theta}_\ell$.}
\end{sproposition}

\begin{proof}
Applying the logic outlined in Appendix A, 
$\hat{\beta}_k$ from Eq.~(\ref{eq:constrained}) is:
\begin{align}\label{v3}
\hat{\beta}_k &= \frac{1}{n_1}\sum_{i: G_i = 1} y_{ik} - \frac{1}{n_0}\sum_{i: G_i = 0} y_{ik} - \frac{1}{T_1-1}\left(\frac{1}{n_1}\sum_{i: G_i = 1, \: t: t<T_1} y_{it} - \frac{1}{n_0}\sum_{i: G_i = 0, \: t: t<T_1} y_{it}\right)
\end{align}

and similarly,

\begin{align}\label{v4}
\hat{\beta}_k^{(e)} - \frac{1}{T_1-1} \sum_{\ell=1}^{T_1-2} \hat{\theta}_\ell&= \frac{1}{n_1}\sum_{i: G_i = 1} y_{ik} - \frac{1}{n_0}\sum_{i: G_i = 0} y_{ik} - \frac{1}{T_1-1}\left(\frac{1}{n_1}\sum_{i: G_i = 1, \: t: t<T_1} y_{it} - \frac{1}{n_0}\sum_{i: G_i = 0, \: t: t<T_1} y_{it}\right)
\end{align}

Noting that the right sides of Eqs.~(\ref{v3}) and (\ref{v4}) are equal, we have:
\begin{align*}
\hat{\beta}_k &= \hat{\beta}_k^{(e)} - \frac{1}{T_1-1} \sum_{\ell=1}^{T_1-2} \hat{\theta}_\ell
\end{align*}

As $\hat{\beta}_k - \hat{\beta}_k^{(e)}$ does not depend on $k$, the average (or any particular one) can be written:
\begin{align*}
\hat{\beta} - \hat{\beta}^{(e)} &= -\frac{1}{T_1-1} \sum_{\ell=1}^{T_1-2} \hat{\theta}_\ell
\end{align*}

\end{proof}

\begin{corollary}[Alternative event study models]
Assume that we modify our event study to specify the expanded model: 
\begin{align*}
\textnormal{\textbf{Expanded: }}y_{it} = \sum_{k=T_1}^T \beta_{k}^{(e)} G_i\mathbb{I}(t = k) + \alpha_i^{(e)} + \gamma_t^{(e)}  + \sum_{\ell \in \mathcal{P}}\theta_\ell G_i  \mathbb{I}(t = \ell) + \epsilon_{it}^{(e)},
\end{align*}
where $\mathcal{P}$ denotes a set of pre-intervention periods.  Then $\hat{\beta} - \hat{\beta}^{(e)} = -\frac{1}{T_1-1} \sum_{\ell \in \mathcal{P}}\hat{\theta}_\ell$.  In this alternative, we estimate $\theta_\ell$ only for $\ell \in \mathcal{P}$ and let the complement serve as the pooled reference period in the expanded model, rather than the last pre-treatment period as in the traditional event study.
\end{corollary}

\begin{proof}
The proof follows by the same logic as above in Supplement Proposition \ref{sprop:event-study}.
\end{proof}

\section{Appendix D. No covariance condition}

We next introduce a correct model specification:
\begin{align}\label{eq:correct}
\textbf{Correct: \: } \mathbf{y} = \mathbf{X} \bm{\beta}^{(w)} + \mathbf{Z} \bm{\theta}^{(w)} + \mathbf{W} \bm{\gamma}  + \bm{\epsilon}^{(w)} ,
\end{align}
which may add additional terms to the expanded model in Eq.~(\ref{eq:gen_expanded}).

\begin{sproposition}[Misspecification]
\label{cor:misspec}
Assume that the correct model specification follows Eq.~(\ref{eq:correct}) and has  $\epsilon_{it}^{(w)} \overset{i.i.d.}{\sim} N\left(0,\sigma^2_{(w)}\right)$. Further assume that reduced and expanded models are specified as in Eq.~(\ref{eq:gen_reduced}) and Eq.~(\ref{eq:gen_expanded}), with $\hat{\beta}$ and $\hat{\beta}^{(e)}$ denoting corresponding OLS ATT estimators.
Then,
$Cov \left(\hat{\beta}, \hat{\beta} - \hat{\beta}^{(e)}\right)=0$.
\end{sproposition}
\begin{proof}
Following the logic in Lemma \ref{lemma:bias}:

\begin{align*}
Cov\left(\hat{\bm\beta}, \hat{\bm \beta}^{(e)}\right) &= Cov\left(\left(\bX' \bX\right)^{-1} \bX' \mathbf{y},  \left(\bX' \mathbf M_Z \bX \right)^{-1} \mathbf{X}' \mathbf M_Z \mathbf{y}\right) \\ 
&= \left(\bX' \bX\right)^{-1} \bX ' Var(\mathbf{y}) \left(\left(\bX' \mathbf M_Z \bX \right)^{-1} \mathbf{X}' \mathbf M_Z \right)' \\
&= \sigma_{(w)}^2\left(\bX' \bX\right)^{-1} \bX ' \mathbf M_Z \mathbf{X} \left(\bX' \mathbf M_Z \bX \right)^{-1} \\
&=   \sigma_{(w)}^2\left(\bX' \bX\right)^{-1},
\end{align*}
where $Var(\mathbf y)=\sigma_{(w)}^2 \mathbf{I}$, not $\sigma_{(e)}^2 \mathbf{I}$.  Similarly, we have:
\begin{align*}
Var(\hat{\bm \beta})&=  \left(\bX' \bX\right)^{-1} \left(\bX' Var(\mathbf{y}) \bX\right) \left(\bX' \bX\right)^{-1}  \\
&=  \sigma^2_{(w)}\left(\bX' \bX\right)^{-1}
\end{align*}

Because $Cov\left(\hat{\bm\beta}, \hat{\bm \beta}^{(e)}\right)=Var(\hat{\bm \beta})$, $Cov\left(\hat{\beta}, \hat{\beta}-\hat{\beta}^{(e)}\right)=0$. The scalar result follows by taking linear combinations of parameter estimates. 
\end{proof}

\begin{sproposition}[Difference between $Cov\left(\bm{\hat{\beta}}, \bm{\hat{\beta}}^{(e)}\right)$ and $Var\left(\bm{\hat{\beta}}\right)$]
\label{sprop:diff} Assume that the correct model specification follows Eq.~(\ref{eq:correct}), with $\bm \epsilon^{(w)}\sim N(0, \bm \Omega)$, where $\bm \Omega$ is an $nT \times nT$ matrix. Further assume that reduced and expanded models are specified as in Eq.~(\ref{eq:gen_reduced}) and Eq.~(\ref{eq:gen_expanded}), with $\bm{\hat{\beta}}$ and $\bm{\hat{\beta}}^{(e)}$ denoting corresponding OLS estimators. Then, 
\begin{align*}
{\bm \Sigma}^*_{1:p,1:p} - {\bm \Sigma} &= \left(\bX' \bX\right)^{-1} \bX' \bm \Omega \mathbf M_X \mathbf M_Z \bX \left(\bX' \mathbf M_Z \bX \right)^{-1},
\end{align*}
where ${\bm \Sigma}^*_{1:p,1:p} = Cov\left(\bm{\hat{\beta}}, \bm{\hat{\beta}}^{(e)}\right)$, the covariance matrix between $\bm{\hat{\beta}}$ and $\bm{\hat{\beta}}^{(e)}$; ${\bm \Sigma} = Var\left(\bm{\hat{\beta}}\right)$, the variance-covariance matrix associated with $\bm{\hat{\beta}}$.
\end{sproposition}

\begin{proof}
We can write ${\bm \Sigma}^*_{1:p,1:p}$ as defined in Proposition \ref{prop:test_inf}:

\begin{align*}
{\bm \Sigma}^*_{1:p,1:p} &= \left(\bX' \bX\right)^{-1} \bX' \bm \Omega \mathbf M_Z \bX \left(\bX' \mathbf M_Z \bX \right)^{-1} \\
&= \left(\bX' \bX\right)^{-1} \bX' \bm \Omega \bX \left(\bX' \mathbf M_Z \bX \right)^{-1} - \left(\bX' \bX\right)^{-1} \bX' \bm \Omega (\mathbf I - \mathbf M_Z) \bX \left(\bX' \mathbf M_Z \bX \right)^{-1},
\end{align*}
where the second line comes from adding and subtracting $\left(\bX' \bX\right)^{-1} \bX' \bm \Omega \bX \left(\bX' \mathbf M_Z \bX \right)^{-1}$.

Noting that $\mathbf M_Z = \mathbf{I} - \mathbf{Z}(\mathbf{Z}' \mathbf{Z})^{-1}\mathbf{Z}'$ and applying the Woodbury matrix identity formula, $(\bm A - \bm B)^{-1} = \bm A^{-1} + \bm A^{-1}\bm B (\bm A - \bm B)^{-1}$ to $\left(\bX' \mathbf M_Z \bX \right)^{-1}$, we can rewrite the first term:
\begin{align*}
\left(\bX' \bX\right)^{-1} \bX' \bm \Omega \bX \left(\bX' \mathbf M_Z \bX \right)^{-1} &= \left(\bX' \bX\right)^{-1} \bX' \bm \Omega \bX \left(\bX' \bX \right)^{-1} + \\ &\left(\bX' \bX\right)^{-1} \bX' \bm \Omega \bX \left(\bX' \bX \right)^{-1} \bX '\mathbf{Z}(\mathbf{Z}' \mathbf{Z})^{-1}\mathbf{Z}' \bX \left(\bX' \mathbf M_Z \bX \right)^{-1} \\
&= \bm \Sigma + \left(\bX' \bX\right)^{-1} \bX' \bm \Omega \bX \left(\bX' \bX \right)^{-1} \bX' \mathbf{Z}(\mathbf{Z}' \mathbf{Z})^{-1}\mathbf{Z}' \bX \left(\bX' \mathbf M_Z \bX \right)^{-1}
\end{align*}

Substituting back in, collecting terms, and setting $\mathbf M_X = \mathbf I - \bX \left(\bX' \bX \right)^{-1}  \bX'$, the projection onto the orthogonal complement of the span of $\mathbf{X}$, we obtain:
\begin{align*}
{\bm \Sigma}^*_{1:p,1:p} &= \bm \Sigma +  \left(\bX' \bX\right)^{-1} \bX' \bm \Omega \left(\mathbf{I} - \mathbf{M}_X \right)\left(\mathbf{I} - \mathbf{M}_Z \right) \bX \left(\bX' \mathbf M_Z \bX \right)^{-1} - \\ &\left(\bX' \bX\right)^{-1} \bX' \bm \Omega \left(\mathbf{I} - \mathbf{M}_Z \right) \bX \left(\bX' \mathbf M_Z \bX \right)^{-1} \\
&=  \bm \Sigma -  \left(\bX' \bX\right)^{-1} \bX' \bm \Omega  \mathbf M_X (\mathbf I - \mathbf M_Z) \bX \left(\bX' \mathbf M_Z \bX \right)^{-1} \\
{\bm \Sigma}^*_{1:p,1:p} - \bm \Sigma &= \left(\bX' \bX\right)^{-1} \bX' \bm \Omega  \mathbf M_X \mathbf M_Z \bX \left(\bX' \mathbf M_Z \bX \right)^{-1} \text{ because } \mathbf M_X \bX = 0, 
\end{align*}
which completes the proof.  

\end{proof}

\subsection{Special cases}

From Supplement Proposition \ref{sprop:diff}, the magnitude of test-induced bias with non-$i.i.d.$ error structures depends on both $\bm{\Omega}$ and $\mathbf{\tilde{X}} = \mathbf M_X \mathbf M_Z \bX$. Let $\bm \Sigma^B = ({\bm \Sigma}^*_{1:p,1:p} - {\bm \Sigma})$. For the content that follows in this subsection, assume that $\bX$ is structured according to Eq.~(\ref{eq:constrained}):
\begin{align*}
\bX &= \begin{bmatrix} \mathbf{trt}_{T_1} & \ldots &\mathbf{trt}_{T} & \mathbf{u}_1 & \ldots & \mathbf{u}_n & \mathbf{t}_2 & \ldots & \mathbf{t}_T \end{bmatrix},
\end{align*}
where $\mathbf{trt}_{k}$ corresponds to the dummy variable vector indicating $G_i = 1$ and $t=k$, $\mathbf{u}_i$ to the dummy variable vector indicating whether an entry corresponds to unit $i$ and $\mathbf{t}_z$ a dummy variable vector indicating whether an entry corresponds to time $z$.  We refer to its columns (and the corresponding rows of $\bX'$) with the numbers $\{1,\ldots,K, u_1, \ldots, u_n,  t_2, \ldots , t_T\}$ and let $\bX_i$ be the subset of rows of the matrix $\bX$ corresponding to unit $i$. As with other results in this paper, we assume a balanced panel in this subsection per Section 2.1 of the main text.

\begin{sproposition}[Independent errors with unit-specific variance]
\label{sprop:bias_hetero}
Assume that the correct model specification follows Eq.~(\ref{eq:correct}) and that errors are independent with unit-specific variance (i.e., $\epsilon_{it}\sim N(0, \sigma^2_i)$ and all $\epsilon_{it}$ independent, such that $\bm \Omega$ is diagonal). Further assume that reduced and expanded models are specified as in Eq.~(\ref{eq:constrained}) and Eq.~(\ref{eq:gen_expanded}) respectively (with $\hat{\beta}$ and $\hat{\beta}^{(e)}$ denoting corresponding OLS ATT estimators) and that the expanded model is specified such that $\mathbf{\tilde{X}}_i = \mathbf{\tilde{X}}_j$ for $i,j$ with shared treatment status.\footnote{The condition that $\mathbf{\tilde{X}}_i = \mathbf{\tilde{X}}_j$ for $i,j$ with shared treatment status applies to both expanded models with a linear slope coefficient Eq.~(\ref{eq:general}) and event studies Eq.~(\ref{eq:event-study}) as well as for other common tests (e.g., an expanded model with unit-specific slopes) under the balanced panel setup introduced in Section 2.1 of the main text.
Intuitively, it requires that the expanded model test the same trend difference across all treated units. 
} 
Then, the entries of  $\bm \Sigma^B$ corresponding to treatment effects $\beta_k$ are 0, i.e., $\bm \Sigma^B_{1:K,1:K} = 0$, and thus 
$Cov \left(\hat{\beta}, \hat{\beta} - \hat{\beta}^{(e)}\right)=0$.
\end{sproposition}
\begin{proof}
Let $\bm \Sigma^B = ({\bm \Sigma}^*_{1:p, 1:p} - {\bm \Sigma})$ and $\mathbf{\tilde{X}} = \mathbf M_X \mathbf M_Z \bX$. Because errors are independent and heteroskedastic only across groups, we can write per Supplement Proposition \ref{sprop:diff}:
\begin{align*}
\bm{\Sigma}^B &= \left(\bX' \bX\right)^{-1} \bX' \bm \Omega  \mathbf{\tilde{X}}\left(\bX' \mathbf M_Z \bX \right)^{-1} \\
&= \left(\bX' \bX\right)^{-1} \left(\sum_{i=1}^n  \sigma_i^2  \bX_i'   \mathbf{\tilde{X}}_i \right) \left(\bX' \mathbf M_Z \bX \right)^{-1}
\end{align*}
where $\bX_i$ is the subset of rows of the matrix $\bX$ corresponding to unit $i$. 

Consider the inner term $\sum_{i=1}^n \sigma^2_i \bX_i' \mathbf{\tilde{X}}_i$.  Recall that $\bX' \mathbf{\tilde{X}} = \sum_{i=1}^n \bX_i' \mathbf{\tilde{X}}_i=0$ by construction. Both of these matrices are of dimension $p \times p$, and we structure $\mathbf X$ as:
\begin{align*}
\bX &= \begin{bmatrix} \mathbf{trt}_{T_1} & \ldots &\mathbf{trt}_{T} & \mathbf{u}_1 & \ldots & \mathbf{u}_n & \mathbf{t}_2 & \ldots & \mathbf{t}_T \end{bmatrix}
\end{align*}
As above, we refer to its columns (and the corresponding rows of $\bX'$) as $\{1,\ldots, K, u_1, \ldots, u_n,  t_2, \ldots , t_T\}$. We also denote entries of $\mathbf{\tilde{X}}$ as $\tilde{x}_{it,p}$, where $it$ specifies row and $p$ the column. We can then consider 3 types of rows of the $p \times p$ matrix $\sum_{i=1}^n \sigma^2_i \bX_i' \mathbf{\tilde{X}}_i$.
\begin{enumerate}
\item \textbf{Rows corresponding to post-treatment indicators $(1,\ldots,K)$.} \\
We first consider the first $K$ rows, each corresponding to a post-treatment indicator row $k$. \\

In $\sum_{i=1}^n \bX_i' \mathbf{\tilde{X}}_i$ (without $\sigma_i^2$), these take the form:

\begin{align*}
\begin{bmatrix} \sum_{i \in \mathcal{N}_1} \tilde{x}_{ik,1}, \hdots , \sum_{i \in \mathcal{N}_1} \tilde{x}_{ik,p}  \end{bmatrix} = \begin{bmatrix} 0, \hdots, 0 \end{bmatrix}
\end{align*}

Under the assumption that $\mathbf{\tilde{X}}_i = \mathbf{\tilde{X}}_j$ for $i,j$ with shared treatment status, we can simplify:

\begin{align*}
\begin{bmatrix} n_1 \tilde{x}_{n_1^* k,1}, \hdots , n_1 \tilde{x}_{n_1^* k,p}  \end{bmatrix} = \begin{bmatrix} 0, \hdots, 0 \end{bmatrix},
\end{align*}

where $\tilde{x}_{i k,\ell} = \tilde{x}_{n_1^* k,\ell}$ for all treated units (i.e., $n_1^*$ indicates a representative treated unit). Extending this to $\sum_{i=1}^n  \sigma_i^2 \bX_i' \mathbf{\tilde{X}}_i$, we have: 

\begin{align*}
\begin{bmatrix}  \tilde{x}_{n_1^* k,1} \sum_{i\in \mathcal{N}_1} \sigma_i^2, \hdots , \tilde{x}_{n_1^* k,p} \sum_{i\in \mathcal{N}_1} \sigma_i^2 \end{bmatrix} = \begin{bmatrix} 0, \hdots, 0 \end{bmatrix}
\end{align*}

\item \textbf{Rows corresponding to unit fixed effects.} \\ 
In $\sum_{i=1}^n \bX_i' \mathbf{\tilde{X}}_i$, these rows take the form, when corresponding to unit fixed effect $i$:

\begin{align*}
\begin{bmatrix} \sum_{t=1}^T \tilde{x}_{it,1}, \hdots , \sum_{t=1}^T\tilde{x}_{it,p}  \end{bmatrix} = \begin{bmatrix} 0, \hdots, 0 \end{bmatrix}
\end{align*}

Therefore in $\sum_{i=1}^n  \sigma_i^2 \bX_i' \mathbf{\tilde{X}}_i$, these rows must also be 0:

\begin{align*}
\begin{bmatrix} \sigma^2_i \sum_{t=1}^T \tilde{x}_{it,1}, \hdots , \sigma^2_i \sum_{t=1}^T\tilde{x}_{it,p}  \end{bmatrix} = \begin{bmatrix} 0, \hdots, 0 \end{bmatrix}
\end{align*}

\item \textbf{Rows corresponding to time fixed effects.}

In $\sum_{i=1}^n \bX_i' \mathbf{\tilde{X}}_i$, these rows take the form, when corresponding to time fixed effect $t$:

\begin{align*}
\begin{bmatrix} \sum_{i=1}^n \tilde{x}_{it,1}, \hdots , \sum_{i=1}^n\tilde{x}_{it,p} \end{bmatrix} = \begin{bmatrix} 0, \hdots, 0 \end{bmatrix}
\end{align*}

In $\sum_{i=1}^n  \sigma_i^2 \bX_i' \mathbf{\tilde{X}}_i$, these take the form:

\begin{align*}
\begin{bmatrix} \sum_{i=1}^n \sigma_i^2 \tilde{x}_{it,1}, \hdots , \sum_{i=1}^n \sigma_i^2  \tilde{x}_{it,p} \end{bmatrix}
\end{align*}

\begin{itemize}
\item For $t \geq T_1:$ Because we assume that $\mathbf{\tilde{X}}_i = \mathbf{\tilde{X}}_j$ for $i,j$ with shared treatment status, each column $p$ of $\mathbf{\tilde X}$ has shared values $a_{t,p}$ for treated units and $b_{t,p}$ for controls. Because $\mathbf{\tilde X}$ is orthogonal to all reduced model regressors, orthogonality to the treatment indicator for time $t$ forces $a_{t,p} = 0$, and orthogonality to the time fixed effect for time $t$ then forces $b_{t,p} = 0$. Hence every row of $\mathbf{\tilde X}$ in post-treatment periods is the zero vector.
\item For $t < T_1:$ These may not be zero.

\end{itemize}
\end{enumerate}

We therefore consider entries of $\left(\bX' \bX\right)^{-1}$, where $\left(\bX' \bX\right)^{-1}_{i,j} = adj(\bX' \bX)/det(\bX' \bX)$ and $adj(\bX' \bX)_{i,j} = (-1)^{i+j} {M}_{ij}$, where $M_{i,j}$ is the determinant of the submatrix formed by removing the $i$th row and $j$th column.  Because $M_{i,j}$ is a determinant, it is zero if columns are linearly dependent.  Indeed, if $i\in \{t_2,\ldots,t_{T_1-1}\}$, $j \in \{1,\ldots,K\}$, and $\mathbf{X}$ is structured as detailed above, submatrix columns are linearly dependent.  Therefore, corresponding entries of $\left(\bX' \bX\right)^{-1}_{i,j}$ are 0.  

As a result, when $\left(\bX' \bX\right)^{-1}$ is multiplied by the second term, the rows $1,\ldots,K$ of the resultant product are 0, and thus the upper left $K \times K$ entries of $\bm \Sigma^B$ are 0, $\bm \Sigma^B_{1:K,1:K} = 0$. Applying Supplement Proposition \ref{sprop:diff}, $Cov\left(\hat{\beta}, \hat{\beta}^{(e)}\right) = Var\left(\hat{\beta}\right)$, and thus $Cov\left(\hat{\beta}, \hat{\beta}-\hat{\beta}^{(e)}\right)=0$.

\end{proof}

\begin{sproposition}[Errors with unit-specific variance and constant error correlation within units]
Assume that the correct model specification follows Eq.~(\ref{eq:correct}) and that errors have unit-specific variance, are independent across units, and have constant error correlation within units (i.e., $\epsilon_{it}\sim N(0, \sigma_i^2)$ and $Cor(\epsilon_{ij}, \epsilon_{ik}) = \rho_{i}$ for $j,k \in \{1,\ldots,T\}$). Further assume that reduced and expanded models are specified as in Eq.~(\ref{eq:constrained}) and Eq.~(\ref{eq:gen_expanded}) respectively (with $\hat{\beta}$ and $\hat{\beta}^{(e)}$ denoting corresponding OLS ATT estimators) and that the expanded model is specified such that $\mathbf{\tilde{X}}_i = \mathbf{\tilde{X}}_j$ for $i,j$ with shared treatment status. Then, the entries of  $\bm \Sigma^B$ corresponding to treatment effects $\beta_k$ are 0, i.e., $\bm \Sigma^B_{1:K,1:K} = 0$, and thus 
$Cov \left(\hat{\beta}, \hat{\beta} - \hat{\beta}^{(e)}\right)=0$.
\end{sproposition}
\begin{proof}

Recall that:
\begin{align*}
\bm{\Sigma}^B &= \left(\bX' \bX\right)^{-1} \bX' \bm \Omega  \mathbf{\tilde{X}}\left(\bX' \mathbf M_Z \bX \right)^{-1} \\
\end{align*}

Because units are independent,
\begin{align*}
\bm \Omega \;=\;\mathrm{diag}(\bm\Omega_1,\dots,\bm\Omega_n),
\qquad
\bm\Omega_i
=\sigma_i^2\left((1-\rho_i)\,\mathbf I_T+\rho_i\,\mathbf J_T\right),
\end{align*}
where \(\mathbf J_T\) is the \(T\times T\) all-ones matrix.  Therefore, again denoting rows of $\mathbf X$ corresponding to unit $i$ as $\mathbf X_i$, we have:
\begin{align*}
\mathbf X'\,\bm \Omega\, \mathbf{\tilde X}
&=\sum_{i=1}^n \mathbf X_i'\, \bm \Omega_i\,\mathbf{\tilde X}_i
=\sum_{i=1}^n\sigma_i^2\left((1-\rho_i)\,\mathbf{X}_i'\mathbf{\tilde X}_i
  +\rho_i\,\mathbf X_i'\,\mathbf J_T\,\mathbf{\tilde X}_i\right)\,
\end{align*}

We note the following:

\begin{enumerate}
\item \textbf{The second term in the above sum $\rho_i\,\bX_i'\,\mathbf J_T\,\mathbf{\tilde X}_i$ is $\mathbf 0$ for all $i$.} \\
Because \(\mathbf{\tilde X}=\mathbf M_X \mathbf M_Z \mathbf X\) is orthogonal to the unit‐fixed‐effect columns of \(\mathbf X\), $\mathbf1_T' \mathbf{\tilde X}_i=\mathbf{0}$, and therefore $\mathbf J_T\,\mathbf{\tilde X}_i=\mathbf1_T \mathbf1_T' \mathbf{\tilde X}_i=\mathbf{0}$.

\item \textbf{The remaining terms take a similar form to Supplement Proposition \ref{sprop:bias_hetero}:} 

\begin{align*}
\bm{\Sigma}^B &= \left(\bX' \bX\right)^{-1} \bX' \bm \Omega  \mathbf{\tilde{X}}\left(\bX' \mathbf M_Z \bX \right)^{-1} \\
&= \left(\bX' \bX\right)^{-1} \left(\sum_{i=1}^n  \sigma_i^2 (1-\rho_i) \bX_i'   \mathbf{\tilde{X}}_i \right) \left(\bX' \mathbf M_Z \bX \right)^{-1} 
\end{align*}
\end{enumerate}
Applying logic from Supplement Proposition \ref{sprop:bias_hetero}, we know the upper left $K \times K$ entries of $\bm \Sigma^B$, $\bm \Sigma^B_{1:K,1:K} = 0$, and as a result, applying Supplement Proposition \ref{sprop:diff}, $Cov\left(\hat{\beta}, \hat{\beta}^{(e)}\right) = Var\left(\hat{\beta}\right)$, and thus $Cov\left(\hat{\beta}, \hat{\beta}-\hat{\beta}^{(e)}\right)=0$.
\end{proof}

\section{Appendix E. Test-induced distortions}

\begin{assumption}[No covariance condition]\label{assump:no_cov}
\emph{Assume that the correct model specification follows Eq.~(\ref{eq:correct}), with $\bm \epsilon^{(w)}\sim N(0, \bm \Omega)$, where $\bm \Omega$ is an $nT \times nT$ matrix, and that the reduced and expanded models are specified as in Eq.~(\ref{eq:gen_reduced}) and Eq.~(\ref{eq:gen_expanded}), with $\hat{\beta}$ and $\hat{\beta}^{(e)}$ denoting corresponding OLS ATT estimators. Further assume that the combination of error structure and model specifications yields $Cov\left(\hat{\beta}, \hat{\beta} - \hat{\beta}^{(e)}\right) = 0$.}
\end{assumption}

\begin{proposition}[Reduced model test-induced distortions under Assumption \ref{assump:no_cov}]\label{prop:no_bias}
\textit{Under Assumption \ref{assump:no_cov}, if we conduct a non-inferiority test with a threshold $\delta$, rejecting the null if $\hat{\beta} - \hat{\beta}^{(e)} < \delta^*$, where $\delta^*= z_{\alpha}{\sigma}_{\hat{\beta}-\hat{\beta}^{(e)}} + \delta$,
then there is no distortion in the reduced model ATT induced by conditioning on the test result}:  
\[\mathbb{E}\left(\hat{\beta} \middle | \hat{\beta} - \hat{\beta}^{(e)} < \delta^* \right) - \mathbb{E}\left(\hat{\beta} \right)= 0\] and likewise,
\[Var\left(\hat{\beta} \middle | \hat{\beta} - \hat{\beta}^{(e)} < \delta^* \right) - Var\left(\hat{\beta} \right)= 0.\]
\end{proposition}

\begin{proof}
Under Assumption \ref{assump:no_cov} and Proposition \ref{prop:test_inf},  $\hat{\beta}$ and $\hat{\beta}-\hat{\beta}^{(e)}$ are jointly Gaussian as linear functions of Gaussian errors, and thus, zero covariance implies independence. Therefore, $\mathbb{E}\left(\hat{\beta} \middle | \hat{\beta} - \hat{\beta}^{(e)} < \delta^* \right) = \mathbb{E}\left(\hat{\beta} \right)$ and $Var\left(\hat{\beta} \middle | \hat{\beta} - \hat{\beta}^{(e)} < \delta^* \right) = Var\left(\hat{\beta} \right)$.
\end{proof}

\begin{corollary}[Equivalence tests]
\textit{Under Assumption \ref{assump:no_cov}, if we conduct an equivalence test with a threshold $|\delta|$, rejecting the null if $\hat{\beta} - \hat{\beta}^{(e)} \in (\delta^*_L, \delta^*_U) = \left(z_{1-\alpha}{\sigma}_{\hat{\beta}-\hat{\beta}^{(e)}} - |\delta|, z_{\alpha}{\sigma}_{\hat{\beta}-\hat{\beta}^{(e)}} + |\delta|\right)$, then there is no distortion in the reduced model ATT induced by conditioning on the test result:}
\[\mathbb{E}\left(\hat{\beta} \Bigg| \hat{\beta} - \hat{\beta}^{(e)} \in(\delta_L^*,\delta_U^*) \right)-\mathbb{E}\left(\hat{\beta}\right) = 0\]
and likewise,
\[Var \left(\hat{\beta} \Bigg| \hat{\beta} - \hat{\beta}^{(e)} \in(\delta_L^*,\delta_U^*) \right)-Var \left(\hat{\beta}\right) = 0.
\]
\end{corollary}
\begin{proof}
 The result follows by applying the logic in Proposition \ref{prop:no_bias}.
\end{proof}

\begin{proposition}[Reduced model test-induced bias]
\label{prop:bias_non_iid}
Assume setup and reduced and expanded model estimators as in Proposition \ref{prop:test_inf}.  If we conduct a non-inferiority test with a threshold $\delta$, rejecting the null if $\hat{\beta} - \hat{\beta}^{(e)} < \delta^*$, where $\delta^*= z_{\alpha}{\sigma}_{\hat{\beta}-\hat{\beta}^{(e)}} + \delta$, then the expectation of the reduced model ATT estimator conditional on passing the test may differ from its unconditional expectation: 
\begin{align*}
\mathbb{E}\left(\hat{\beta} \middle | \hat{\beta} - \hat{\beta}^{(e)} < \delta^* \right) - \mathbb{E}\left(\hat{\beta}\right) &= -
\frac{Cov\left(\hat{\beta}, \hat{\beta} - \hat{\beta}^{(e)}\right)}{{\sigma}_{\hat{\beta}-\hat{\beta}^{(e)}}}
\frac{\phi\left(z_{\alpha} + \frac{\delta - \beta + \beta^{(e)}}{
        {\sigma}_{\hat{\beta}-\hat{\beta}^{(e)}}}\right)}
     {\Phi\left(z_{\alpha} + \frac{\delta - \beta + \beta^{(e)}}{
        {\sigma}_{\hat{\beta}-\hat{\beta}^{(e)}}}\right)},
\end{align*}
where $\phi$ and $\Phi$ are the probability density function and cumulative distribution function of a standard normal, respectively.
\end{proposition}

\begin{proof}
Let $\mathbb{E}\left(\hat{\beta}\right)=\beta$, the unconditional expected value of the reduced model effect estimate (but not necessarily an unbiased ATT) and $\mathbb{E}\left(\hat{\beta}^{(e)}\right)=\beta^{(e)}$. The distortion in the expected value of the reduced model treatment effect estimates induced only by the selection on pre-trends can be characterized by a truncated bivariate normal distribution:\citep{lee_exact_2016, roth_pretest_2022}

\begin{align*}
\mathbb{E}\left(\hat{\beta} - \beta \middle | \hat{\beta} - \hat{\beta}^{(e)} < \delta^* \right) = \frac{Cov\left(\hat{\beta}, \hat{\beta} - \hat{\beta}^{(e)}\right)}{Var\left(\hat{\beta}-\hat{\beta}^{(e)}\right)} \left(\mathbb{E}\left(\hat{\beta} - \hat{\beta}^{(e)} \middle | \hat{\beta} - \hat{\beta}^{(e)} < \delta^*\right) - \left({\beta} - {\beta}^{(e)}\right) \right)
\end{align*}
We substitute in $\mathbb{E}\left(\hat{\beta} - \hat{\beta}^{(e)} \middle | \hat{\beta} - \hat{\beta}^{(e)} < \delta^*\right)$, also following a truncated normal distribution: 

\begin{align*}
\mathbb{E}\left(\hat{\beta} - \beta \middle | \hat{\beta} - \hat{\beta}^{(e)} < \delta^* \right) &= \frac{Cov\left(\hat{\beta}, \hat{\beta} - \hat{\beta}^{(e)}\right)}{Var\left(\hat{\beta}-\hat{\beta}^{(e)}\right)} \Bigg[ \left({\beta} - {\beta}^{(e)}\right) - \\ &\sqrt{Var\left(\hat{\beta}-\hat{\beta}^{(e)}\right)} \frac{\phi\left(z_\alpha + \frac{\delta -\beta+\beta^{(e)}}{{\sigma}_{\hat{\beta}-\hat{\beta}^{(e)}}}\right)}{\Phi\left(z_\alpha + \frac{\delta - \beta+\beta^{(e)}}{{\sigma}_{\hat{\beta}-\hat{\beta}^{(e)}}}\right)}   - \left({\beta} - {\beta}^{(e)}\right) \Bigg] \\
&=-\frac{Cov\left(\hat{\beta}, \hat{\beta} - \hat{\beta}^{(e)}\right)}{{\sigma}_{\hat{\beta}-\hat{\beta}^{(e)}}}  \frac{\phi\left(z_\alpha + \frac{\delta -\beta+\beta^{(e)}}{{\sigma}_{\hat{\beta}-\hat{\beta}^{(e)}}}\right)}{\Phi\left(z_\alpha + \frac{\delta - \beta+\beta^{(e)}}{{\sigma}_{\hat{\beta}-\hat{\beta}^{(e)}}}\right)}
\end{align*}
 
\end{proof}

\begin{proposition}[Expanded model test-induced bias]
\label{prop:bias_expanded}
Assume setup and reduced and expanded model estimators as in Proposition \ref{prop:test_inf}. If we conduct a non-inferiority test with a threshold $\delta$, rejecting the null if 
$\hat{\beta} - \hat{\beta}^{(e)} < \delta^*$, where $\delta^*=  z_{\alpha}{\sigma}_{\hat{\beta}-\hat{\beta}^{(e)}} + \delta$, then the expectation of the expanded model ATT estimator conditional on passing the test may differ from its unconditional expectation:
\begin{align*}
\mathbb{E}\left(\hat{\beta}^{(e)} \middle | \hat{\beta} - \hat{\beta}^{(e)} < \delta^* \right) - \mathbb{E}\left(\hat{\beta}^{(e)}\right) &=
-\frac{Cov\left(\hat{\beta}^{(e)}, \hat{\beta} - \hat{\beta}^{(e)}\right)}{{\sigma}_{\hat{\beta}-\hat{\beta}^{(e)}}}
\frac{\phi\left(z_{\alpha} + \frac{\delta - \beta + \beta^{(e)}}{
        {\sigma}_{\hat{\beta}-\hat{\beta}^{(e)}}}\right)}
     {\Phi\left(z_{\alpha} + \frac{\delta - \beta + \beta^{(e)}}{
        {\sigma}_{\hat{\beta}-\hat{\beta}^{(e)}}}\right)}. 
        \end{align*}
\end{proposition}

\begin{proof}
The result follows by the same logic as Proposition \ref{prop:bias_non_iid}, applying the bivariate normal truncation formula to $\hat{\beta}^{(e)}$ conditioning on $\hat{\beta} - \hat{\beta}^{(e)} < \delta^*$.
\end{proof}

\section{Appendix F. Non-inferiority test power} \label{app:NI.power}

\begin{remark} \label{finite.sample.power}
Under standard regularity conditions, OLS estimators are asymptotically normal:
\[
\frac{\hat{\beta}_k - \beta_k}{{\sigma_{\hat{\beta}_k}}}
\xrightarrow{d} N(0, 1),
\]
where $\sigma^2_{\hat{\beta}_k} = {Var}\left(\hat{\beta}_k \mid \mathbf{X}\right)$ is the $(k,k)$ element of
$(\mathbf{X}'\mathbf{X})^{-1} \mathbf{X}'\mathbf{\Omega}\mathbf{X} (\mathbf{X}'\mathbf{X})^{-1}$. Consistent variance estimators satisfy
$\hat{\sigma}^2_{\hat{\beta}_k} \xrightarrow{p} \sigma^2_{\hat{\beta}_k}$.
In this section, we discuss power evaluated under this asymptotic normal approximation, treating
$\sigma^2_{\hat{\beta}_k}$ as known.
\end{remark}

\begin{sproposition}[Non-inferiority power heuristic]\label{prop:NI.heuristic}
\textit{Assume conditions in Proposition \ref{prop:test_inf}. If a one-sided Wald test has power $p$ evaluated under an asymptotic normal approximation to reject $H_0: \beta-\beta^{(e)} \leq 0$ given $\beta-\beta^{(e)} = \beta^* - \beta^{(e)*}$ at level $\alpha$, then the non-inferiority formulation will have power $p$ to reject $H_0: \beta - \beta^{(e)}\geq \beta^* - \beta^{(e)*}$, given no violation exists.}
\end{sproposition}
\begin{proof}
Suppose we want to perform a Wald test on $\beta-\beta^{(e)}$ from Equation~(\ref{eq:gen_expanded}). We specify a one-sided test with $H_0: \beta-\beta^{(e)} \leq 0$ versus the alternative $H_A: \beta-\beta^{(e)} > 0$. Using the asymptotic distribution of the test statistic under the null, the critical value should be $\Phi^{-1}(1-\alpha)$ to control type I error at $\alpha$. Then assuming $\hat{\beta}-\hat{\beta}^{(e)}{\sim} N\left(\beta-\beta^{(e)}, \sigma^2_{\hat{\beta}-\hat{\beta}^{(e)}}\right)$ (see Remark \ref{finite.sample.power}) and $\beta-\beta^{(e)} = \beta^*-\beta^{(e)*}$, we define power ($p$), the probability that the test statistic exceeds the critical value:
$$
p = Pr\left(\frac{\hat{\beta}-\hat{\beta}^{(e)}}{\sigma_{\hat{\beta}-\hat{\beta}^{(e)}}} > \Phi^{-1}(1-\alpha) \Bigg | \beta-\beta^{(e)} = \beta^*-\beta^{(e)*} \right)
$$
We subtract $\frac{\beta-\beta^{(e)}}{\sigma_{\hat{\beta}-\hat{\beta}^{(e)}}}$ from each side:
$$
p = Pr\left(\frac{\hat{\beta}-\hat{\beta}^{(e)}}{\sigma_{\hat{\beta}-\hat{\beta}^{(e)}}} - \frac{\beta-\beta^{(e)}}{\sigma_{\hat{\beta}-\hat{\beta}^{(e)}}} > \Phi^{-1}(1-\alpha) - \frac{\beta-\beta^{(e)}}{\sigma_{\hat{\beta}-\hat{\beta}^{(e)}}} \Bigg | \beta-\beta^{(e)} = \beta^*-\beta^{(e)*} \right)
$$
Then, because $\hat{\beta}-\hat{\beta}^{(e)} {\sim} N \left(\beta^*-\beta^{(e)*}, \sigma^2_{\hat{\beta}-\hat{\beta}^{(e)}}\right)$, we can rewrite our expression as:
$$
p = 1-\Phi \left(\Phi^{-1}(1-\alpha) - \frac{\beta^*-\beta^{(e)*}}{\sigma_{\hat{\beta}-\hat{\beta}^{(e)}}}\right)
$$
Using $1-\Phi(x)=\Phi(-x)$, and $-\Phi^{-1}(\alpha)=\Phi^{-1}(1-\alpha)$ for $\alpha \in [0,1]$, we rearrange as:
$$
p = \Phi \left(\Phi^{-1}(\alpha) +  \frac{\beta^*-\beta^{(e)*}}{\sigma_{\hat{\beta}-\hat{\beta}^{(e)}}}\right) 
$$
This means we can write $\beta^*-\beta^{(e)*}$, the value at which power $p$ is achieved as:
$$
\beta^*-\beta^{(e)*} = \sigma_{\hat{\beta}-\hat{\beta}^{(e)}} \left(\Phi^{-1}(p) - \Phi^{-1}(\alpha)\right)
$$
That is, if $\beta-\beta^{(e)}= \beta^*-\beta^{(e)*} = \sigma_{\hat{\beta}-\hat{\beta}^{(e)}} \left(\Phi^{-1}(p) - \Phi^{-1}(\alpha)\right)$, we will have power $p$ to reject the null that $\beta-\beta^{(e)}\leq 0$.

Now suppose we conduct a non-inferiority test. Following the reasoning above, we formulate a test with a bound based on the $\beta^*-\beta^{(e)*}$ from the main study. That is, we wish to test whether we can rule out differences (relative to the reference of $0$) at least as big as the treatment effect we are powered to detect. The hypotheses are therefore $H_0: \beta-\beta^{(e)}\geq \beta^*-\beta^{(e)*}$ versus $H_A: \beta-\beta^{(e)}< \beta^*-\beta^{(e)*}$. We write a test statistic $\frac{\hat{\beta}-\hat{\beta}^{(e)}-\left(\beta^*-\beta^{(e)*}\right)}{\sigma_{\hat{\beta}-\hat{\beta}^{(e)}}}$. We reject violations when the test statistic is smaller than a critical value. As before, we determine the critical value by assuming that the null is true and controlling the Type I error rate at $\alpha$, which yields a critical value of $\Phi^{-1}(\alpha)$.

We are interested in the power of the test when the parallel trends assumption is exactly met, that is, when $\beta-\beta^{(e)}=0$. By definition, this is:

$$
P\left(\frac{\hat{\beta}-\hat{\beta}^{(e)}-\left(\beta^*-\beta^{(e)*}\right)}{\sigma_{\hat{\beta}-\hat{\beta}^{(e)}}} < \Phi^{-1}(\alpha) \Bigg | \beta-\beta^{(e)}= 0 \right) = P\left(\frac{\hat{\beta}-\hat{\beta}^{(e)}}{\sigma_{\hat{\beta}-\hat{\beta}^{(e)}}} < \Phi^{-1}(\alpha) + \frac{\beta^*-\beta^{(e)*}}{\sigma_{\hat{\beta}-\hat{\beta}^{(e)}}} \Bigg | \beta-\beta^{(e)}= 0 \right)
$$
Because we assume $\beta-\beta^{(e)}= 0$, we then have $\frac{\hat{\beta}-\hat{\beta}^{(e)}}{\sigma_{\hat{\beta}-\hat{\beta}^{(e)}}} {\sim} N(0,1)$. Plugging in the standard normal cumulative distribution function and our expression for $\beta^*-\beta^{(e)*}$ from above, we obtain

$$
\Phi\left( \Phi^{-1}(\alpha) + \frac{\beta^*-\beta^{(e)*}}{\sigma_{\hat{\beta}-\hat{\beta}^{(e)}}}\right) = \Phi\left( \Phi^{-1}(\alpha) + \left(\Phi^{-1}(p) - \Phi^{-1}(\alpha)\right)\right) = p.
$$
\end{proof}

\begin{corollary}[Equivalence test power]
Assume conditions in Proposition \ref{prop:test_inf}. If a Wald non-inferiority test conducted as in Supplement Proposition \ref{prop:NI.heuristic} has power $p$ to rule out $H_0: \beta - \beta^{(e)} \geq |\beta^*-\beta^{(e)*}|$ given $\beta-\beta^{(e)}=0$, then a Wald equivalence test has power less than or equal to $p$ to reject $H_0: |\beta-\beta^{(e)}| \geq |\beta^*-\beta^{(e)*}|$.
\end{corollary}

\begin{proof}
As detailed in Supplement Proposition \ref{prop:NI.heuristic}, the non-inferiority rejection event for ruling out
$H_0:\beta-\beta^{(e)} \ge |\beta^*-\beta^{(e)*}|$
is:
\[
A
=
\left\{
\frac{\hat{\beta}-\hat{\beta}^{(e)}}{\sigma_{\hat{\beta}-\hat{\beta}^{(e)}}}
<
\Phi^{-1}(\alpha) + \frac{|\beta^*-\beta^{(e)*}|}{\sigma_{\hat{\beta}-\hat{\beta}^{(e)}}}
\right\}
\]
A Wald equivalence test rejects
$H_0:|\beta-\beta^{(e)}| \ge |\beta^*-\beta^{(e)*}|$
only if it rejects both one-sided null hypotheses
$H_{0U}:\beta-\beta^{(e)} \ge |\beta^*-\beta^{(e)*}|$
and
$H_{0L}:\beta-\beta^{(e)} \le -|\beta^*-\beta^{(e)*}|$.
In terms of the same standardized statistic, the additional rejection event is:
\[
B
=
\left\{
\frac{\hat{\beta}-\hat{\beta}^{(e)}}{\sigma_{\hat{\beta}-\hat{\beta}^{(e)}}}
>
\Phi^{-1}(1-\alpha)-\frac{|\beta^*-\beta^{(e)*}|}{\sigma_{\hat{\beta}-\hat{\beta}^{(e)}}}
\right\}
\]
Therefore the equivalence test rejection event is $A \cap B$, and since
$A \cap B \subseteq A$,
\[
P\left(A \cap B \mid \beta-\beta^{(e)}=0\right)
\le
P\left(A \mid \beta-\beta^{(e)}=0\right)
=
p,
\]
and the power of the Wald equivalence test is less than or equal to $p$.
\end{proof}

\normalsize

\begin{proposition}[Non-inferiority difference-in-differences power]\label{prop:DiD.NI.heuristic.alt}
\textit{Assume setup and reduced and expanded model estimators as in Proposition \ref{prop:test}. If a non-inferiority test has power $p$ evaluated under an asymptotic normal approximation to rule out violations of parallel trends equal to or larger than $\beta^*_k$ (i.e., to reject $H_0: \beta_k- \beta_k^{(e)} \geq \beta_k^*$, with $\beta_k^*>0$) in a Wald test at level $\alpha$, and assuming no violation exists $(\bm \theta = \mathbf{0})$, then $p > p_e$, where $p_e$ is the power to detect $\beta_k^{(e)} = \beta_k^*$ (likewise evaluated) in a one-sided Wald test at level $\alpha$ in an expanded model.}
\end{proposition}
\begin{proof}
We wish to compare the power to detect an effect of size $\beta^{(e)}_k = \beta_k^*$ to the power to rule out a difference $\beta_k - \beta^{(e)}_k \geq \beta_k^*$, assuming $\bm \theta = \mathbf{0}$. We consider the following asymptotic normal approximations (see Remark \ref{finite.sample.power}), noting that because $\bm \theta = \mathbf{0}$, we have $\beta_k^{(e)} = \beta_k$:
\begin{align*}
  \hat{\beta}_k &{\sim} N\left(\beta_k, \sigma^2_{\hat{\beta}_k}\right) \\
  \hat{\beta}^{(e)}_k &{\sim} N\left(\beta_k, \sigma^2_{\hat{\beta}^{(e)}_k}\right)
\end{align*}
Let $\mathbf{X}_{-\mathbf{trt}_k}$ denote the matrix of the inputs in the reduced model other than $\mathbf{trt}_k = G_i\mathbb{I}(t = k)$ and $\mathbf{Z}$ indicate the additional columns in the expanded model.  In what follows, $\sigma^2$ and $R^2$ terms denote their population analogs (i.e., probability limits of the corresponding sample quantities).  Through standard regression algebra, the asymptotic variances satisfy:
\begin{align*}
  \sigma^2_{\hat{\beta}_k} &= \frac{\sigma^2_\varepsilon }{SST_{\mathbf{trt}_k}\left(1-R^2_{\mathbf{trt}_k | \mathbf{X}_{-\mathbf{trt}_k}}\right)} \\
  \sigma^2_{\hat{\beta}^{(e)}_k} &= \frac{\sigma^2_\varepsilon }{SST_{\mathbf{trt}_k}\left(1-R^2_{\mathbf{trt}_k | \mathbf{X}_{-\mathbf{trt}_k}, \mathbf{Z}}\right)} \\
  &= \frac{1-R^2_{\mathbf{trt}_k| \mathbf{X}_{-\mathbf{trt}_k}}}{1-R^2_{\mathbf{trt}_k | \mathbf{X}_{-\mathbf{trt}_k}, \mathbf{Z}}} \sigma^2_{\hat{\beta}_k} \\
  &= \kappa \sigma^2_{\hat{\beta}_k}, \quad \text{where } \kappa = \frac{1-R^2_{\mathbf{trt}_k | \mathbf{X}_{-\mathbf{trt}_k}}}{1-R^2_{\mathbf{trt}_k | \mathbf{X}_{-\mathbf{trt}_k}, \mathbf{Z}}}
\end{align*}
The second equality uses the fact that when $\bm \theta = \mathbf{0}$, adding $\mathbf{Z}$ to the model does not change the population residual variance.  Further, because $Cov\left(\hat{\beta}_k, \hat{\beta}^{(e)}_k\right) = \sigma^2_{\hat{\beta}_k}$ per Lemma \ref{lemma:bias},\citep{clogg_statistical_1995} we have:
\begin{align*}
 \hat{\beta}_k - \hat{\beta}^{(e)}_k {\sim} N\left(0, (\kappa-1) \sigma^2_{\hat{\beta}_k}\right)
\end{align*}
Note that $\kappa \geq 1$ always, because adding regressors cannot decrease the regression $R^2$ (i.e., $R^2_{\mathbf{trt}_k\mid \mathbf{X}_{-\mathbf{trt}_k}, \mathbf{Z}} \geq R^2_{\mathbf{trt}_k\mid \mathbf{X}_{-\mathbf{trt}_k}}$). 
Following the logic in Supplement Proposition \ref{prop:NI.heuristic}, power to detect an effect of $\beta^{(e)}_k = \beta^*_k$ at level $\alpha$ in a one-sided test is:
\begin{align*}
  p_e &= \Phi\left( \Phi^{-1}(\alpha) + \frac{\beta^*_k}{\sqrt{\kappa} \sigma_{\hat{\beta}_k}} \right) 
\end{align*}
Likewise the power to rule out a difference ${\beta}_k-{\beta}_k^{(e)} \geq \beta^*_k$ at level $\alpha$ is:
\begin{align*}
  p = \Phi\left( \Phi^{-1}(\alpha) + \frac{\beta^*_k}{\sqrt{\kappa-1} \sigma_{\hat{\beta}_k}} \right)
\end{align*}

Because $\sqrt{\kappa - 1} < \sqrt{\kappa}$, we know $p > p_e$ for $\kappa > 1$. (When $\kappa=1$, the reduced and expanded estimators coincide, and no test is needed.)
\end{proof}

The assumption $\beta_k^* > 0$ is set because we posit that researchers aim to show that violations are small in magnitude. With a negative violation, the same logic applies with a non-inferiority null hypothesis of $H_0: \beta_k- \beta_k^{(e)} \leq \beta_k^*$. Furthermore, tests for main effects in reduced and expanded models would typically be two-sided, further reducing their power relative to a direction-specific non-inferiority test.

Last, for completeness, note that the power to detect an effect in the reduced model at level $\alpha$ in a one-sided test is:
\begin{align*}
  p_r = \Phi\left( \Phi^{-1}(\alpha) + \frac{\beta^*_k}{\sigma_{\hat{\beta}_k}} \right)
\end{align*}
  If $\kappa < 2$, indicating low added explanatory power for $\mathbf{trt}_k$ from $\mathbf{Z}$, then $p > p_r$.

\section{Appendix G. Additional results for the ACA dependent coverage re-analysis}

\begin{figure}[ht]
\centerline{\includegraphics[width=1.0\textwidth]{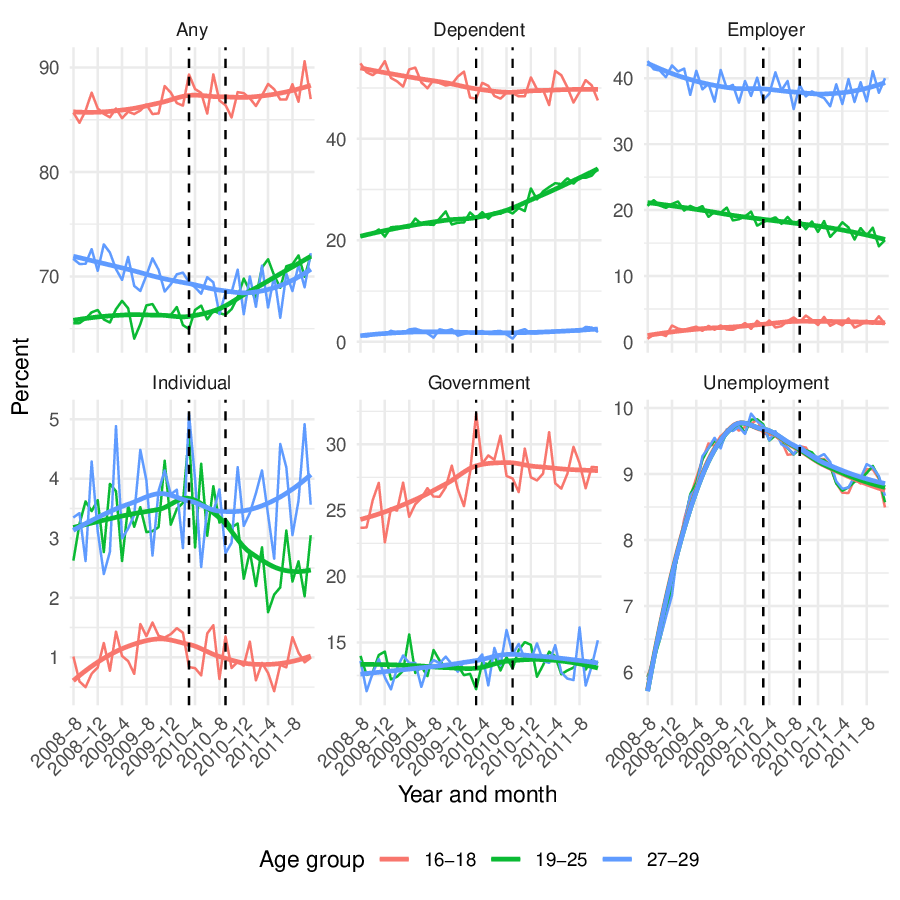}}
\caption{Monthly insurance coverage and unemployment by age group. Dashed vertical lines mark the beginnings of the enactment and implementation periods.}\label{fig:ue_trends}
\end{figure}

During most of the pre-period of this study, the US experienced rapid growth in unemployment due to a recession (see the bottom right plot of Figure~\ref{fig:ue_trends}).
According to the Federal Reserve, this recession lasted from Jan 2008 to June 2009.\citep{fed_recessions}

The three age groups were affected differently by this rising unemployment.
During the period of rising unemployment, both the treated group (ages 19-25) and the older comparison group (ages 27-29) lost employer coverage, while the younger comparison group (ages 16-18) lost dependent coverage (presumably as their parents became unemployed).
Those in the younger comparison group were eligible for Medicaid as dependents and therefore received government insurance. Those in the treated group (even in the pre-period) often gained dependent coverage (perhaps due to a combination of voluntary coverage of people in these age groups by employers and state laws already enacted prior to the ACA). 
The older comparison group was more likely to become uninsured, having access to neither dependent coverage nor Medicaid.

\begin{figure}[ht]
\centerline{\includegraphics[width=1.0\textwidth]{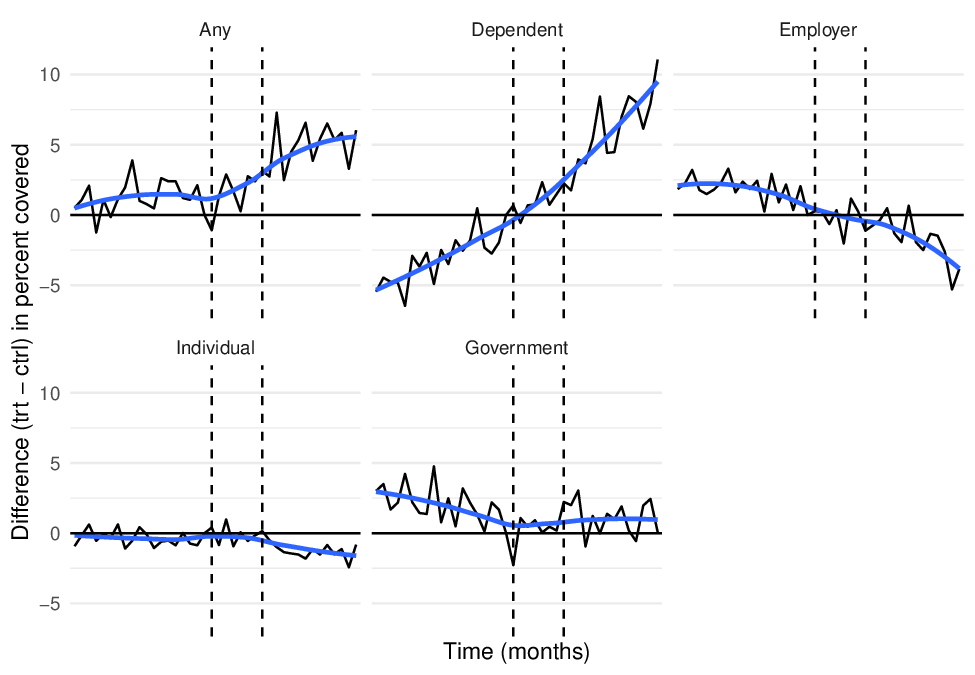}}
\caption{Differential monthly insurance coverage trends in treatment versus control. Dashed vertical lines mark the beginnings of the enactment and implementation periods.}\label{fig:pre-period_diff_coverage}
\end{figure}

Why didn't the original authors' evaluations of pre-intervention trends (including those reported in their Appendix Table A1) suggest violations? First, although the two comparison age cohorts had different trends, their average trends were more similar to the treated group (Figure \ref{fig:ue_trends}). Although some visual differences may still be apparent (see Figure \ref{fig:pre-period_diff_coverage}, which expands on Figure 1 of Akosa Antwi et al. by including the rest of the insurance variables), there were likely two additional factors at play. As noted in the main text, conventional tests have limited power and may have been underpowered to detect effects. This may have been exacerbated by the fact that test models were fit to pre-intervention data and tested for differential trends using the same control variables as their analytic specification, which included an interaction term between unemployment and treatment group that was highly correlated with the interaction between time and treatment group in the pre-intervention period.

\begin{figure}
\centerline{\includegraphics[width=1.0\textwidth]{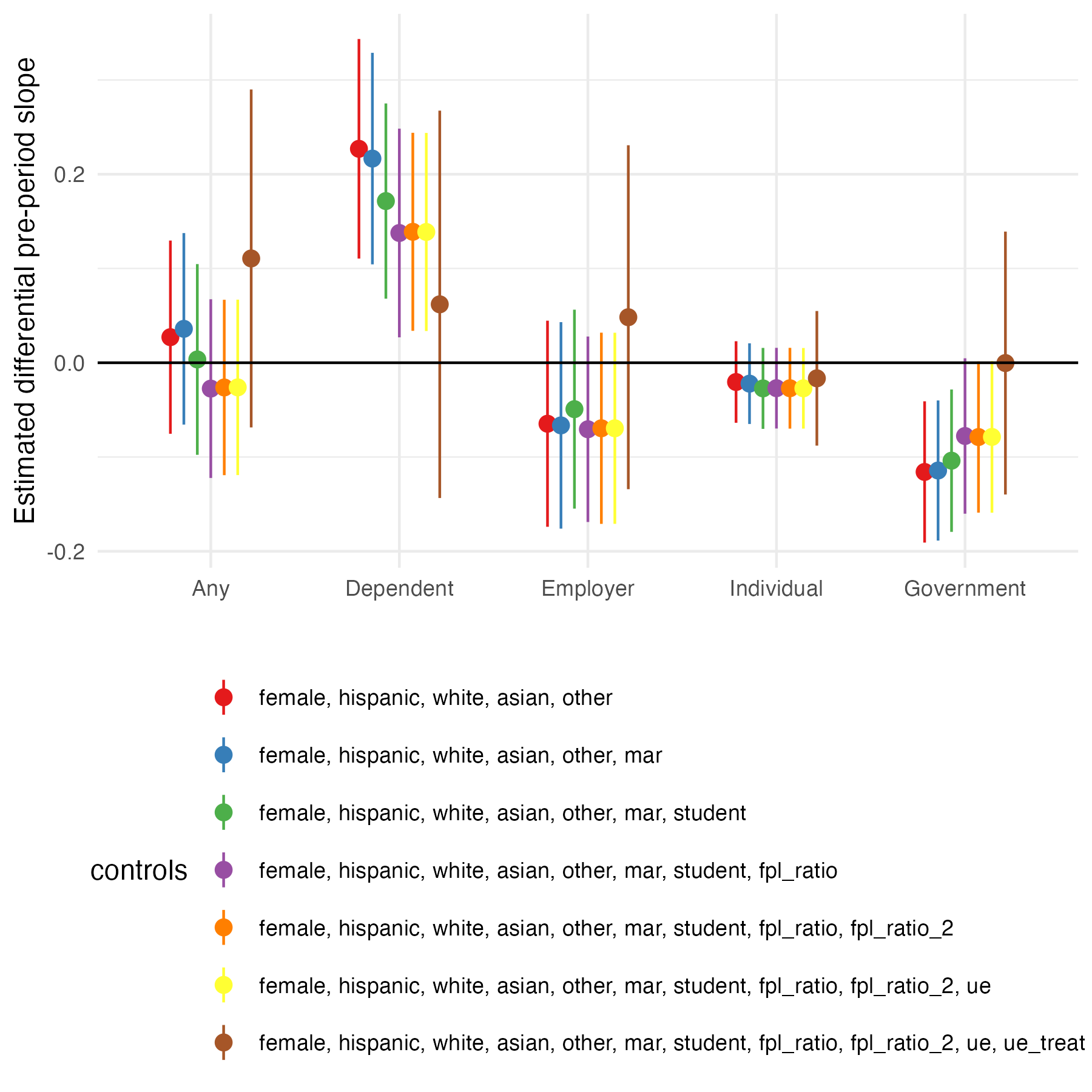}}
\caption{Estimated differential pre-period slopes from the model in Eq.~(15) of the main text, using different sets of control variables (estimated only on pre-intervention data). mar = married; fpl\_ratio = household income expressed as a ratio of the federal poverty limit; ue = state-month unemployment; ue\_treat = state-month unemployment interacted with treatment group indicator}\label{fig:pre-period_ins_slopes_by_control_vars}
\end{figure}

Figure~\ref{fig:pre-period_ins_slopes_by_control_vars} illustrates the impact of this collinearity.
It plots the pre-period slopes ($\theta$) estimated by fitting the model with a differential linear trend to pre-period data only, varying the included control variables in $\mathbf{X}_{it}$.
With the interaction between treatment group and unemployment in the model, the added differential trend parameter was nearly collinear, roughly doubling the standard errors of the estimated slope difference.
This made it more likely that we would pass a conventional trends test, but fail a non-inferiority test.

For completeness, we used the original authors' model specifications (with all control variables in the regression) to replicate their results for the conventional tests of parallel pre-trends (Table~\ref{tab:ptt_replication_results}) and estimated treatment effects (Tables~\ref{tab:replication_trt_effects}-\ref{tab:replication_trt_effects2}). However, we also show results without the treatment/unemployment interaction, which has greater non-inferiority test power.

\begin{table}[ht]
\caption{Conventional tests of parallel trends in the pre-implementation period. These were estimated by fitting the model in Eq.~(15) of the main text, using the specification and all control variables from the original analysis.\citep{akosa_antwi_effects_2013} 
The estimates (and the state-clustered robust standard errors) represent monthly differential slopes on the percentage point scale. SE = standard error}
\label{tab:ptt_replication_results}
\begin{center}
\begin{tabular}{lrcc}
\hline
Outcome & Estimate (SE) \\
\hline
Any & 0.11 (0.09)\\
Dependent & 0.06 (0.1)\\
Employer & 0.05 (0.09)\\
Individual & -0.02 (0.04)\\
Government & 0.01 (0.07)\\
\hline
\end{tabular}
\end{center}
\end{table}

\begin{table}[ht]
\begin{center}
\caption{Estimated effects of the ACA dependent coverage provision on insurance coverage (replication). These were estimated by fitting the reduced model in Eq.~(13) of the main text, using the original specification and all control variables from the original analysis.\citep{akosa_antwi_effects_2013} The estimates (and their state-clustered robust standard errors) represent differential changes on the percentage point scale, averaged over the post-implementation period. SE = standard error}\label{tab:replication_trt_effects}
\begin{tabular}{lr}
\hline
Outcome & Estimate (SE) \\
\hline
Any & 3.18 (0.74)\\
Dependent & 7.02 (0.69)\\
Employer & -3.12 (0.60)\\
Individual & -0.80 (0.23)\\
Government & -0.25 (0.57)\\
\hline
\end{tabular}
\end{center}
\end{table}

\begin{table}[ht]
\begin{center}
\caption{Estimated effects of the ACA dependent coverage provision on insurance coverage. These were estimated by fitting the reduced model in Eq.~(13) of the main text, using all control variables from the original analysis with month-year fixed effects but changing the specification to include saturated treatment effects (and omitting linear trend variables).\citep{akosa_antwi_effects_2013} The estimates (and their state-clustered robust standard errors) represent differential changes on the percentage point scale, averaged over the post-implementation period. SE = standard error}\label{tab:replication_trt_effects2}
\begin{tabular}{lr}
\hline
Outcome & Estimate (SE) \\
\hline
Any & 3.16 (0.75)\\
Dependent & 6.99 (0.71)\\
Employer & -3.10 (0.6)\\
Individual & -0.80 (0.23)\\
Government & -0.23 (0.57)\\
\hline
\end{tabular}
\end{center}
\end{table}

\clearpage

\clearpage
\newpage

\newpage

\bibliographystyle{wileyNJD-AMA} 
\bibliography{wileyNJD-AMA}